\documentclass[10pt,dvips,twoside]{article}
\usepackage{amstext,amsmath,amssymb,amsfonts}
\usepackage{cite}
\usepackage{graphicx}
\usepackage{epsfig}
\usepackage[dvips,usenames]{color}
\usepackage{color}
\usepackage{rotating}

\usepackage[all]{xy}
\usepackage{tabularx}
\usepackage{fancyhdr}

\usepackage{latexsym}

\newtheorem{theo}{Theorem}[section]

\newtheorem{lem}[theo]{Lemma}
\newtheorem{prop}[theo]{Proposition}
\newtheorem{cor}[theo]{Corollary}

\newtheorem{definition}[theo]{Definition}

\newtheorem{example}[theo]{Example}
 
 \numberwithin{equation}{section}

\newtheorem{remark}[theo]{Remark}

\newcommand{\betheo}{\begin{theo}$\!\!\!${\bf } }
\newcommand{\entheo}{\end{theo}}

\newcommand{\becor}{\begin{cor}$\!\!\!$  }
\newcommand{\encor}{\end{cor}}

\newcommand{\belem}{\begin{lem}$\!\!\!$  }
\newcommand{\enlem}{\end{lem}}

\newcommand{\beprop}{\begin{prop}}
\newcommand{\enprop}{\end{prop}}

\newcommand{\bedefi}{\begin{definition}$\!\!\!$ \rm }
\newcommand{\findefi}{ \end{definition}}

\newcommand{\beex}{\begin{example}$\!\!\!$ \rm }
\newcommand{\enex}{ \end{example}}

\newcommand{\berem}{\begin{remark}$\!\!\!$ \rm }
\newcommand{\enrem}{ \end{remark}}

\newcommand{\bea}{\begin{eqnarray}}
\newcommand{\ena}{\end{eqnarray}}

\newcommand{\beano}{\begin{eqnarray*}}
\newcommand{\enao}{\end{eqnarray*}}

\newcommand{\bee}{\begin{enumerate}}
\newcommand{\ene}{\end{enumerate}}

\newcommand{\bei}{\begin{itemize}}
\newcommand{\eni}{\end{itemize}}

\newcommand{\betab}{\begin{tabular}}
\newcommand{\entab}{\end{tabular}}

\newcommand{\CN}{\mathbb{C}}

\newcommand{\IN}{\mathbb{I}}
\newcommand{\NN}{\mathbb{N}}
\newcommand{\PN}{{\mathbb P}}
\newcommand{\RN}{\mathbb{R}}
\newcommand{\ZN}{\mathbb{Z}}

\newcommand{\noi}{\noindent}

\renewcommand{\leq}{\leqslant}
\renewcommand{\geq}{\geqslant}

\def\dag{\dagger}

\newcommand{\Id}{1\!\!1}

\begin{document}

 \setcounter{section}{0}
\setcounter{equation}{0}
\setcounter{figure}{0}
\setcounter{table}{0}
\setcounter{footnote}{0}


\vspace*{10mm}

\begin{center}
{\bf\Large  On Hilbert-Schmidt operator formulation of noncommutative quantum mechanics} 
\vspace{10pt}
\end{center}



\begin{center}
\vspace*{10pt}

{\large\sc Isiaka Aremua $\!^{\rm a,b}$, Ezinvi Balo\"{i}tcha $\!^{\rm b}$, 
    Mahouton Norbert Hounkonnou $\!^{\rm b}$  and  Komi Sodoga $\!^{\rm a,b}$}
\\[3mm]
$^{\rm a}$ \textsl{\small  Universit\'e de Lom\'e, Facult\'e  des Sciences, D\'epartement de Physique,\\
 Laboratoire de Physique des Mat\'eriaux et de M\'ecanique Appliqu\'ee,
 02 BP 1515 Lom\'e, Togo  \\
\hspace*{3mm}E-mail: iaremua@univ-lome.tg, ksodoga@univ-lome.tg}
\\[2mm]
$^{\rm b}$ \textsl{\small University of Abomey-Calavi, International Chair in Mathematical Physics   and Applications (ICMPA), 072 B.P. 050  Cotonou, Benin \\
\hspace*{3mm}E-mail : ezinvi.baloitcha@cipma.uac.bj,
norbert.hounkonnou@cipma.uac.bj
}
\end{center}

\vspace{15pt}

\begin{quote}
 
This work gives value to the importance of Hilbert-Schmidt operators in the formulation of  noncommutative quantum theory. 
A system of charged particle in a constant magnetic field is investigated in this framework.

\end{quote}

\thispagestyle{plain}
\fancyfoot{}


\renewcommand\headrulewidth{0.5pt}

\section*{Introduction}
\label{sect_intro}

The theory of Hilbert-Schmidt operators plays a key role in the formulation of the noncommutative quantum mechanics. In the past three decades,
the von Neumann algebras \cite{vonneumann, vonneumann1} underwent a vigorous growth after the discovery of a natural infinite family of 
pairwise nonisomorphic factors, and the advent of Tomita-Takesaki theory \cite{takesaki} and   Connes noncommutative geometry \cite{connes1} 
iniated with  the classification theorems for von Neumann algebras and structure of extensions of $C^{*}$-algebras \cite{connes}. 
The modular theory of von Neumann algebras was created by M. Tomita \cite{tomita} in 1967 and perfectioned  by M. Takesaki around 1970. 

\noi From physical point of view, a charged  particle interacting with a constant magnetic field  is  one of the important problems in
quantum mechanics described by the Hamiltonian $\mathcal H = \frac{1}{2M}({\bf p} + \frac{e}{c}{\bf A})^{2}$, 
inspired by condensed matter physics, quantum optics, etc.  The Landau problem \cite{landau}
is related to the motion of a charged particle on the flat plane xy in
the presence of a constant magnetic field along the $z$-axis. In metals,  the electrons occupy many Landau levels \cite{goerbig} 
 $E_{n} = \hbar \omega_{c} (n+\frac{1}{2})$, each level being infinitely 
degenerate, with $\omega_{c} = eB/Mc$, the cyclotron frequency, which are those of the one-dimensional harmonic oscillator, and correspond 
to the kinetic
energy levels of electrons.

\noi This  physical model represents an interesting application \cite{ali-bagarello} of the Tomita-Takesaki modular
theory \cite{tomita, takesaki}. Taking into account the sense of the magnetic field, one obtains a pair of commuting Hamiltonians.   Both
these Hamiltonians can be written in terms of two pairs of mutually commuting oscillator-type
creation and annihilation operators, which then generate two mutually commuting  von Neumann algebras, commutants of each other. 
 The associated von Neumann algebra of observables
displays a modular structure in the sense of the Tomita-Takesaki theory, with
the algebra and its commutant referring to the two orientations of the magnetic
field.

\noi  Hilbert spaces, at the mathematical side, realize  the skeleton of quantum theories. 
Coherent states
(CS), defined as a specific overcomplete family of vectors in the Hilbert space describing 
quantum phenomena \cite{sch, klauder-skagerstam, perelomov, ali-antoine-gazeau, gazbook09}, constitute an important tool of investigation.
In the studies and understanding of   noncommutative geometry, 
CS were  
proved to be useful objects \cite{grosse-presnajder}.  
Based on the approach developed in \cite{scholtz}, Gazeau-Klauder CS were constructed in
 noncommutative quantum mechanics \cite{ben-scholtz}. Besides, in  studying, in the noncommutative plane \cite{a-hk}, the 
 behavior of an electron in an external uniform electromagnetic background coupled
to a harmonic potential, matrix vector coherent states (MVCS)  as
well as quaternionic vector coherent states (QVCS) were  
constructed and discussed.

\noi Our present contribution paper  is organized as follows:
 
\bei
\item
First,  we formulate the Hilbert-Schmidt operators and the Tomita-Takesaki modular theories in the framework of
noncommutative quantum mechanics.  
\item  Detailed proofs are given for main frequently used statements in the study of  modular theory and Hilbert-Schmidt operators.  
As application,  a construction of CS from the thermal state is achieved as in a previous 
work \cite{ali-bagarello}.  Relevant properties are discussed.  Then, a light is put on  the Wigner map as an
interplay between the  noncommutative quantum mechanics formalism \cite{scholtz} and  the  modular theory based on Hilbert-Schmidt operators. 
 
\item Finally, the motion of a charged particle on the flat plane xy in
the presence of a constant magnetic field along the $z$-axis with a harmonic potential is studied.
\eni

\section{von Neumann algebras: modular theory, Hilbert-Schmidt operators and coherent states}
This section recapitulates fundamental notions and main ingredients of the modular theory used in the
sequel. More details on these mathematical structures and their applications may be found in a series of
works
 \cite{vonneumann, vonneumann1, tomita, takesaki, bratelli, prugovecki, connes, ali-bagarello, ali-bagarello-honnouvo, bertozzini}
(and  references therein), widely exploited to write this review section.

\subsection{Basics on von Neumann algebras}

\noi In this paragraph, $\mathfrak{H}$ denotes a Hilbert space over $\CN.$ $\mathfrak{H}$ is assumed to be separable, of dimension $N, $ 
which could be finite or infinite. Denote by $\mathcal L(\mathfrak H)$ { the $C^{*}$-algebra} of all bounded operators on $\mathfrak{H}.$ 
{ The following definitions are in order}:

\bedefi 

\noi Let $\mathfrak G$ be an algebra. A mapping $A \in \mathfrak G \mapsto A^{*} \in \mathfrak G$ is called an {\it involution}, or {\it adjoint operation}, of the algebra $\mathfrak{G}, $ if it has the
following properties:
\begin{enumerate}
\item $A^{**} = A$
\item $(AB)^{*} = B^{*}A^{*},$ { with $A, B \in \mathfrak G, \, A^{*}, B^{*} \in \mathfrak G$}
\item $(\alpha A + \beta B)^{*} = \bar{\alpha}A^{*} +  \bar{\beta}B^{*}, \, \alpha, \beta \in \CN $.

\noi ($\bar{\alpha}$ is the complex conjugate of $\alpha.$)
\end{enumerate}

\findefi


\bedefi $^{*}$-algebra 

\noi An algebra with an involution is called  {\it $^{*}$-algebra} and a subset $\mathfrak{B}$ of $\mathfrak{G}$ is called self-adjoint if $A \in \mathfrak{B}$ implies that $A^{*} \in \mathfrak{B}.$
\findefi

The algebra $\mathfrak{G}$ is a normed algebra if to each $A \in \mathfrak{G}$ there is associated a real number $||A||, $ the norm of $A, $ satisfying the requirements
\bee
\item $||A|| \geq 0$ and $||A|| = 0$ if, and only if, $||A||= 0, $
\item $||\alpha A|| = |\alpha | ||A||,$
\item $||A + B|| \leq ||A|| + ||B||,$
\item $||AB|| \leq ||A|| ||B||.$
\ene
The third of these conditions is called the
 triangle inequality and the fourth the
product inequality. The norm defines a metric topology on $\mathfrak{G}$ which is referred to as the uniform topology. The neighborhoods of an element $A \in \mathfrak{G}$ in this topology are given  by 
\bea
\mathcal U(A; \varepsilon) = \{B; B \in \mathfrak{G}, ||B-A||< \varepsilon\}, 
\ena
where $\varepsilon > 0.$ If $\mathfrak{G}$ is complete with respect to the uniform topology, then it is called a Banach algebra. A normed algebra with involution which is complete and has the property $||A|| = ||A^{*}||$ is called a {\it Banach $^{*}$-algebra.} Then, follows the definition:
\bedefi 
\noi A $C^{*}$-algebra is a Banach $^{*}$-algebra $\mathfrak{G}$ with the property
\bea
||A^{*}A|| = ||A||^2
\ena 
for all $A \in \mathfrak{G}.$
\findefi
{ Before going further, let us deal, in the following, with some notions about representations and states.}
\bedefi $^{*}$-Morphism between two $^{*}$-algebras 
\noi Let $\mathfrak{G}$ and $\mathfrak{B}$ be two $^{*}$-algebras. The $^{*}$-morphism between $\mathfrak{G}$ and $\mathfrak{B}$ is given by the mapping $\pi:  A \in \mathfrak{G}\mapsto  \pi(A) \in \mathfrak{B} $, 
satisfying:

\bee
\item $\pi(\alpha A + \beta B) = \alpha \pi(A) + \beta \pi (B)$
\item $\pi (AB) = \pi(A)\pi(B)$
\item $\pi(A^{*}) = \pi(A)^{*}$
\ene
for all $A, B \in \mathfrak{G}, \alpha \in \CN.$

\findefi

\berem
\noi Each $^{*}$-automorphism $\pi$ between two $^{*}$-algebras $\mathfrak{G}$ and $\mathfrak{B}$ is positive because if $A \geq 0,$ then $A = B^{*}B$ for some $B \in \mathfrak{G}$. Hence, 
\bea
\pi(A) = \pi(B^{*}B) = \pi(B)^{*}\pi(B) \geq 0.
\ena
\enrem

\bedefi Representation of a $C^{*}$-algebra 

\noi A {\it representation of a $C^{*}$-algebra} $\mathfrak{G}$ is defined to be a pair $(\mathfrak{H}, \pi)$, where $\mathfrak{H}$  is a complex Hilbert space and $\pi$ is a   $^{*}$-morphism of $\mathfrak{G}$ into $\mathcal{L}(\mathfrak{H})$. The representation $(\mathfrak{H}, \pi)$ is said to be {\it faithful} if, and only if, $\pi$ is a  $^{*}$-isomorphism between $\mathfrak{G}$ and $\pi(\mathfrak{G})$, i.e., if, and only if, $\mbox{ker} \, \pi = \{0\}.$
\findefi

Each representation $(\mathfrak{H}, \pi)$ of a $C^{*}$-algebra $\mathfrak{G}$ defines a faithful representation of the quotient algebra $\mathfrak{G}_{\pi} = \mathfrak{G}/\mbox{ker}  \, \pi.$

Then, follows the proposition on the criteria for faithfulness:

\beprop\label{crit00} (\cite{bratelli}, p.44)

\noi Let $(\mathfrak{H}, \pi)$ be a representation of the $C^{*}$-algebra $\mathfrak{G}.$ The representation is faithful if,  and only if,  it satisfies each of the following
equivalent conditions:
\bee
\item $\mbox{ker} \,   \pi = \{0\}$;
\item $||\pi(A)|| = ||A||$ for all $A \in \mathfrak{G}$; 
\item $\pi(A) > 0$ for all $A > 0$.
\ene
\enprop

\noi The Proof of this proposition is achieved by the following proposition:

\beprop\label{proproof00} (\cite{bratelli}, pp.42-43)

\noi Let $\mathfrak{G}$ be a Banach $^{*}$-algebra with identity, $\mathfrak{B}$ a $C^{*}$-algebra, and $\pi$ a 
$^{*}$-morphism of $\mathfrak{G}$ into $\mathfrak{B}.$ Then $\pi$ is continuous and 
\bea
||\pi(A)|| \leq ||A||
\ena
for all $A \in \mathfrak{G}.$ Moreover, if $\mathfrak{G}$ is a $C^{*}$-algebra then the range $\mathfrak{B}_{\pi} 
= \{\pi(A); A \in \mathfrak{G}\}$ of $\pi$ is a $C^{*}$-subalgebra of $\mathfrak{B}$.
\enprop

\noi {\bf Proof.} (see \cite{bratelli} p.43)

\noi First assume $A = A^{*}.$ Then since $\mathfrak{B}$ is  a $C^{*}$-algebra and $\pi(A) \in \mathfrak{B}$, one has 

\bea
||\pi(A)||= \mbox{sup}\left\{|\lambda|; \lambda \in \sigma(\pi(A))\right\}
\ena
\noi by Theorem 2.2.5(a) (see \cite{bratelli} p.29). Next, define $P = \pi(\Id_{\mathfrak{G}})$ where $\Id_{\mathfrak{G}}$ denotes the identity of $\mathfrak{G}$. It follows from the definition of $\pi$ that $P$ is a projection in $\mathfrak{B}$.

\noi Hence replacing $\mathfrak{B}$ by
the $C^{*}$-algebra $P\mathfrak{B}P$, the projection $P$ becomes the identity $\Id_{\mathfrak{B}}$ of the new algebra $\mathfrak{B}$. 
Moreover, $\pi(\mathfrak{G}) \subseteq \mathfrak{B}$. Now it follows from the definitions of a morphism and of the spectrum that $\sigma_{\mathfrak{B}}(\pi(A)) \subseteq \sigma_{\mathfrak{G}}(A).$ Therefore, 
\bea
||\pi(A)||\leq \mbox{sup}\left\{|\lambda|; \lambda \in \sigma_{\mathfrak{G}}(A)\right\} \leq ||A||
\ena
{
\noi by  the following Proposition:

\beprop (\cite{bratelli} p.26)

\noi Let $A$ be an element of a Banach algebra with identity and define the spectral radius $\rho(A)$ of $A$ by 
\bea
\rho(A)  = \mbox{sup}\{|\lambda|; \lambda \in \sigma_{\mathfrak{G}}(A)\}.
\ena
\noi It follows that 

\bea
\rho(A) = \lim_{n \rightarrow \infty}||A^{n}||^{1/n} = \mbox{inf}_{n}||A^{n}||^{1/n} \leq ||A||.
\ena
\noi In particular, the limit exists. Thus the spectrum of $A$ is a nonempty compact set.
\enprop

\noi {\bf Proof.} (see \cite{bratelli} p.26)

\noi Let $|\lambda|^{n} > ||A^{n}||$ for some $n  > 0.$ As each $m \in \ZN$ can be decomposed  as $m=pn +q$ with $p,q \in \ZN$  and $0 \leq q < n$ one again establishes that the series 
\bea
\lambda^{-1}\sum_{m \geq 0}\left(\frac{A}{\lambda}\right)^{m}
\ena 
is Cauchy in the uniform topology and defines $(\lambda \Id - A)^{-1}.$ Therefore, 
\bea
\rho(A) \leq ||A^{n}||^{1/n}
\ena
\noi for all $n  > 0, $ and consequently
\bea
\rho(A) \leq \mbox{inf}_{n} ||A^{n}||^{1/n} \leq \lim_{n \rightarrow \infty}{\mbox{inf}}||A^{n}||^{1/n}.
\ena
\noi Thus to complete the proof it sufficies to establish that $\rho(A) \geq r_{A}, $ where 
\bea
r_{A} = \lim_{n \rightarrow \infty}\mbox{sup} ||A^{n}||^{1/n}.
\ena

\noi There are two cases.

\noi Firstly, assume $0 \in r_{\mathfrak{G}}(A), $ i.e., $A$ is invertible. Then $1 = ||A^n A^{-n}|| \leq ||A^n|| \, ||A^{-n}||$ and hence $1 \leq r_A r_{A^{-1}}.  $  This implies $r_A > 0.$ Consequently, if $r_A = 0$ one must have $0 \in r_{\mathfrak{G}}(A)$ and 
$\rho(A) \geq r_A.$

\noi Secondly, we may assume $r_A > 0.$
 We will need the following observation. 
 \noi If $A_n$ is any sequence of elements such that $R_n = (\Id - A_n)^{-1}$ exists then $\Id - R_n = - A_n(\Id- A_n)^{-1}$ and $A_n = -(\Id - R_n)(\Id -(\Id -R_n))^{-1}. $ Therefore, $||\Id -R_n|| \rightarrow 0$ is equivalent to $||A_n|| \rightarrow 0$ by power series expansion.
 
\noi Define $S_A = \{\lambda; \lambda \in \CN, |\lambda| \geq r_A\}. $ We assume that $S_A \subseteq r_{\mathfrak{G}}(A)$ and obtain a contradiction. Let $\omega$ be a primitive nth root of unity. By assumption
 
\bea
R_n(A; \lambda) = n^{-1}\sum_{k=1}^{n}\left(\Id - \frac{\omega^k A}{\lambda}\right)^{-1}
\ena
\noi is well defined for all $\lambda \in S_A. $ But an elementary calculation shows that 
\bea
R_n(A; \lambda) = \left(\Id - \frac{A^n}{\lambda^n}\right)^{-1}.
\ena
Next one has the continuity estimate
\bea
 \left| \left| \left(\Id - \frac{\omega^k A}{r_A}\right)^{-1}  - \left(\Id - \frac{\omega^k A}{\lambda}\right)^{-1}\right|\right| &=& \left|\left|\left(\Id - \frac{\omega^k A}{r_A}\right)^{-1}\omega^k A\left(\frac{1}{\lambda} - \frac{1}{r_A}\right)\left(\Id - \frac{\omega^k A}{\lambda}\right)^{-1}\right|\right|\cr
& &  \leq|\lambda - r_A|\, ||A|| \; \mbox{sup}_{\gamma \in S_A}||(\gamma \Id - A)^{-1}||^{2},
\ena
 
\noi which is uniform in $k. $ The supremum   is finite since $\lambda \mapsto ||(\lambda \Id - A)^{-1}||$ is continuous on $r_{\mathfrak{G}}(A)$ and for $|\lambda| > ||A||$ one has 

\bea
||(\lambda \Id - A)^{-1}|| \leq |\lambda|^{-1} \sum_{n \geq 0} ||A||^{n}/|\lambda|^{n}=
(|\lambda| - ||A||)^{-1}.
\ena
\noi It follows then that for each $\varepsilon  > 0$  there is a $\lambda  > r_A$ such that 

\bea
\left|\left|\left(\Id - \frac{A^{n}}{r^{n}_{A}}\right)^{-1}  - \left(\Id - \frac{A^{n}}{\lambda^{n}}\right)^{-1}\right|\right| < \varepsilon
\ena
\noi uniformly in $n. $ But $||A^n||/\lambda^{n} \rightarrow 0$ and by the above observation 
$||(\Id-A^n/\lambda^n)^{-1} - \Id|| \rightarrow 0. $ This implies that $||(\Id - A^n/r^n_A)^{-1} - \Id|| \rightarrow 0$ and $||A^n||/r^n_A 
\rightarrow 0$ by another application of the same observation. This last statement contradicts, however, the definition of $r_A$ and hence
the proof is complete. 
$\hfill{\square}$

\noi Finally, if $A$ is not selfadjoint one can combine this inequality with the $C^{*}$-norm property and the product inequality to deduce that 
\bea
||\pi(A)||^{2} = ||\pi(A^{*}A)|| \leq ||A^{*}A|| \leq ||A||^{2}.
\ena
Thus $||\pi(A)|| \leq ||A||$ for all $A \in \mathfrak{G}$ and $\pi$ is continuous.

\noi The range $\mathfrak{B}_{\pi}$ is a $^{*}$-subalgebra of $\mathfrak{B}$ by definition and to deduce that it is a $C^{*}$-subalgebra  we must prove that it is closed, under the assumption that $\mathfrak{G}$ is a $C^{*}$-algebra.

\noi Now introduce the kernel ker $\pi$ of $\pi$ by 
\bea
\mbox{ker}\, \pi = \{A \in \mathfrak{G}; \pi(A) = 0\}
\ena
\noi then ker $\pi$ is closed two-sided $^{*}$-ideal. Given $A \in \mathfrak{G}$ and $B \in \mbox{ker}\, \pi$ then $\pi(AB) = \pi(A)\pi(B)= 0, \pi(BA) = \pi(B)\pi(A) = 0, $ and $\pi(B^{*})= \pi(B) =0.$ The closedness follows from the estimate $||\pi(A)||   \leq ||A||. $ Thus we can form the quotient algebra $\mathfrak{G}_{\pi}= \mathfrak{G}/\mbox{ker}\, \pi$ and $\mathfrak{G}_{\pi}$ is a $C^{*}$-algebra. The elements of $\mathfrak{G}_{\pi}$ are the classes $\hat{A} = \{A + I; I \in \mbox{ker}\, \pi\}$ and the morphism $\pi$ induces a morphism $\hat{\pi}$ from $\mathfrak{G}_{\pi}$ onto $\mathfrak{B}_{\pi}$ by the defintion $\hat{\pi}(\hat{A}) = \pi(A).$ The kernel of $\hat{\pi}$ is zero by construction and hence $\hat{\pi}$ is an isomorphism  between $\mathfrak{G}_{\pi}$ and  $\mathfrak{B}_{\pi}$. Therefore, one can define  a morphism $\hat{\pi}^{-1}$ from the $^{*}$-algebra  $\mathfrak{B}_{\pi}$ onto the $C^{*}$-algebra  $\mathfrak{G}_{\pi}$
by $\hat{\pi}^{-1}(\hat{\pi}(\hat{A})) = \hat{A}$ and then applying the first statement of the proposition to  $\hat{\pi}^{-1}$ and $\hat{\pi}$ successively one obtains 
 \bea
||\hat{A}|| = ||\hat{\pi}^{-1}(\hat{\pi}(\hat{A}))||\leq ||\hat{\pi}(\hat{A})|| \leq ||\hat{A}||.
\ena
\noi Thus $||\hat{A}|| = ||\hat{\pi}(\hat{A})|| = ||{\pi}(A)||.$  Consequently, if $\pi(A_n)$ converges uniformly  in $\mathfrak{B}$ to an element $A_{\pi}$ then $\hat{A_n}$  converges in 
$\mathfrak{G}_{\pi}$  to an element $\hat{A}$ and $A_{\pi} = \hat{\pi}(\hat{A}) = \pi(A)$ where $A$ is any element of the equivalence  class $\hat{A}. $ Thus $A_{\pi} \in \mathfrak{B}_{\pi}$ and $\mathfrak{B}_{\pi}$ is closed.
$\hfill{\square}$
}

\noi {\bf Proof of Proposition \ref{crit00}} (see \cite{bratelli} p.44).

\noi The equivalence of condition (1) and faithfulness is by definition. Prove that (1) $\Rightarrow$ (2) $\Rightarrow$ (3) $\Rightarrow$ (1).

\noi (1) $\Rightarrow$ (2) Since ker$\, \pi = \{0\}$,  we can define a morphism $\pi^{-1}$ from the range of $\pi$ into $\mathfrak{G}$ by $\pi^{-1}(\pi(A))= A$ and then applying Proposition \ref{proproof00} to $\pi^{-1}$ and $\pi$ sucessively one has 
\bea
||A|| = ||\pi^{-1}(\pi(A))|| \leq ||\pi(A)|| \leq ||A||.
\ena
\noi (2) $\Rightarrow$ (3) If $A > 0$ then $||A|| > 0$ and hence $||\pi(A)|| > 0, $
or $\pi(A) \neq 0. $ But $\pi(A) \geq 0 $ by Proposition \ref{proproof00} and therefore $\pi(A) > 0.$
\noi (3) $\Rightarrow$ (1) If condition (1) is false then there is  a $B \in$ ker  $\pi$ with $B \neq 0$ and $\pi(B^{*}B) = 0. $ But $||B^{*}B|| \geq 0$ and as $||B^{*}B||  = ||B||^2$ one has 
$B^{*}B  > 0.$ Thus condition (3) is false.
$\hfill{\square}$

\bedefi Cyclic representation of a $C^{*}$-algebra 
\noi A {\it cyclic representation of a $C^{*}$-algebra} $\mathfrak{G}$ is defined to be a triplet $(\mathfrak{H}, \pi, \Omega)$, 
where $(\mathfrak{H}, \pi)$  is a representation of $\mathfrak{G}$ and $\Omega$ is a vector in $\mathfrak{H}$ which is cyclic for $\pi$, 
in $\mathfrak{H}$.
\noi $\Omega$ is called {\it cyclic vector} or {\it cyclic vector for} $\pi$.
\noi If $\mathfrak{K}$ is a closed subspace of $\mathfrak{H}$ then $\mathfrak{K}$ is called a 
{\it cyclic subspace} for $\mathfrak{H}$ whenever the set 
\bea
\left\{\sum_{i}\pi(A_i)\psi_i; A_i \in \mathfrak{G}, \psi_i \in \mathfrak{K}\right\}
\ena
is dense in $\mathfrak{H}.$
\findefi
\bedefi\label{statedefi00} State over a $C^{*}$-algebra 
\noi A linear functional $\omega$ over the $C^{*}$-algebra $\mathfrak{G}$ is defined to be positive if 
\bea
\omega(A^{*}A) \geq 0
\ena
for all $A \in \mathfrak{G}. $ A positive linear functional $\omega$ over a $C^{*}$-algebra $\mathfrak G$ with $||\omega||=1$ is
called a {\it state}.
\findefi
\berem
\bee
\item  Every positive element of a $C^{*}$-algebra is of the form $A^{*}A$ and hence  positivity of $\omega$ is equivalent to  $\omega$ being 
positive on positive elements.
\item Considering a representation $(\mathfrak{H}, \pi)$ of the  $C^{*}$-algebra $\mathfrak{G}$, taking $\Omega \in \mathfrak{H}$ being a
nonzero vector and define $\omega_{\Omega}$ by 
\bea
\omega_{\Omega}(A) = (\Omega, \pi(A)\Omega)
\ena
for all $A \in \mathfrak{G}$.  It follows that  $\omega_{\Omega}$ is a linear function over
$\mathfrak{G}, $ it is also positive since 
\bea
\omega_{\Omega}(A^{*}A) = ||\pi(A)\Omega||^2 \geq 0.
\ena
\ene
\item $||\omega_{\Omega}|| = 1$ whenever $||\Omega|| = 1$ and then, $\pi$ is nondegenerate. In this case $\omega_{\Omega}$ is a state, 
and is usually called {\it vector state} for the representation $(\mathfrak{H}, \pi)$.
\item 
\enrem
\bedefi 
\noi The cyclic representation $(\mathfrak{H}_\omega , \pi_\omega , \Omega_\omega)$, constructed from the state 
$\omega$ over the $C^{*}$-algebra $\mathfrak{G}$ , is defined as the {\it canonical cyclic representation of $\mathfrak{G}$ associated
with $\omega$.}
\findefi 
\noi Next it will be demonstrated  that the notions of purity of a state $\omega$ and irreducibility of the representation 
associated with $\omega$  are intimately related.
%
%
\betheo\label{repthem00}(\cite{bratelli} p.57)
\noi Let $\omega$ be a state over the $C^{*}$-algebra $\mathfrak{G}$ and $(\mathfrak{H}_{\omega}, \pi_\omega, \Omega_\omega )$ the 
associated cyclic representation. The following conditions are equivalent:
\bee
\item
$(\mathfrak{H}_\omega , \pi_\omega )$ is irreducible;
\item
$\omega$ is pure;
\item
$\omega$ is an extremal point of the set $E_{\mathfrak{G}}$ of states over $\mathfrak{G}$.
\noi Furthermore, there is one-to-one correspondence
\bea
\omega_T (A)= (T\Omega_\omega, \pi_\omega(A)\Omega_\omega)
\ena
\noi between positive functionnals $\omega_T,$ over $\mathfrak{G}$, majorized by $\omega$ and positive operators $T$ in the 
commutant $\pi'_\omega$, of $\pi_\omega ,$ with $||T|| \leq 1.$
\ene
\entheo
\noi {\bf Proof.}(see \cite{bratelli} pp.57-58)
\noi $(1)\Rightarrow(2)$ Assume that $(2)$ is false. Thus there exists a positive functional $\rho$ such that $\rho (A^* A)\leq \omega (A^* A)$ for all $A\in \mathfrak{G}$. But applying the Cauchy-Schwarz inequality one then has
\beano
|\rho (B^* A)|^2 &\leq &\rho (B^* B)\rho (A^* A)\cr
&\leq &\omega (B^* B )\omega (A^* A)\cr
&=&||\pi_\omega (B)\Omega_\omega ||^2 ||\pi_\omega (A)\Omega_\omega||^2.
\enao
\noi Thus $\pi_\omega (B)\Omega_\omega \times \pi_\omega (A)\Omega_\omega \longmapsto \rho (B^*A)$ is a densely defined, bounded, sesquilinear functional, over $\mathfrak{H}_\omega \times \mathfrak{H}_\omega$, and there exists a unique bounded operator $T$, on $\mathfrak{H}_\omega$, such that
\beano
(\pi_\omega (B)\Omega_\omega, T\pi_\omega (A)\Omega_\omega)=\rho (B^* A).
\enao
\noi As $\rho$ is not a multiple of $\omega$ operator $T$ is not a multiple of the identity. Moreover,
\beano
0 &\leq &  \rho (A^* A)\cr
&=& (\pi_\omega (A)\Omega_\omega , T\pi_\omega (A)\Omega_\omega)\cr
&\leq & \omega (A^* A) = (\pi_\omega (A)\Omega_\omega , \pi_\omega (A)\Omega_\omega)
\enao
and hence $0\leq T \leq \Id.$ But 
\beano
(\pi_\omega (B)\Omega_\omega , T\pi_\omega (C)\pi_\omega (A)\Omega_\omega)&=&\rho (B^{*}CA) \cr
&=& \rho ((C^* B)^* A)= (\pi_\omega (B)\Omega_\omega , \pi_\omega (C)T\pi_\omega (A)\Omega_\omega)
\enao
and therefore $T\in \pi'_\omega .$ Thus condition $(1) $ is false.
\noi $(2)\Rightarrow (1)$ Assume that $(1)$ is false. If $T\in \pi'_\omega$ then $T^* \in \pi'_\omega$ and $T+T^* ,\,\,(T-T^* )/i$ are also elements of the commutant. Thus there exists a selfadjoint element $S$ of $\pi'_\omega $ which is not a multiple  of the identity. Therefore there exists a spectral projector $P$ of $S$ such that $0<P<\Id $ and $P\in \pi'_\omega$. Consider the functional
\beano
\rho(A)= (P\Omega_\omega, \pi_\omega(A)\Omega_\omega).
\enao
\noi This is certainly positive since
\beano
\rho(A^* A)= (P\pi_\omega (A)\Omega_\omega, P\pi_\omega (A)\Omega_\omega)\geq 0.
\enao
Moreover,
\beano
\omega(A^* A)-\rho(A^* A)&=& (\pi_\omega (A)\Omega_\omega, (\Id-P)\pi_\omega (A)\Omega_\omega)\cr
&\geq & 0.
\enao
\noi Thus $\omega$ majorizes $\rho$. It is verified  that $\rho$ is not a multiple of $\omega$ and hence $(2)$ is false.
\noi This proves the equivalence of the first two conditions stated in the theorem and simultaneously establishes the correspondence described by the last statement.

\noi The equivalence of conditions (2) and (3) is performed as follows. Suppose that $\omega$
 is an extremal point of $E_{\mathfrak{G}}$ and $\omega \neq 0.$ Then, we must have $||\omega|| = 1.$ Thus $\omega$ is a state and we must deduce that it is pure. Suppose the contrary; then there
is a state $\omega_1 \neq \omega$ and a $\lambda$ with $0 < \lambda < 1$ such that  $\omega \geq \lambda \omega_1. $ 
Define $\omega_2$ by $\omega_2 = (\omega- \lambda \omega_1)/(1-\lambda); $ then $||\omega_2|| = 
(||\omega|| - \lambda||\omega_1||)/(1-\lambda)=1$  and $\omega_2$ is also a state. But $\omega =
\lambda \omega_1 + (1-\lambda)\omega_2$ and $\omega$  is not extremal, which is a contradiction. 
$\hfill{\square}$


\noi In the following, some notions on von Neumann algebra are provided. 
 To specify the Hilbert space upon which a von Neumann algebra $\mathfrak{A}$ acts, one often uses the notation 
 $\{\mathfrak{A}, \mathfrak{H}\}$ to denote the von Neumann algebra $\mathfrak{A}.$

\bedefi
von Neumann algebra 

\noi Let $\mathfrak{H}$ be a Hilbert space. For each subset $\mathfrak{A}$ of $\mathcal L(\mathfrak H)$, let $\mathfrak{A'}$ denote the set
of all bounded operators on $\mathfrak{H}$ commuting with every operator in $\mathfrak{A}$.  Clearly, $\mathfrak{A'}$ is a Banach algebra of operators containing the identity operator $I_{\mathfrak H}$ on $\mathfrak H.$
\bei
\item[(i)] 
A {\it von Neumann algebra} 
is a $^{*}$-subalgebra  $\mathfrak A$ of $\mathcal L(\mathfrak H)$ 
such that $\mathfrak A = \mathfrak A''$.

\item[(ii)] $\mathfrak A'$ denotes 
the {\it commutant} of $\mathfrak A$, the 
set of all elements in $ \mathcal L(\mathfrak H)$ which commute 
with every element of $\mathfrak A$.
\item[(iii)]
A von Neumann algebra always contains the 
identity 
operator  $I_{\mathfrak H}$ on $\mathfrak H$. It is called a {\it {factor}} if $\mathfrak A \cap \mathfrak A' = \mathbb C I_{\mathfrak H}$.

\item[(iv)] If a subset $\mathcal{S}$ of $ \mathcal L(\mathfrak H)$ is invariant under the  $^{*}$-operation, then $\mathcal{S''}$, the
double commutant of $\mathcal{S}$, is the smallest von Neumann algebra containing
$\mathcal{S}$, and it is called the von Neumann algebra {\it generated} by $\mathcal{S}$.
\eni
\findefi
We also have the following definition: 
\bedefi 
\noi A {von Neumann algebra} $\mathfrak{A} \subset \mathcal L(\mathfrak H)$ is a $C^{*}$-algebra acting on
the Hilbert space $\mathfrak H$ that is closed under the {\it weak-operator topology}:
$A_n \stackrel{n\rightarrow +\infty}{\longrightarrow} A$ iff $\langle \xi|A_n \eta \rangle 
\stackrel{n\rightarrow +\infty}{\longrightarrow} \langle \xi|A \eta \rangle, \forall \xi, \eta \in \mathfrak H, $
or equivalently under the $\sigma$-{\it weak topology}:
\noi $A_n \stackrel{n\rightarrow +\infty}{\longrightarrow} A$ iff for all sequences $(\xi_k), 
(\zeta_k)$ in $\mathfrak H$ such that $\displaystyle\sum_{k=1}^{+\infty}||\xi_k||^2 <+\infty$ and 
$\displaystyle\sum_{k=1}^{+\infty}||\zeta_k||^2 <+\infty$ we have $\displaystyle\sum_{k=1}^{+\infty}\langle \xi_k|A_n\zeta_k\rangle 
\stackrel{n\rightarrow +\infty}{\longrightarrow} 
\displaystyle\sum_{k=1}^{+\infty}\langle \xi_k|A \zeta_k\rangle.$
\findefi

\noi { Case of $\mathcal L(\mathfrak H)$} 
\bei
\item[(i)] $\mathcal L(\mathfrak H)$ is a von Neumann algebra and even a factor since $\mathcal L(\mathfrak H)' = \CN \Id \quad (\Id = I_{\mathfrak H})$.
\item[(ii)] The Hilbert space adjoint operation defines an involution on $\mathcal L(\mathfrak H)$ and with respect to these operations and this norm,  
$\mathcal L(\mathfrak H)$ is a $C^{*}$-algebra. In particular, the $C^{*}$-norm property follows from $||A||^{2} \leq ||A^{*}||||A|| = ||A||^{2}$.
 
\item[(iii)] Any uniformly closed subalgebra $\mathfrak M$ of $\mathcal L(\mathfrak H)$ which is self-adjoint is also a $C^{*}-$algebra.

\noi Next, it comes the following definition:

\bedefi Closure-Orthogonal projection  

\noi If $\mathfrak M$ is a subset of $\mathcal L(\mathfrak H)$ and $\mathfrak{K}$ is a subset of $\mathfrak{H}, $ 
$[\mathfrak{M}\mathfrak{K}]$ denotes the closure of the linear span of elements of the form $A\xi, $ where $A \in \mathfrak{M}$ and
$\xi \in \mathfrak{K}.$  $[\mathfrak{M}\mathfrak{K}]$ also denotes  the orthogonal projection onto $[\mathfrak{M}\mathfrak{K}].$
\findefi

\item[(iv)] A $^{*}$-subalgebra  $\mathfrak M \subseteq \mathcal L(\mathfrak H)$ is said to be nondegenerate if $[\mathfrak M \mathfrak H] 
= \mathfrak H$.  


\item[(v)] If $\mathfrak M \subseteq \mathcal L(\mathfrak H)$ contains the identity operator, then, it is automatically nondegenerate. 

\noi A nondegenerate $^{*}$-algebra contains the identity operator;
If a subalgebra of $\mathcal L(\mathfrak H)$ is invariant under the *-operation, then it is called a $^{*}$-subalgebra of $\mathcal L(\mathfrak H)$
 or a $^{*}$-algebra of operators on $\mathfrak H$.

 We have the following proposition (see \cite{takesaki} pp.72-73):

\beprop (\cite{takesaki}, p.72)

The subset $\mathfrak{M}$ of $\mathcal L(\mathfrak H)$ is a von Neumann algebra on $\mathfrak{H}$.
\enprop

\noi {\bf Proof.} (see \cite{takesaki}, p.72)

\noi Let $\{\mathfrak{M}_i, \mathfrak{H}_i\}_{i \in I}$ be a family of von Neumann algebras. Let $\mathfrak{H}$ denote the direct sum $\displaystyle\sum^{\oplus}_{i \in I}\mathfrak{H}_i$ of Hilbert spaces 
$\{\mathfrak{H}_i\}_{i \in I}. $ Each vector $\xi = \{\xi_i\}_{i \in I}$ in $\mathfrak{H}$ is denoted by $\displaystyle\sum^{\oplus}_{i \in I}\xi_{i}. $ For each bounded sequence $\{x_i\}_{i \in I}$ in 
$\displaystyle\prod_{i \in I}\mathfrak{M}_i, $ one defines an operator $x$ on $\mathfrak{H}$ by
\bea
x \sum^{\oplus}_{i \in I}\xi_i = \sum^{\oplus}_{i \in I}x_i \xi_i
\ena
\noi Then, $x$ is a bounded operator on $\mathfrak{H} $ denoted by $\displaystyle\sum^{\oplus}_{i \in I} x_i.$ Let $\mathfrak{M}$ be the set of all such $x.$

\noi Particularly, taking $ 
\mathfrak{M}$ as a subset of  $\mathcal L(\mathfrak H)$,  the proof is completed. 
$\hfill{\square}$

\eni

\bedefi\label{vecyclicsepar} Cyclic and separating vector 

\noi The modular theory of von Neumann algebras is such that to every von Neumann algebra
$\mathfrak M \subset \mathcal L(\mathfrak H), $ and to every vector $\xi \in \mathfrak H$ that is {\it cyclic} 
\bea
\overline{(\mathfrak M \xi)} = \mathfrak H
\ena
i.e. the set  $\{A\Phi; A \in \mathfrak M\}$ 
($\mathfrak M$ denoting a set of bounded operators on $\mathfrak H$)  is dense in $\mathfrak H;$ and {\it separating} i.e. 
for $\mathcal A \in \mathfrak M, $
\bea
A\xi = 0 \Rightarrow A = 0.
\ena

\findefi

\noi Moreover, a vector $\psi \in \mathfrak{H}$ is said separating for a von Neumann algebra 
$\mathfrak{A}$ if $A\psi = B\psi, \, A, B \in \mathfrak{A}, $ if and only if $A = B.$ 


{

\noi  We have the following definitions:

\noi \bedefi Separating subset 

\noi Let $\mathfrak A$ be a von Neumann algebra on a Hilbert space $\mathfrak{H}. $ A subset $\mathfrak{K} \subseteq \mathfrak{H}$ is 
separating for $\mathfrak{A}$ if for any $A \in \mathfrak{A}, \, A\xi = 0$ for all $\xi \in \mathfrak{K}$ implies $A = 0.$
\findefi

\bedefi  Cyclic and separating subset of a von Neumann algebra 

\noi Let $\{\mathfrak{A}, \mathfrak{H}\}$ be a von Neumann algebra. A subset $\mathfrak{M}$ of $\mathfrak{H}$ is called 
{\it separating} (resp.{\it cyclic}) for $\mathfrak{A}$ if $a\xi = 0, \, a \in \mathfrak{A}, $ for every $\xi \in \mathfrak{M}$ 
implies $a=0$ (resp. the smallest invariant subspace $[\mathfrak{A}\mathfrak{M}]$ under $\mathfrak{A}$ containing $\mathfrak{M}$ in
the whole space $\mathfrak{H}$).
\findefi

\noi Recall that a subset $\mathfrak{K} \subseteq \mathfrak{H}$ is cyclic for $\mathfrak M$ if $[\mathfrak{M}\mathfrak{K}] = \mathfrak{H}.$
There  is a dual relation  between the properties of cyclic for the algebra and separating for the
commutant.

We have the following propositions:

\beprop\label{propos00} (\cite{bratelli} p.85)

\noi Let $\mathfrak{A}$ be a von Neumann algebra on $\mathfrak{H}$ and $\mathfrak{K} 
\subseteq \mathfrak{H}$ a subset. The following conditions are equivalent:
\bei
\item[(1)] $\mathfrak{K}$ is cyclic for $\mathfrak{A}$;
\item[(2)] $\mathfrak{K}$ is separating for $\mathfrak{A'}.$
\eni
\enprop

\noi {\bf Proof.} (see \cite{bratelli} p.85)

\noi (1) $\Rightarrow$ (2) Assume that $\mathfrak{K}$ is cyclic for $\mathfrak{A}$ and choose $A' \in \mathfrak{A'}$ such that $A'\mathfrak{K} = \{0\}. $ Then, for any $B \in \mathfrak{A}$ and $\xi \in \mathfrak{K}, \, A'B\xi = BA'\xi = 0, $ hence $A'[\mathfrak{A}\mathfrak{K}] = 0$ and $A' = 0.$

\noi (2) $\Rightarrow$ (1) Suppose that $\mathfrak{K}$ is separating for $\mathfrak{A'}$ and set $P' = [\mathfrak{A}\mathfrak{K}].$ $P'$ is then a projection in $\mathfrak{A'}$ and $(\Id -P')\mathfrak{K} = \{0\}. $ Hence $\Id - P' = 0$ and $[\mathfrak{A}\mathfrak{K}] = \mathfrak{H}.$
$\hfill{\square}$

\bedefi {The weak and $\sigma$-weak topologies
\noi If $\xi, \eta  \in \mathfrak{H}, $ then $A  \mapsto |(\xi, A\eta)|$ is a seminorm on $\mathcal{L}(\mathfrak{H})$. 
The locally convex topology on $\mathcal{L}(\mathfrak{H})$ defined by these
seminorms is called the {\it weak topology}. The seminorms defined by the vector states $A \mapsto |(\xi, A\xi)|$ suffice to 
define this topology because $\mathfrak{H}$ is
complex and one has the polarization identity

\bea
4(\xi, A\eta) = \sum_{n=0}^{3}i^{-n}(\xi  + i^n \eta, A(\xi + i^n \eta)).
\ena

\noi Let $\{\xi_n\}, \{\eta_n\}$ be two sequences from  $\mathfrak{H}$ such that

\bea
\sum_{n}||\xi_n||^2 < \infty, \qquad \sum_{n}||\eta_n||^2 < \infty.
\ena

\noi Then for $A \in \mathcal{L}(\mathfrak{H})$

\bea
|\sum_{n}(\xi_n, A\eta_n)| &\leq & \sum_{n} ||\xi_n|| \, ||A|| \, ||\eta_n|| \cr
& \leq & ||A|| \left(\sum_{n} ||\xi_n||^2\right)^{1/2}\left(\sum_{n} ||\eta_n||^2\right)^{1/2}\cr
& \leq &  \infty.
\ena
\noi Hence $A \mapsto |\displaystyle\sum_{n}(\xi_n, A\eta_n)|$ is a seminorm on $ \mathcal{L}(\mathfrak{H}).$ The locally convex topology
on $ \mathcal{L}(\mathfrak{H})$ induced by these seminorms is called the $\sigma$-weak topology.
}
\findefi
\noi {\bf Notations: }
In  the sequel, 
\bei 
\item $\mathfrak{A}_{+}$ denotes the {\it positive part} of the von Neumann algebra $\mathfrak{A}$ or the set of positive elements of the von Neumann algebra $\mathfrak{A}$;
\item $\mathfrak{A}_{*}$ denotes the {\it predual of a von Neumann algebra}. It is the space of all $\sigma$-weakly continuous linear functionals on $\mathfrak{A}$;
\item $\mathfrak{L}(\mathfrak{H})_{1}$ denotes the unit ball of $\mathfrak{L}(\mathfrak{H})$. $\mathfrak{L}(\mathfrak{H})_{1}$ is norm dense in the unit ball of the norm closure of $\mathfrak{L}(\mathfrak{H})$, and it is taken as a $C^{*}$-algebra (see \cite{bratelli} p.74).
\eni
}
{

\bedefi

\noi Let $\varphi: \mathfrak{A} \rightarrow \CN$ be a bounded linear functional on $\mathfrak{A}, 
$ which is denoted by $\langle \varphi; A \rangle, \, A \in \mathfrak{A}.$

\noi $\varphi$ is called a state on this algebra if it also satisfies the two conditions: 

\bei
\item[(a)] $\langle \varphi; A^{*}A\rangle \geq 0, \quad \forall A \in \mathfrak{A}$
\item[(b)] $\langle \varphi; I_{\mathfrak{H}}\rangle  = 1.$
\eni

\noi The state $\varphi$ is called a {\it vector state} if there exists a vector $\phi \in \mathfrak{H}$ such that 
\bea
\langle \varphi; A\rangle = \langle \phi|A \phi \rangle, \quad \forall A \in \mathfrak{A}.  
\ena
\noi Such a state is also {\it normal}.

\findefi

}

\bedefi\label{impdefi00} 

\noi A state $\omega$ on a von Neumann algebra $\mathfrak{A}$ is {\it faithful} if $\omega(A) > 0$ for all nonzero $A \in \mathfrak{A}_+.$
\findefi

\berem\label{impdefi01} (see \cite{bratelli}, Example 2.5.5 p.85)
\noi Let $\mathfrak{A} = \mathcal{L}(\mathfrak{H})$ with $\mathfrak{H}$ separable. 
Every normal state $\omega$ over $\mathfrak{A}$ is of the form
\bea\label{impdefi03}
\omega(A) = \mbox{Tr}(\rho A),
\ena
\noi where $\rho$ is a density matrix. If $\omega$ is faithful then $\omega(E) > 0$ for each rank one projector, i.e., $||\rho^{1/2}\psi||> 0$ for each $\psi \in \mathfrak{H} \setminus \left\{0\right\}. $
 Thus $\rho$ is invertible (in the densely defined
self-adjoint operators on $\mathfrak{H}$). Conversely, 
if $\omega$ is not faithful then $\omega(A^{*}A) = 0$ for some
nonzero $A$ and hence  $||\rho^{1/2}A^{*}\psi|| = 0$ for all $\psi \in \mathfrak{H}, $ i.e., 
$\rho$ is not invertible. This
establishes that $\omega$ is faithful if, and only if, $\rho$ is invertible.
\enrem
\belem\label{lem00} (\cite{bratelli} p.76)
Let $\{A_\alpha\}$ be an increasing set in $\mathcal L(\mathfrak{H})_+$ with an upper bound in 
$\mathcal L(\mathfrak{H})_+$. Then  $\{A_\alpha\}$ has a least upper bound (l.u.b.) A, and the 
net converges $\sigma$-strongly to $A.$
\enlem
\noi {\bf Proof.}   (see \cite{bratelli} p.76)
Let $\mathfrak{K}_{\alpha}$ be the weak closure of the set of $A_\beta$ with $\beta > \alpha. 
$ Since $\mathcal L(\mathfrak{H})_1$ is weakly compact, there exists an element A in $\bigcap_\alpha \mathfrak{K}_\alpha. $ For all $A_\alpha$ the set of $B \in \mathcal L(\mathfrak{H})_+$ such that $B \geq A_\alpha$ is $\sigma$-weakly closed and contains $\mathfrak{K}_\alpha, $ hence $A \geq A_\alpha.$ Thus, $A$ majorizes $\{A_\alpha\}$ and lies in the weak closure of  $\{A_\alpha\}$. If $B$ is another operator majorizing $\{A_\alpha\}$, then it majorizes its weak closure; thus $B\geq A$ and $A$ is the least upper bound of 
$\{A_\alpha\}. $ Finally, if $\xi \in \mathfrak{H}$ then 
\bea
||(A-A_\alpha)\xi||^2 &\leq & ||A-A_\alpha||\, ||(A-A_\alpha)^{1/2}\xi||^{2} \cr
&\leq & ||A||(\xi, (A-A_\alpha)\xi)\crcr
&    {\longrightarrow}_{\small{\alpha}} & 0.
\ena
\mbox Since the strong and $\sigma$-strong topology coincide on $\mathcal L(\mathfrak{H})_1,$ this ends the proof.
$\hfill{\square}$
 
\beprop\label{Traceprop00}(\cite{bratelli} p.68)
Let Tr be the usual trace on $\mathcal{L}(\mathfrak{H})$, and let $\mathcal{T}(\mathfrak{H})$ be the Banach space of trace-class 
operators on $\mathfrak{H}$ equipped with the trace norm
$T\mapsto \mbox{Tr}(|T|) = ||T||_{\tiny{\mbox{Tr}}}$. Then it follows that $\mathcal{L}(\mathfrak{H})$ is the dual 
$\mathcal{T}(\mathfrak{H})^{*}$ of $\mathcal{T}(\mathfrak{H})$ by the duality 
\bea
A \times T \in \mathcal{L}(\mathfrak{H}) \times \mathcal{T}(\mathfrak{H})\mapsto \mbox{Tr}(AT).
\ena
The weak$^{*}$ topology on $\mathcal{L}(\mathfrak{H})$ arising from this duality is just the $\sigma$-weak topology.
\enprop
\noi {\bf Proof.}(see \cite{bratelli} pp.68-69)
\noi Due to the inequality $|\mbox{Tr}(AT)| \leq ||A||\, ||T||_{\tiny{\mbox{Tr}}}, \, \mathcal{L}(\mathfrak{H})$ is the subspace of $\mathcal{T}(\mathfrak{H})^{*}$
by the duality described in the proposition. Conversely, assume $\omega \in \mathcal{T}(\mathfrak{H})^{*}$ and consider a rank one operator $E_{\varphi, \psi}$ defined for $\varphi, \psi \in \mathfrak{H}$ by 
\bea
E_{\varphi, \psi} \chi = \varphi(\psi, \chi).
\ena
\noi One has $E^{*}_{\varphi, \psi} = E_{\psi, \varphi}$ and $E_{\varphi, \psi} E_{\psi, \varphi} = ||\psi||^{2}E_{\varphi, \varphi}.$  Hence 
\bea
||E_{\varphi, \psi}||_{\tiny{\mbox{Tr}}} = ||\psi||\mbox{Tr}(E_{\varphi, \varphi})^{1/2} = ||\psi||||\varphi||.
\ena
\noi It follows that 
\bea
|\omega(E_{\varphi, \psi})|\leq ||\omega||\, ||\varphi||\, ||\psi||.
\ena
Hence there exists, by the Riesz representation theorem, an $A \in \mathcal{L}(\mathfrak{H})$ with $||A|| \leq ||\omega||$ such that 
\bea
\omega(E_{\varphi, \psi}) = (\psi, A \varphi).
\ena
\noi Consider $\omega_{0} \in  \mathcal{T}(\mathfrak{H})^{*}$ defined by 
\bea
\omega_{0}(T) = \mbox{Tr}(AT)
\ena
then 
\bea
\omega_{0}(E_{\varphi, \psi}) &=& \mbox{Tr}(AE_{\varphi, \psi})\cr
&=& (\psi, A\varphi) \cr
&=& \omega(E_{\varphi, \psi}).
\ena
Now for any $T \in \mathcal{T}(\mathfrak{H})$ there exist bounded sequences $\{\psi_{n}\}$ and 
 $\{\varphi_{n}\}$ and a sequence $\{\alpha_n\}$  of complex numbers such that 
 \bea
 \sum_{n}|\alpha_n| < \infty
 \ena
and 
\bea\label{Teq00}
T = \sum_{n}\alpha_n E_{\varphi_n, \psi_n}.
\ena
The latter series converges with respect to the trace norm and hence
\bea
\omega(T) &=& \sum_{n}\alpha_n \omega(E_{\varphi_n, \psi_n})\cr
&=& \sum_{n}\alpha_n \omega_0(E_{\varphi_n, \psi_n}) = \omega_{0}(T)= \mbox{Tr}(AT).
\ena
\noi Thus $\mathcal{L}(\mathfrak{H})$ is just the dual of $\mathcal{T}(\mathfrak{H}).$ 
\noi The weak$^{*}$ topology on $\mathcal{L}(\mathfrak{H})$ arising from this duality is given by the seminorms
\bea
A \in \mathcal{L}(\mathfrak{H}) \mapsto |\mbox{Tr}(AT)|.
\ena
Now, for $T$ as in (\ref{Teq00}), one has 
\bea
\mbox{Tr}(AT) &=& \sum_{n}\alpha_n \mbox{Tr}(E_{\varphi_n, \psi_n}A)\cr
&=& \sum_{n}\alpha_n (\psi_n, A\varphi_n).
\ena
Thus the seminorms are equivalent to the seminorms defining the $\sigma$-weak topology.
$\hfill{\square}$\\\\
It follows the theorem below:
\betheo\label{normdens00} (\cite{bratelli}, p.76)
\noi Let $\omega$ be a state on a von Neumann algebra $\mathfrak{A}$ acting on a Hilbert 
space 
$\mathfrak{H}. $ The following conditions are equivalent:
\bei
\item[(1)] $\omega$ is normal;
\item[(2)] $\omega$ is $\sigma$-weakly continuous; 
\item[(3)] there exists a density matrix $\rho, $ i.e., a positive trace-class operator $\rho$ on $\mathfrak{H}$ with $Tr(\rho) = 1, $ such
that 
\bea
\omega(A) = Tr(\rho A).
\ena
\eni
\entheo
\noi {\bf Proof.}  (See \cite{bratelli} pp.76 to 78).
\noi (3) $\Rightarrow$ (2) follows from Proposition \ref{Traceprop00} and (2) $\Rightarrow$ (1) from Lemma \ref{lem00}. Next show (2) $\Rightarrow$ (3). If $\omega$ 
is $\sigma$-weakly continuous there exist sequences $\{\xi_n\}, \, \{\eta_n\}$ of vectors such that $\displaystyle\sum_n ||\xi_n||^2 < \infty, \, \displaystyle\sum_n ||\eta_n||^2 < \infty, $ and $\omega(A) = \displaystyle\sum_n (\xi_n, A \eta_n). $ Define $\tilde{\mathfrak{H}} = \displaystyle\bigoplus_{n=1}^{\infty}\mathfrak{H}_n$ and introduce a representation $\pi$ of $\mathfrak{A}$ on $\tilde{\mathfrak{H}}$ by 
 $\pi(A)(\displaystyle\bigoplus_{n}\psi_n)  = \displaystyle\bigoplus_{n}(A \psi_n). $ Let $\xi = \displaystyle\bigoplus_{n} \xi_n, \, \eta = \displaystyle\bigoplus_{n} \eta_n$ and then $\omega(A) = (\xi, \pi(A)\eta).$
\noi Since $\omega(A)$ is real for $A \in \mathfrak{A}_{+}$ (with $\mathfrak{A}_{+}$ denoting  the {\it positive part} of the von Neumann algebra $\mathfrak{A}$ or the set of positive elements of the von Neumann algebra $\mathfrak{A}$), we have 
\bea
4\omega(A) &=& 2(\xi, \pi(A)\eta) + 2(\xi, \pi(A^{*})\eta) \cr
&=&2(\xi, \pi(A)\eta) + 2(\eta, \pi(A)\xi)\cr
&=& (\xi + \eta, \pi(A)(\xi + \eta))- (\xi - \eta, \pi(A)(\xi - \eta)) \cr
&\leq & (\xi + \eta, \pi(A)(\xi + \eta)).
\ena
\noi Hence, by Theorem \ref{repthem00} there exists a positive $T \in \pi(\mathfrak{A})'$ with $0 \leq T \leq \Id/2$ such that 
\bea
(\xi, \pi(A)\eta) &=& (T(\xi + \eta), \pi(A)T(\xi + \eta))\cr
&=& (\psi, \pi(A)\psi).
\ena 
\noi Now $\psi \in \tilde{\mathfrak{H}}$  has the form $\psi = \displaystyle\bigoplus_{n} \psi_n, $ and therefore
\bea
\omega(A) = \sum_n (\psi_n, A \psi_n).
\ena
\noi The right side of this relation can be used to extend $\omega$ to a $\sigma$-weakly continuous positive linear functional $\tilde{\omega}$ on $\mathcal{L}(\mathfrak{H}). $
Since $\tilde{\omega}(\Id) = 1, $ it is a state. Thus, by Proposition \ref{Traceprop00} there exists a trace-class operator $\rho$ with $Tr(\rho) = 1$ such that 
\bea
\tilde{\omega}(A) = Tr (\rho A).
\ena
Let $P$ be the rank one projector with range $\xi$; then 
\bea
(\xi, \rho \xi) = Tr (P \rho P ) = Tr(\rho P) = \tilde{\omega}(P) \geq 0. 
\ena
\noi Thus $\rho$ is positive. 
\noi Turn now to the proof of (1) $\Rightarrow$ 2. Assume that $\omega$ is a normal state on $ \mathfrak{A}. $ Let $\{B_{\alpha}\}$ be an increasing net of elements in $\mathfrak{A}_{+}$
 such that $||B_{\alpha}|| \leq 1$ for all $\alpha$ and such that $A \mapsto \omega(AB_{\alpha})$ is $\sigma$-strongly continuous for all $\alpha.$ One can use Lemma \ref{lem00} to define $B$ by 
\bea
B = {\mbox{l.u.b.}}_{\alpha} \, B_{\alpha} = \sigma\mbox{-strong} \lim_{\alpha} B_{\alpha}.
\ena
\noi Then $0 \leq B \leq \Id$ and $B \in \mathfrak{A}. $ But for all $A \in \mathfrak{A}$ we have 
\bea
|\omega(AB - AB_{\alpha})|^2 &=& |\omega(A(B-B_\alpha)^{1/2}(B-B_\alpha)^{1/2})|^2 \cr
&\leq & \omega(A(B-B_\alpha)A^{*})\omega(B-B_\alpha)\cr
&\leq & ||A||^2\omega(B-B_\alpha).
\ena
\noi Hence
\bea
||\omega(\cdot B)- \omega(\cdot B_\alpha)|| \leq (\omega(B-B_\alpha))^{1/2}.
\ena 
\noi But $\omega$ is normal. Therefore $\omega(B-B_\alpha) \rightarrow 0$ and $\omega(\cdot B_\alpha)$ tends to  $\omega(\cdot B)$ in norm. As $\mathfrak{A}_{*}$ is a Banach space, $\omega(\cdot B) \in \mathfrak{A}_{*}.$ Now,  applying Zorn's lemma, we can find a maximal element $P \in \mathfrak{A}_{+} \cap \mathfrak{A}_{1}$  such that $A \mapsto \omega(AP)$ is $\sigma$-strongly continuous. If $P = \Id$ the theorem is proved. Assume {\it ad absurdum} 
that $P \neq \Id.$ Put $P' = \Id - P$ and choose $\xi \in \mathfrak H$ such that  $\omega(P') < (\xi, P'\xi). $ If $\{B_\alpha\}$ is an increasing net in $\mathfrak{A}_{+}$ such that $B_\alpha \leq P', $ $\omega(B_\alpha) \geq (\xi, B_\alpha \xi), $ and $B = {\mbox{l.u.b.}}_{\alpha} \, B_{\alpha} = \sigma$-strong $\lim_{\alpha} B_{\alpha}, $ then $B \in \mathfrak{A}_{+}, B \leq P', $ and $\omega(B) = \mbox{sup}\, \omega(B_\alpha) \geq \mbox{sup}\,(\xi, B_\alpha \xi) = (\xi, B \xi). $ Hence, by Zorn's lemma, there exists a maximal $B \in \mathfrak{A}_{+}$ such that $B \leq P'$ and $\omega(B) \geq (\xi, B \xi). $ Take $Q = P'-B. $ Then, $Q \in \mathfrak{A}_{+}, Q \neq 0$, since $\omega(P') < (\xi, P'\xi), $ and if $A \in \mathfrak{A}_{+}, \, A \leq Q, \, A \neq 0, $ then $\omega(A) < (\xi, A\xi)$ by the maximality of $B.$

\noi For any $A \in \mathfrak{A}$ one has 
\bea
QA^{*}AQ \leq ||A||^2Q^2 \leq ||A||^2 ||Q|| Q. 
\ena 
\noi Hence $(QA^{*}AQ)/ ||A||^2||Q|| \leq Q$ and $\omega(QA^{*}AQ)< (\xi, QA^{*}AQ\xi)$. Combining this with the Cauchy-Schwartz inequality one finds
\bea
|\omega(AQ)|^2 &\leq &  \omega(\Id)\omega(QA^{*}AQ) \cr
& < & (\xi, QA^{*}AQ\xi) = ||AQ\xi ||^2.
\ena
\noi Thus both $A \mapsto \omega(AQ)$ and $A \mapsto \omega(A(P+Q))$ are $\sigma$-strongly continuous. Since $P + Q \leq \Id, $ this contradicts the maximality of $P.$
$\hfill{\square}$
\beprop\label{normfaith00} (\cite{bratelli} p.86)
\noi Let $\mathfrak{A}$ be a von Neumann algebra on a Hilbert space $\mathfrak{H}.$
\noi Then the following four conditions are equivalent:
\bei
\item[(1)] $\mathfrak{A}$ is $\sigma$-finite;
\item[(2)] there exists a countable subset of $\mathfrak{H}$ which is separating for $\mathfrak{A}$;
\item[(3)] there exists a faithful normal state on $\mathfrak{A}$; 
\item[(4)] $\mathfrak{A}$ is isomorphic with a von Neumann algebra $\pi(\mathfrak{A})$ 
which admits a separating and cyclic vector.
\eni
\enprop
\noi {\bf Proof.} (see \cite{bratelli} p.86)
\noi (1) $\Rightarrow$ (2) Let $\{\xi_{\alpha}\}$ be a maximal family of vectors in $\mathfrak{H}$ 
such that $[\mathfrak{A'}\xi_{\alpha}]$ and $[\mathfrak{A'}\xi_{\alpha'}]$ are orthogonal whenever $\alpha \neq \alpha'. $ Since $[\mathfrak{A'}\xi_{\alpha}]$ is a projection in $\mathfrak{A}$ (in fact the smallest projection in $\mathfrak{A}$ containing $\xi_{\alpha}), $
$\{\xi_{\alpha}\}$ is countable. But by the maximality, 
\bea
\sum_{\alpha}[\mathfrak{A'}\xi_{\alpha}] = \Id.
\ena
\noi Thus $\{\xi_{\alpha}\}$ is cyclic for $\mathfrak{A'}.$ Hence $\{\xi_{\alpha}\}$ is separating for $\mathfrak{A}$ by Proposition 
\ref{propos00}.
\noi (2) $\Rightarrow$ (3) Choose a sequence $\xi_n$ such that the set $\{\xi_n\}$ is separating for $\mathfrak{A}$ and such that $\displaystyle \sum_{n}||\xi_n||^2 = 1. $ Define $\omega$ by
\bea
\omega(A) = \sum_{n}(\xi_n, A\xi_n).
\ena
\noi $\omega$ is $\sigma$-weakly continuous, hence normal, by using Theorem \ref{normdens00}. If $\omega(A^{*}A) = 0$ then $0 = (\xi_n, A^{*}A\xi_n) = ||A\xi_n||^2$ for all $n, $ hence $A = 0.$ 
\noi (3) $\Rightarrow$ (4) Let $\omega$ be a faithful normal state on $\mathfrak{A}$ and $(\mathfrak{H}, \pi, \Omega)$ the corresponding
cyclic representation. Since $\pi(\mathfrak{A})$ is a von Neumann algebra, if $\pi(A)\Omega = 0$ for an $A \in \mathfrak{A}$ then $\omega(A^{*}A) = ||\pi(A)\Omega||^2 = 0, $ hence $A^{*}A = 0$ and $A = 0.$ This proves that $\pi$ is faithful and $\Omega$ separating for $\pi(\mathfrak{A}).$
\noi (4) $\Rightarrow$ (1) Let $\Omega$ be the separating (and cyclic) vector for $\pi(\mathfrak{A}), $ and let $\{E_\alpha\}$ be a family of mutually orthogonal projections in
$\mathfrak{A}. $ Set $E = \displaystyle\sum_{\alpha} E_{\alpha}. $ Then
\bea
||\pi(E)\Omega||^2 &=& (\pi(E)\Omega, \pi(E)\Omega) \cr
&=& \sum_{\alpha, \alpha'}(\pi(E_\alpha)\Omega, \pi(E_{\alpha'})\Omega) \cr
&=& \sum_{\alpha} ||\pi(E_\alpha)\Omega||^2
\ena
by Lemma \ref{lem00}.
\noi Since $\displaystyle\sum_{\alpha}||\pi(E_\alpha)\Omega||^2 <  +\infty, $  only a countable number of
the $\pi(E_\alpha)\Omega$ is nonzero, and thus the same is true for the $E_{\alpha}.$
$\hfill{\square}$
 
\subsection{Hilbert space of Hilbert-Schmidt operators}
\noi Here we recall  some definitions provided in \cite{ali-bagarello, ali-bagarello-honnouvo, prugovecki} (and references therein).
\bedefi The trace of a linear operator
\noi A linear operator $A$ defined on the separable Hilbert space $\mathfrak{H}$ is said to be of {\it trace class} if the series 
$\displaystyle\sum_{k}\langle e_k|A e_k\rangle$ converges and has the same value in any orthonormal basis $\{e_k\}$ of $\mathfrak{H}.$
\noi The sum
\bea
\mbox{Tr} A = \sum_{k}\langle e_k|A e_k\rangle
\ena
\noi is called the {\it trace} of $A.$
\findefi
\bedefi Trace norm  
\noi Consider the class of Hilbert-Schmidt operators. For every such operator $A$, the {\it trace norm} is given by 
\bea\label{seri00}
\mbox{Tr} [\sqrt{A^{*}A}]  = \mbox{Tr} [
\sum_{k}|e_{k}\rangle \lambda_{k}\langle e_{k}|] = \sum_{k}\lambda_{k} < +\infty.
\ena
\findefi
\berem
If $A$ is any operator of trace class, then $A^{*}$ is also of  trace class:
\bea
\mbox{Tr}  A^{*} = \sum_{k}\langle e_k|A^{*} e_k\rangle = \sum_{k}\langle e_k|A e_k\rangle^{*} = (\mbox{Tr}  A)^{*}.
\ena
\enrem
\bedefi  Hilbert-Schmidt operator 
\noi Given a  bounded operator, having the decomposition $A = \displaystyle\sum_{k}|\phi_{k}\rangle\lambda_{k}\langle \phi_{k}|$, 
where $\{\phi_{k}\}$ is an othonormal basis of  $\mathfrak H$, and $\lambda_{1}, \lambda_{2}, \dots$  positive numbers, A 
is called a   {\it Hilbert-Schmidt operator}  if 
\bea
\mbox{Tr} [AA^{*}]  = 
\sum_{k}\langle \phi_{k}|A^{*}A \phi_{k}\rangle = \sum_{k}\lambda^{2}_{k} < +\infty.
\ena
\findefi
\berem
\noi If the series (\ref{seri00}) is infinte, its convergence implies that $\lambda_k \rightarrow 0$ when $k \rightarrow + \infty.$
Consequently, $\lambda^2_k \geq \lambda_k$ for sufficiently large value of $k. $ Hence, $\displaystyle\sum_k \lambda^2_k$ converges when 
$\displaystyle\sum_k \lambda_k$ converges. This shows that any completely continuous operator $A$ satisfying (\ref{seri00}) is a Hilbert-Schmidt operator.
\enrem
\bedefi Hilbert-Schmidt norm  
\noi For any Hilbert-Schmidt operator $A, $ the quantity
\bea
||A||_2 = \sqrt{\mbox{Tr}[A^{*}A]}
\ena
\noi exists, and is called {\it Hilbert-Schmidt norm} of $A.$
\findefi
\bedefi 
\noi Let $\mathcal B_{2}(\mathfrak H)$, $\mathcal B_{2}(\mathfrak H) \subset \mathcal L(\mathfrak H)$ 
the set of all bounded operators on $\mathfrak H$,
be the Hilbert space of Hilbert-Schmidt 
operators on $\mathfrak H = L^{2}(\RN)$, with the scalar product 
\bea\label{scalprod00}
\langle X|Y \rangle_{2} = Tr[X^{*}Y] = \sum_{k}\langle \Phi_{k}|X^{*}Y\Phi_{k}\rangle,
\ena 
where $\{\Phi_{k}\}^{\infty}_{k=0}$ is an orthonormal basis of $\mathfrak H$.
\findefi
\noi $\mathcal B_{2}(\mathfrak H)\simeq \mathfrak H \otimes \bar{\mathfrak H}$ (where $\bar{\mathfrak H}$ denotes the dual of $\mathfrak H$) 
and basis vectors of $\mathcal B_{2}(\mathfrak H)$ are  given by
\begin{eqnarray}{\label{basis01}}
\Phi_{nl} := |\Phi_{n}\rangle \langle \Phi_{l}|, \quad n,  l = 0,1,2,\dots, \infty.
\end{eqnarray}
\berem In the notation  $B_2(\mathfrak H) \simeq \mathfrak H
\otimes \overline{\mathfrak H}$, $\,\mathfrak H \otimes \overline
{\mathfrak H}$ is taken here as the completude of the algebraic tensor product of
 $\mathfrak H$ by $\overline{\mathfrak H}$  which is a prehilbert space containing finite sums of the type
$ \displaystyle\sum_{j,k=0}^{n} \lambda_{jk} |\phi_j\rangle \otimes |\overline{\phi_k\rangle}$,
where a basis of $\mathfrak H \otimes \overline{\mathfrak H}$ is 
$ \{|\phi_j\rangle  \otimes |\overline{\phi_k\rangle}\}_{j,k=0}^{\infty}. $
Then,  $B_2(\mathfrak H)$, being the Hilbert space of 
Hilbert-Schmidt operators on $\mathfrak H$ is isomorphic to  $\mathfrak H \otimes \overline{\mathfrak H}$, since the separable Hilbert spaces are taken two by two isomorphic each other. 
Setting 
$ |\phi_j\rangle \otimes |\overline{\phi_k\rangle} = |\phi_j\rangle \langle\phi_k|, $
$B_2(\mathfrak H)$ admits for orthonormal basis $\{\phi_{jk}\}_{j,k=0}^{\infty}$ such that 
$\phi_{jk} := |\phi_j\rangle \langle\phi_k| $.
\enrem
\bedefi 
\noi Let  $A$ and  $B$ two operators on  $\mathfrak H$. The operator 
$A \vee B$ is such that 
\begin{eqnarray}\label{opera00}
A \vee B (X) = AXB^{*}, \; X \in B_2(\mathfrak H).
\end{eqnarray}
\findefi
 
\noi For  bounded linear operators $A$ and $B$,  $A \vee B$ defines a linear operator on $B_2(\mathfrak H)$.
\noi Indeed, $\forall A, B \in \mathcal{L}(\mathfrak H),$ (the space of bounded linear operators on $\mathfrak H$), since
$B_2(\mathfrak H)\subset \mathcal{L}(\mathfrak H),$ we get 
 $\forall X \in B_2(\mathfrak H),$
$X \in \mathcal{L}(\mathfrak H)$.
Then,  $AXB^{*} \in \mathcal{L}(\mathfrak H)$, i.e.
$(A \vee B) \in \mathcal{L}(\mathfrak H).$
Thus, $A \vee B$  defines a bounded linear operator on 
$B_2(\mathfrak H)$.

\noi From the scalar product in $\mathcal B_2(\mathfrak H),$
\bea
\langle X|Y \rangle_2 = Tr [X^*Y], \quad X, Y \in B_2(\mathfrak H),
\ena
it comes 
\bea
Tr[X^*(AYB^*)] = Tr[(A^*XB)^*Y]\Rightarrow (A \vee B)^* = A^* \vee B^*.
\ena
Since for  any $X \in \mathcal B_2(\mathfrak H)$, 
\bea
(A_1 \vee B_1)(A_2 \vee B_2)(X) = A_1[(A_2 \vee B_2)(X)]B^{*}_1 = A_1 A_2 X B^{*}_2 B^{*}_1,
\ena
we have
\bea
(A_1 \vee B_1)(A_2 \vee B_2) = (A_1 A_2) \vee (B_1 B_2).
\ena
 
\subsection{Modular theory and Hilbert-Schmidt operators}
This paragraph is devoted to Tomita-Takesaki modular theory \cite{tomita, takesaki, summers} of von Neumann algebras 
\cite{vonneumann, vonneumann1}.  Recall that the origins of Tomita-Takesaki modular theory lie in two unpublished papers of M.
Tomita in 1967 \cite{tomita} and a slim volume  by M. Takesaki.  As one of the
most important contributions in the operator algebras, this theory  finds many applications
in mathematical physics. 

We provide some key ingredients from \cite{bratelli, ali-bagarello, ali-bagarello-honnouvo, bertozzini, takesaki} as needed for this section. 
First, let us deal with some notions from \cite{bratelli}:
\bee
\item  Let $\mathfrak{A}$ be a von Neumann algebra on a Hilbert space $\mathfrak{H}$ and $\mathfrak{A'}$ its commutant. 
Let $\Phi \in \mathfrak{H}$ be a unit vector which is cyclic and separating for $\mathfrak{A}.$  
This is the case, if  $\mathfrak{A}$ is a $\sigma$-finite von Neumann algebra, and by applying  Proposition \ref{propos00}. 
\item The mapping $A \in \mathfrak{A}\mapsto A\Omega \in \mathfrak{H}$ then establishes a one-to-one linear
correspondence between $\mathfrak{A}$ and a dense subspace $\mathfrak{A}\Omega$ of $\mathfrak{H}$. Let $S_0$ and $F_0$ 
two antilinear operators on $\mathfrak{A}\Omega$ and $\mathfrak{A'}\Omega, $ respectively. By Proposition \ref{propos00}, 
$\Omega$ is cyclic and separating for $\mathfrak{A}$ and $\mathfrak{A'}$. Therefore the two antilinear
operators $S_0$ and $F_0$, given by
\bea
S_0 A \Omega &=& A^{*}\Omega, \quad \mbox{for} \quad A \in \mathfrak{A} \cr
F_0 A' \Omega &=& A'^{*}\Omega \quad \mbox{for} \quad A' \in \mathfrak{A'}
\ena
are both well defined on the dense domains on $D(S_0) = \mathfrak{A}\Omega$ and $D(F_0) = \mathfrak{A'}\Omega.$ 
\noi Then follows the definition:
\bedefi\label{defini00} 
Define $S$ and $F$ as the closures of $S_0$ and $F_0, $ respectively, i.e., 
\bea
S = \bar{S}_0, \quad F = \bar{F}_0
\ena
where the bar denotes the closure.  Let $\Delta$ be the unique, positive, selfadjoint operator and $J$ the unique antiunitary operator 
occurring in the polar decomposition 
\bea
S = J \Delta^{1/2}
\ena
of $S$. $\Delta$ is called the {\it modular operator} associated with the pair $\{\mathfrak{A}, \Omega\}$ and $J$ the {\it modular conjugation}. 
\findefi
\noi The following proposition provides connections between $S, F, \Delta$ and $J$:
\beprop (\cite{bratelli} p.89)
The following relations are valid: 
\bea
\Delta &=& FS, \qquad \Delta^{-1} = SF \cr
S &=& J \Delta^{1/2}, \qquad F = J\Delta^{-1/2}\cr
J &=& J^{*}, \qquad J^2 = I_{\mathfrak{H}} \cr
\Delta^{-1/2} &=& J \Delta^{1/2} J.
\ena
\enprop
\noi {\bf Proof.} (see \cite{bratelli} pp.89-90)
\noi $\Delta = S^{*}S = FS, $  and $S= J \Delta^{1/2}$ by Definition \ref{defini00}. Using the fact that for any $\psi \in D(\bar{S}_{0})$ there exists a closed operator $Q$ on $\mathfrak{H}, $with $S^{*}_0 = \bar{F}_0, F^{*}_{0} = \bar{S}_{0}, $ such that 
\bea
Q\Omega = \psi, \quad Q^{*}\Omega = \bar{S}_{0}\psi
\ena
where $\mathfrak{A'}D(Q) \subseteq D(Q), \, QQ' \supseteq Q'Q$ for all $Q' \in \mathfrak{A'}$, 
with $S_0 = S^{-1}_0$,  it follows by closure that $S = S^{-1}, $ and hence 
\bea
J\Delta^{1/2} = S = S^{-1} = \Delta^{1/2} J^{*},
\ena
so that $J^{2}\Delta^{1/2} = J\Delta^{-1/2} J^{*}$. Since $J\Delta^{-1/2} J^{*}$ is a positive operator, and by the uniqueness of the polar decomposition one deduces that
\bea
J^{2} = I_{\mathfrak{H}}
\ena
and then 
\bea
J^{*} = J, \quad \Delta^{-1/2} = J \Delta^{1/2} J.
\ena
But this implies that 
\bea
F = S^{*} = (\Delta^{-1/2} J)^{*} = J \Delta^{-1/2}
\ena
and 
\bea
SF = \Delta^{-1/2} JJ \Delta^{-1/2}  = \Delta^{-1}.
\ena
$\hfill{\square}$

\item The principal result of the Tomita-Takesaki theory \cite{takesaki, bratelli} is that the following relations
\bea
J \mathfrak{A} J  = \mathfrak{A'}, \qquad \Delta^{it}\mathfrak{A}\Delta^{-it} = \mathfrak{A}
\ena
hold for all $t \in \RN$. 
\item 
\bedefi Modular automorphism group 
Let $\mathfrak{A}$ be a von Neumann algebra, $\omega$ a faithful, normal state on $\mathfrak{A}$, 
$(\mathfrak{H}_{\omega}, \pi_\omega, \Omega_\omega)$ the corresponding cyclic representation, and $\Delta$ the modular operator associated 
with the pair $(\omega(\mathfrak{A}),  \Omega_\omega)$. The Tomita-Takesaki theorem establishes the existence of a $\sigma$-weakly
continuous one-parameter group $t \mapsto \sigma^{\omega}_t$ of $^{*}$-automorphisms of $\mathfrak{A}$ through the definition 
\bea\label{auto00}
\sigma^{\omega}_t(A) = \pi^{-1}_\omega(\Delta^{it}\pi_\omega(A)\Delta^{-it}).
\ena
The group $t \mapsto \sigma^{\omega}_t$ is called {\it the modular automorphism group associted with the pair $(\mathfrak{A}, \omega)$}.
\findefi
\item 
\bedefi $C^{*}$-{\it dynamical system} 
\noi A $C^{*}$-dynamical system $(\mathfrak{G}, \alpha)$ is a $C^{*}$-algebra $\mathfrak{G}$ equipped with a group homomorphism 
$\alpha: G \rightarrow \mbox{Aut} (\mathfrak{G})$ that is strongly continuous i.e., $g \mapsto ||\alpha_g(x)||$ is a continuous 
map for all $x \in \mathfrak{G}$. 
\noi A {\it  von Neumann dynamical system} $(\mathfrak{A}, \alpha)$ is a von Neumann algebra acting on the Hilbert space $\mathfrak{H}$ 
equipped with a group homomorphism $\alpha: G \rightarrow \mbox{Aut}(\mathfrak{A})$ that is weakly continuous i.e., 
$g\mapsto \langle \xi|\alpha_{g}(x)\eta\rangle$ is continuous for all $x \in \mathfrak{A}$ and all $\xi, \eta \in \mathfrak{H}.$ 

\findefi
 \bedefi
\noi A state $\omega$ on a one-parameter $C^{*}$-dynamical system $(\mathfrak{G}, \alpha)$ is a $(\alpha, \beta)$-{\it KMS state}, 
for $\beta \in \RN, $ if for all pairs of elements $x,y$ in a norm dense $\alpha$-invariant ${*}$-subalgebra of $\alpha$-analytic 
elements of $\mathfrak{G}$, then $\omega(x \alpha_{i\beta}(y)) = \omega(yx).$
\findefi
\berem
In the case of a von Neumann dynamical system $(\mathfrak{A}, \alpha), $ a $(\alpha, \beta)$-KMS state must be normal
(i.e., for every increasing bounded net of positive elements $x_\lambda \rightarrow x$, we have $\omega(x_\lambda) \rightarrow \omega(x)$). 
Besides, given  $\alpha: \RN \rightarrow \mbox{Aut}(A), $ an element $x \in \mathfrak{G}$ is $\alpha$-{\it analytic} if there exists a 
holomorphic extension of the map $t \mapsto \alpha_t(x)$ to an open horizontal strip $\{z \in \CN| |\mbox{Im}\, z| < r\}$, with $r > 0, $ in 
the  complex plane. The set of $\alpha$-analytic elements is always $\alpha$-invariant (i.e., for all $x$ is analytic, $\alpha(x)$ is 
analytic)$^{*}$-subalgebra of $\mathfrak{G}$ that is norm dense in the $C^{*}$ case and weakly dense in the von Neumann case.
\enrem
\item The modular automorphism group associated with $\omega$ is only the one parameter automorphism group that satisfies 
the Kubo-Martin-Schwinger (KMS)-condition with respect to the state $\omega, $ at inverse temperature $\beta, $ i.e., 
\bea
\omega(\sigma^{\omega}_t(x)) = \omega(x), \quad \forall x \in \mathfrak{A}
\ena
and for all $x,y \in \mathfrak{A}, $ there exists a function $F_{x,y}: \RN \times [0, \beta] \rightarrow \CN$ such that: 
\bea
&&F_{x,y} \quad \mbox{is holomorphic on} \quad  \RN \times ]0, \beta[,  \cr
&& F_{x,y} \quad \mbox{is bounded continuous on} \quad \RN \times [0, \beta], \cr
&& F_{x,y}(t) = \omega(\sigma^{\omega}_t(y)x), \quad t \in \RN, \cr
&& F_{x,y}(i \beta + t) = \omega (x \sigma^{\omega}_t(y)), \quad t \in \RN.
\ena
\ene 
\noi  \subsubsection{Kubo-Martin-Schwinger   (KMS) state}
 \noi Let $\alpha_i, \, i=1, 2, \dots, N$ be a sequence of non-zero, positive numbers, satisfying :  
 $\displaystyle\sum^{N}_{i=1}  \alpha_{i} = 1.$
\noi Let 
\bea\label{vectphi00}
\Phi = \sum^{N}_{i=1} \alpha^{\frac{1}{2}}_i \PN_i = \sum^{N}_{i=1} \alpha^{\frac{1}{2}}_i X_{ii} \in \mathcal{B}_2(\mathfrak{H}) \quad 
\mbox{with} \quad X_{ii} = |\zeta_i\rangle \langle \zeta_i| 
\ena
 $\{\zeta_i\}^{\infty}_{i=0}$ being   an orthonormal basis of $\mathfrak H, $ and the vectors
$\{X_{ij} = |\zeta_i\rangle \langle \zeta_j|, i,j =1,2,\dots, N \}$ forming an orthonormal basis of $\mathcal{B}_2(\mathfrak{H}),$ 
\bea
\langle X_{ij}|X_{kl}\rangle_2  = \delta_{ik} \delta_{lj}.
\ena
In particular, the vectors  
\bea\label{vectphi01}
\mathbb P_{i} = X_{ii} = |\zeta_i\rangle \langle \zeta_i| 
\ena
are one-dimensional projection operators on $\mathfrak H.$ 
Then, we have the following properties:
\bei
\item[(i)] 
\beprop
\noi $\Phi$ defines a vector state $\varphi$ on the von Neumann algebra $\mathfrak{A}_l$ corresponding to the operators given with
$A$ in the left of the identity operator $I_{\mathfrak H}$ on $\mathfrak H, $ i.e.,  
$\mathfrak{A}_l= \left\{A_{l} = A \vee I| A \in \mathcal L(\mathfrak H)\right\}.$
\enprop
\noi {\bf Proof.}
\noi Indeed, for any $A \vee I \in \mathfrak{A}_l, $ since $\mathcal{B}_2(\mathfrak{H}) 
\subset \mathcal{L}(\mathfrak{H})$ and $\mathfrak{A} \subset  \mathcal{L}(\mathfrak{H})$, 
from the Remark \ref{impdefi01} and the equality (\ref{impdefi03}) together,  the state $\varphi$ on $\mathfrak{A}_l$ may be defined by 
\bea\label{densmat00}
\langle \varphi; A \vee I \rangle  = \langle \Phi|(A \vee I)(\Phi)\rangle_2 = \mbox{Tr}[\Phi^{*}A\Phi] = \mbox{Tr}[\rho_\varphi A], \quad \mbox{with} \quad \rho_\varphi = \sum^{N}_{i=1} \alpha_i \PN_i.
\ena
$\hfill{\square}$

\item[(ii)]
\beprop
\noi The state $\varphi$ is faithful and normal.
\enprop
\noi {\bf Proof.}
\noi The state $\varphi$ is normal by Theorem \ref{normdens00}, using the fact that $\rho_\varphi$ is a density matrix and since we have 
\bea 
\langle \varphi; A \vee I \rangle  =  \mbox{Tr}[\rho_\varphi A].
\ena 
Its faithfulness comes from Proposition \ref{normfaith00} by use of the equivalence (2) $\Leftrightarrow$ (3), since $\mathfrak{A}_l \subset \mathcal{L}(\mathfrak{H})$ using the Eq.(\ref{impdefi03}) in the  Remark \ref{impdefi01}, we have,  with $\PN  = |\zeta_i \rangle \langle \zeta_i|, $ 
\bea
\langle \varphi; (A \vee I)^{*}(A \vee I)\rangle &:=& \varphi\left(\left\{(A \vee I)^{*}(A \vee I) \right\}\right)\cr
&=&\mbox{Tr}[\rho_\varphi A^{*}A]  \qquad [\mbox{by}\,  (\ref{impdefi03})]\cr
&=& \sum^{N}_{k=1}\langle \zeta_k|\rho_\varphi A^{*}A|\zeta_k\rangle  \qquad[\mbox{by}\,  (\ref{scalprod00})] \cr
&=& \sum^{N}_{k=1}\langle \zeta_k|\left\{\sum^{N}_{i=1}\alpha_i |\zeta_i\rangle \langle \zeta_i|\right\} A^{*}A|\zeta_k\rangle \qquad[\mbox{by}\,  (\ref{densmat00})] \cr
&=& \sum^{N}_{k=1}\sum^{N}_{i=1}\alpha_i \langle \zeta_i|A^{*}A|\zeta_k\rangle\ \langle \zeta_k|\zeta_i\rangle \cr
&=& \sum^{N}_{i=1}\alpha_i ||A\zeta_i||^2, \quad \alpha_i > 0, \quad A \in \mathcal{L}(\mathfrak{H})
\ena
\noi where the $\{\zeta_i\}^N_{i=1}$ form an orthonormal basis set of $\mathfrak{H}$. $\Phi$ is separating for $\mathfrak{A}_l$,  by use of Theorem \ref{normdens00}, and the relation
\bea
\langle \varphi; (A \vee I)^{*}(A \vee I)\rangle = 0 &\Longleftrightarrow &  \sum^{N}_{i=1}\alpha_i ||A\zeta_i||^2 = 0, \qquad  \forall i =1,2,\dots, N\cr
& \Longleftrightarrow & A \vee I = 0 \Longleftrightarrow A = 0. 
\ena
\noi Thereby,  $\langle \varphi; (A \vee I)^{*}(A \vee I)\rangle = 0$ if and only if $A \vee I = 0.$
$\hfill{\square}$

\item[(iii)]
\beprop
\noi The vector $\Phi$ is cyclic and separating for $\mathfrak{A}_l.$
\enprop
\noi {\bf Proof.}
\noi If $X \in \mathcal{B}_{2}(\mathfrak{H})$ is orthogonal to all $(A \vee I)\Phi, A \in \mathcal{L}(\mathfrak{H}), $ then 
\bea
\mbox{Tr}[X^{*}A\Phi] = \sum_{i=1}^{N}\alpha^{\frac{1}{2}}_i\langle \zeta_i| X^{*}A \zeta_i\rangle = 0, \qquad \forall A \in \mathcal{L}(\mathfrak{H}).
\ena
Taking $A = X_{kl}$, it follows from the above equality, $\langle \zeta_l|X^{*}\zeta_k\rangle  = 0$ and since this holds for all $k, l, $ we get $X = 0.$
Indeed, let $\Phi =  \displaystyle \sum^{N}_{i=1} \alpha^{\frac{1}{2}}_i \PN_i = \displaystyle\sum^{N}_{i=1} \alpha^{\frac{1}{2}}_i X_{ii} = \displaystyle\sum^{N}_{i=1} \alpha^{\frac{1}{2}}_i |\zeta_i\rangle \langle \zeta_i|, $ then
by definition, see Eq.(\ref{scalprod00}),
\bea\label{eqorth00}
\langle X|(A \vee I)\Phi \rangle_{2} &=& \mbox{Tr}\left[X^{*}(A \vee I)\Phi\right]\cr
&=& \mbox{Tr}\left[X^{*}A\Phi I^{*} \right] \qquad   [\mbox{by}\,  (\ref{opera00})]\cr
&=& \mbox{Tr}\left[X^{*}A\Phi\right]\cr
&=& \sum^{N}_{k=1} \langle \zeta_k|X^{*}A\Phi|\zeta_k\rangle \qquad   [\mbox{by} \,  (\ref{scalprod00})]\cr
&=& \sum^{N}_{k=1} \langle \zeta_k|X^{*}A\left\{\sum^{N}_{i=1} \alpha^{\frac{1}{2}}_i |\zeta_i\rangle \langle \zeta_i|\right\}|\zeta_k\rangle  \qquad   [\mbox{by}\,  (\ref{vectphi00})]\cr
&=& \sum^{N}_{k=1}\sum^{N}_{i=1} \alpha^{\frac{1}{2}}_i \langle \zeta_k|X^{*}A\zeta_i\rangle \langle \zeta_i|\zeta_k\rangle \cr
&=& \sum^{N}_{i=1}\alpha^{\frac{1}{2}}_i\langle \zeta_i| X^{*}A \zeta_i\rangle
\ena
\noi such that the orthogonality implies 
\bea\label{cond00}
\langle X|(A \vee I)\Phi \rangle_{2} = 0 \Longrightarrow \sum_{i=1}^{N}\alpha^{\frac{1}{2}}_i\langle \zeta_i| X^{*}A \zeta_i\rangle = 0.
\ena
\noi Now taking  $A = X_{kl} = |\zeta_k\rangle \langle \zeta_l|, $ it follows that 
\bea\label{cond01}
\sum_{i=1}^{N}\alpha^{\frac{1}{2}}_i\langle \zeta_i| X^{*}A \zeta_i\rangle &=& \sum_{i=1}^{N}\alpha^{\frac{1}{2}}_i\langle \zeta_i| X^{*}\left\{|\zeta_k\rangle \langle \zeta_l|\right\}|\zeta_i\rangle\cr
&=& \sum_{i=1}^{N}\alpha^{\frac{1}{2}}_i\langle \zeta_i| X^{*}\zeta_k\rangle \delta_{il}\cr
&=& \alpha^{\frac{1}{2}}_l\langle \zeta_l| X^{*}\zeta_k\rangle.
\ena
From (\ref{cond00}) and (\ref{cond01}) together, it follows 
\bea
\sum_{i=1}^{N}\alpha^{\frac{1}{2}}_i\langle \zeta_i| X^{*}A \zeta_i\rangle = 0  \Longrightarrow \alpha^{\frac{1}{2}}_l\langle \zeta_l| X^{*}\zeta_k\rangle = 0, \quad \forall \alpha_l > 0.
\ena 
Thereby, 
\bea
\langle \zeta_l|X^{*}\zeta_k\rangle  = 0,  \quad \forall k, l \Longrightarrow X = 0.
\ena
Therefore, we have 
\bea
\langle X|(A \vee I)\Phi \rangle_{2} = 0 \Longrightarrow X = 0
\ena
implying that the set $\left\{(A \vee I)\Phi, A \in \mathfrak{A}_l\right\}$ is dense in $\mathcal{B}_{2}(\mathfrak{H})$, proving from the Definition \ref{vecyclicsepar} that $\Phi$ is cyclic for $\mathfrak{A}_l.$
$\hfill{\square}$

\noi The fact that $\Phi$ is separating for $\mathfrak{A}_l$ is obtained through the relation 
\bea
(A \vee I)\Phi = (B \vee I)\Phi \Longleftrightarrow A \vee I = B \vee I, \quad \forall A, B \in \mathfrak{A}_l.
\ena
\noi {\bf Proof.}
\noi Let $A, B \in \mathfrak{A}_l, $ such that $(A \vee I)\Phi = (B \vee I)\Phi,$ and take 
  $X \neq 0,\, X \in \mathcal{B}_{2}(\mathfrak{H})$. We have
\bea
\langle X|\left\{(A \vee I)-(B \vee I)\right\}\Phi \rangle_{2} 
&=& \mbox{Tr}\left[X^{*}\left\{(A \vee I)-(B \vee I)\right\}\Phi\right]\cr
&=& \mbox{Tr}\left[X^{*}(A-B)\Phi I^{*} \right] \qquad   [\mbox{by}\,  (\ref{opera00})]\cr
&=& \sum^{N}_{k=1} \langle \zeta_k|X^{*}(A-B)\left\{\sum^{N}_{i=1} \alpha^{\frac{1}{2}}_i |\zeta_i\rangle \langle \zeta_i|\right\}|\zeta_k\rangle  \qquad   [\mbox{by}\,  (\ref{vectphi00})]\cr
&=& \sum^{N}_{k=1}\sum^{N}_{i=1} \alpha^{\frac{1}{2}}_i \langle \zeta_k|X^{*}(A-B)\zeta_i\rangle \langle \zeta_i|\zeta_k\rangle \cr
&=& \sum^{N}_{i=1}\alpha^{\frac{1}{2}}_i\langle \zeta_i| X^{*}(A-B) \zeta_i\rangle\cr
&=& \sum_{i=1}^{N}\alpha^{\frac{1}{2}}_i\langle \zeta_i| X^{*}(A-B) \zeta_i\rangle. \quad [\mbox{by} \, (\ref{eqorth00})\; \mbox{and}\; (\ref{cond01})]
\ena
Taking $(A \vee I)\Phi = (B \vee I)\Phi$,  the equality $\langle X|\left\{(A \vee I)-(B \vee I)\right\}\Phi \rangle_{2}  = 0$ leads to 
\bea
\langle X|\left\{(A \vee I)-(B \vee I)\right\}\Phi \rangle_{2}  = 0 &\Longleftrightarrow & 
\sum_{i=1}^{N}\alpha^{\frac{1}{2}}_i\langle \zeta_i| X^{*}(A-B) \zeta_i\rangle = 0, \quad \alpha_i > 0 \cr
& \Longleftrightarrow & A \vee I = B \vee I
\ena
\noi which completes the proof.
$\hfill{\square}$

\noi In the same way, $\Phi$ is also cyclic for $\mathfrak{A}_r = \left\{A_{r} = I \vee A| A \in \mathcal L(\mathfrak H)\right\},$ 
which corresponds to the operators given with $A$ in the right of the identity operator $I_{\mathfrak H}$ on $\mathfrak H, $ 
hence separating for $\mathfrak{A}_r, $ i.e. 
$(I \vee A)\Phi = (I \vee B)\Phi \Longleftrightarrow I \vee A = I \vee B.$
\eni

 \noi Then, starting to the above setup, to the pair 
$\{\mathfrak{A}, \varphi\}$ is  associated:
\bei
\item a {\it one parameter unitary group} $t \mapsto \Delta^{-\frac{i}{t}\beta}_{\varphi} \in \mathcal{L}(\mathfrak{H})$
\item  and a {\it conjugate-linear isometry} $J_{\varphi}: \mathfrak{H} \rightarrow \mathfrak{H}$
that:
\bea
\Delta^{\frac{i}{t}\beta}_{\varphi}\mathfrak{A}\Delta^{-\frac{i}{t}\beta}_{\varphi} = \mathfrak{A}, \; t \in \RN,
\ena
\bea
J_{\varphi} \mathfrak{A} J_{\varphi} = \mathfrak{A'},
\ena
\bea
J_{\varphi} \circ J_{\varphi} = I_{\mathfrak{H}}, \quad J_{\varphi} \circ \Delta^{\frac{i}{t}\beta}_{\varphi} = \Delta^{-\frac{i}{t}\beta}_{\varphi}
\circ J_{\varphi}.
\ena
\eni

\noi Denote the automorphisms by $\alpha_{\varphi}(t)$, and deal with operators $A \in \mathfrak{A}$ with  $\mathfrak{A} \subset \mathcal{L}(\mathfrak{H})$. Then,  taking into account the Definition \ref{impdefi00} and the Remark \ref{impdefi01},  from the expresssion (\ref{auto00}),   the automorphisms, in this case, satisfy the following relation:
\bea\label{auto01}
\alpha_{\varphi}(t)[A] = \Delta^{\frac{i}{t}\beta}_{\varphi}A \Delta^{-\frac{i}{t}\beta}_{\varphi}, \qquad \forall A \in \mathfrak{A}.
\ena
\noi The KMS condition with respect to the automorphism group $\alpha_\varphi(t), t \in \RN, $ is obtained for any two 
$A, B \in \mathfrak{A}, $ such that the function 
\bea
F_{A, B}(t) = \langle \varphi; A\alpha_\varphi(t)[B]\rangle
\ena
has an extension to the strip $\{z = x+iy | t \in \RN, y \in [0, \beta]\} \subset \CN$ such that $F_{A,B}(z)$ is analytic in the 
strip $(0,\beta)$ and continuous on its boudaries. In addition, it also satisfies the boundary condition, {\it at an inverse temperature 
$\beta$}
\bea
\langle \varphi; A\alpha_\varphi(t +i\beta)[B]\rangle = \langle \varphi; \alpha_\varphi(t)[B]A\rangle, \quad t \in \RN.
\ena
\noi Setting  the generator of the one-parameter group 
by ${\bf H}_{\varphi}$, the operators $\Delta^{-\frac{i}{t}\beta}_{\varphi}$ verify the relation
\bea
\Delta^{-\frac{i}{t}\beta}_{\varphi} = e^{it {\bf H}_{\varphi}} \quad \mbox{and} \quad 
\Delta_{\varphi} = e^{-\beta {\bf H}_{\varphi}}.
\ena
\noi Before introducing the von Neumann algebra generated by the unitary
operators, let us consider the following:
\bedefi\label{wignermap}  
\noi Consider the unitary operator 
$U(x,y)$ on $\mathfrak H  =  L^2(\RN)$ given by 
\begin{eqnarray}\label{map0}
(U(x,y)\Phi)(\xi) =  e^{-i x \left(\xi - y/2 \right)}\Phi\left(\xi -y \right),
\end{eqnarray}
$x,y, \xi \in \RN$, with $U(x,y) = e^{-i (xQ + y P)}$, where $Q, P$ are the usual position and momentum operators given on 
$\mathfrak H  =  L^2(\RN)$, with $[Q, P] = i \IN_{\mathfrak H},  $ and the Wigner transform,  given by
\begin{eqnarray}{\label{map1}}
&& \mathcal W: \mathcal B_{2}(\mathfrak H) \rightarrow L^{2}(\RN^{2}, dxdy) \cr
\cr
&&(\mathcal WX)(x,y) = \frac{1}{(2\pi)^{1/2}}Tr[(U(x,y))^{*}X],  
\end{eqnarray}
where $X \in \mathcal B_{2}(\mathfrak H), x,
 y \in \RN$. $\mathcal W$ is unitary.  
\findefi  
\noi Indeed, given $X_{1},X_{2} \in \mathcal B_{2}(\mathfrak H)$,
\begin{eqnarray}
\int_{\RN^{2}}\overline{(\mathcal WX_{2}(x,y))}(\mathcal WX_{1}(x,y))dxdy = 
\langle X_{2}|X_{1} \rangle_{2} = \langle X_{2}|X_{1} \rangle_{\mathcal B_{2}(\mathfrak H)}.
\end{eqnarray}
\noi On $\tilde{\mathfrak H} = L^2(\RN^2, dxdy)$, $\forall(x,y) \in \RN^2$, consider  the operators
\begin{eqnarray}\label{unitop}
U_1(x,y) &=& \mathcal{W}\left[U(x,y) \vee I_{\mathfrak H}\right]
\mathcal{W}^{-1}, \cr
 U_2(x,y) &=& \mathcal{W}\left[I_{\mathfrak H}
\vee U(x,y)^{*}\right]\mathcal{W}^{-1}
\end{eqnarray}
\noi and let  $\mathfrak{A}_i, i = 1,2$, be the von Neumann algebra generated by the unitary operators 
\cite{takesaki} $\{U_i(x,y)/(x,y) \in \RN^2\}$.   Then, it follows that:
\beprop (\cite{ali-bagarello})
\bei
\item[(i)]
The algebra $\mathfrak{A}_1$ is the  commutant of the algebra 
$\mathfrak{A}_2$ (i.e. each element of $\mathfrak{A}_1$ commutes with every element of $\mathfrak{A}_2$) and vice versa with a factor, i.e, 
\bea
\mathfrak{A}_1 \cap
\mathfrak{A}_2 = {\CN}I_{\tilde{\mathfrak H}}.
\ena
Considering  the antiunitary operator $J_{\beta}$ (i.e. $\langle J\phi|J\psi \rangle  = \langle \psi|\phi \rangle, \forall \phi, \psi \in
 \mathfrak H = L^{2}(\RN)$) such that:
\begin{eqnarray}
J_{\beta}\Psi_{nl} = \Psi_{ln}, \quad J^2_{\beta} = I_{\mathfrak H},
\quad J_{\beta}\Phi_{\beta} = \Phi_{\beta}, \nonumber
\end{eqnarray}
it comes
\begin{eqnarray}\label{antunit}
 J_{\beta}\mathfrak{A}_1J_{\beta} = \mathfrak{A}_2.
\end{eqnarray}
{\it The relation (\ref{antunit})
and the property (i) provide the modular structure of the triplet 
 $\{\mathfrak{A}_1,
\mathfrak{A}_2, J_{\beta}\}$.} 
\item [(ii)] The map \cite{ali-bagarello}
\bea
 S_{\beta}: \tilde{\mathfrak H}\rightarrow {\tilde{\mathfrak H}},
\;\; S_{\beta}\left[U_1(x,y)\Phi_{\beta}\right]
= U_1(x,y)^{*}\Phi_{\beta}, \nonumber
\ena
is closable and has the polar decomposition
\bea\label{antunit00}
S_{\beta} = J_{ \beta}\Delta^{\frac{1}{2}}_{\beta},
\ena
where $J_{\beta}$ is the antiunitary operator, with 
\bea\label{antunit01}
J_{\beta}\Psi_{nl} = \Psi_{ln}, \quad J^2_{\beta} = I_{\mathfrak H},
\quad J_{\beta}\Phi_{\beta} = \Phi_{\beta},
\ena
\bea
J_{\beta}\mathfrak{A}_1J_{\beta} = \mathfrak{A}_2.
\ena
\eni
\enprop
\noi  Indeed, $J_{\beta}$ is by definition an antiunitary operator. Then,  it is self-adjoint, symmetric and consequently closable.   
$\Delta^{\frac{1}{2}}_{\beta}$ also being self-adjoint by definition,   is   closable too.  
From (\ref{antunit00}), the map  $S_{\beta}$ given as the product of two closable operators, is then closable.\\

\noi {\bf Proof of (\ref{antunit00})}. The proof is achieved as follows:
\noi The vectors  $\Psi_{jk}, \, j,k= 0,1,2,\cdots,\infty$,
form an orthonormal  basis of $\tilde{\mathfrak H} = L^2(\RN^2, dxdy)$.
We have $\,$
$\Phi_{\beta} = \displaystyle\sum_{i=0}^{\infty}\lambda^{\frac{1}{2}}_i\Psi_{ii}.$
Applying  $U_1(x,y)$ to both sides leads to
\bea
U_1(x,y)\Phi_{\beta} = \sum_{i=0}^{\infty}\lambda^{\frac{1}{2}}_i
U_1(x,y)\Psi_{ii}. \nonumber
\ena
Since
$\displaystyle\sum_{j,k=0}^{\infty}|\Psi_{jk}\rangle \langle\Psi_{jk}|
 = I_{\tilde{\mathfrak H}}$, 
\noi we get
\bea\label{sigma}
U_1(x,y)\Phi_{\beta} = \sum_{i=0}^{\infty}
\lambda^{\frac{1}{2}}_i U_1(x,y)\Psi_{ii}
= \sum_{i,j,k=0}^{\infty}
\lambda^{\frac{1}{2}}_i\langle\Psi_{jk}
|U_1(x,y)\Psi_{ii}\rangle_{\tilde{\mathfrak H}}
\Psi_{jk}. \nonumber
\ena
\noi From the relations
\bea \phi_{nl} = |\phi_n\rangle \langle\phi_l|\quad  \mbox{and} \quad
 \mathcal{W}\phi_{nl} = \Psi_{nl},\;  n,l=0,1,2,\cdots,\infty \nonumber
 \ena
we get
\bea
\mathcal{W}\phi_{jk} = \mathcal{W}(|\phi_j\rangle \langle\phi_k|) = \Psi_{jk},
\, \forall j,k. \nonumber
\ena
\noi Using the fact that  $\phi_i,\; i=0,1,2,\cdots,\infty,$
form a basis of  $\mathfrak H = L^{2}(\RN)$, we have
\bea
\langle\Psi_{jk}|U_1(x,y)\Psi_{ii}\rangle_{\tilde{\mathfrak H}} &=&
Tr\left[|\phi_k\rangle \langle\phi_j|U(x,y)|\phi_i\rangle_{\tilde{\mathfrak H}}\langle\phi_i|\right] \cr
&=& \sum_{l=0}^{\infty}\langle\phi_l|\phi_k\rangle \langle\phi_j|U(x,y)|\phi_i\rangle_{\tilde{\mathfrak H}}\delta_{il} \cr
&=&\delta_{ik}\langle\phi_j|U(x,y)|\phi_i\rangle_{\tilde{\mathfrak H}}\cr
&=&\delta_{ik}\overline{\langle\phi_i|U(x,y)|\phi_j\rangle_{\tilde{\mathfrak H}}}\cr
&=&(2\pi)^{\frac{1}{2}}\delta_{ik}\overline{\mathcal{W}(|\phi_j\rangle \langle\phi_i|)(x,y)}.
\nonumber
\ena 
Thus
$\langle\Psi_{jk}|U_1(x,y)\Psi_{ii}\rangle_{\tilde{\mathfrak H}}
= (2\pi)^{\frac{1}{2}} \delta_{ik}\overline{\Psi_{ji}(x,y)}$.
From (\ref{sigma}), it comes
\bea\label{omega}
&&U_1(x,y)\Phi_{\beta} = (2\pi)^{\frac{1}{2}}\sum_{i,j,k=0}^{\infty}
\lambda^{\frac{1}{2}}_i\overline{\Psi_{ji}(x,y)}\Psi_{jk}\delta_{ki}
\ena
\noi i.e. 
\bea\label{phibeta00}
U_1(x,y)\Phi_{\beta} = (2\pi)^{\frac{1}{2}}\sum_{i,j=0}^{\infty}
\lambda^{\frac{1}{2}}_i\overline{\Psi_{ji}(x,y)}\Psi_{ji}.
\ena
\noi Let us calculate $U_1(x,y)^{*}\Phi_{\beta}$.  We have
\bea\label{mu}
U_1(x,y)^{*}\Phi_{\beta} = \sum_{j=0}^{\infty}\lambda^{\frac{1}{2}}_j
U_1(x,y)^{*}\Psi_{jj} = \sum_{i,j,k=0}^{\infty}\lambda^{\frac{1}{2}}_j
\langle\Psi_{ik}|U_1(x,y)^{*}\Psi_{jj}\rangle_{\tilde{\mathfrak H}}\Psi_{ik},
\ena
where
\bea\label{pion}
\langle\Psi_{ik}|U_1(x,y)^{*}\Psi_{jj}\rangle_{\tilde{\mathfrak H}} &=& 
Tr\left[|\phi_k\rangle \langle\phi_i|U(x,y)^{*}|\phi_j\rangle_{\tilde{\mathfrak H}} \langle\phi_j|\right] \cr
&=&\sum_{l=0}^{\infty} \langle\phi_l|\phi_k\rangle \langle\phi_i|U(x,y)^{*}|\phi_j\rangle_{\tilde{\mathfrak H}}\delta_{jl} \cr
&=& \langle\phi_j|\phi_k\rangle \langle\phi_i|U(x,y)^{*}|\phi_j\rangle_{\tilde{\mathfrak H}}\cr
&=&\delta_{jk}\left(\langle\phi_j|U(x,y)|\phi_i\rangle_{\tilde{\mathfrak H}}\right)^{*}  \cr
&=& (2\pi)^{\frac{1}{2}}\delta_{kj} \left(\mathcal{W}(|\phi_i\rangle \langle\phi_j|)\right)^{*}(x,y) \cr
&=& (2\pi)^{\frac{1}{2}}\delta_{kj}\Psi^{*}_{ij}(x,y) \cr
&=& (2\pi)^{\frac{1}{2}}\delta_{kj}\Psi_{ji}(x,y).
\ena
\noi Putting  (\ref{pion}) in (\ref{mu}) leads to
\bea\label{muon}
U_1(x,y)^{*}\Phi_{\beta} = (2\pi)^{\frac{1}{2}}\sum_{i,j,k=0}^{\infty}
\lambda^{\frac{1}{2}}_j\Psi_{ji}(x,y)\Psi_{ik}\delta_{kj}
=(2\pi)^{\frac{1}{2}} \sum_{i,j=0}^{\infty}\lambda^{\frac{1}{2}}_j
\Psi_{ji}(x,y)\Psi_{ij}.
\ena
From (\ref{omega}), we have
\bea
U_1(x,y)\Phi_{\beta} = (2\pi)^{\frac{1}{2}}\sum_{i,j=0}^{\infty}
\lambda^{\frac{1}{2}}_i\overline{\Psi_{ji}(x,y)}\Psi_{ji}.\nonumber
\ena
Applying  $S_{\beta}$ to both sides of the equality gives
\bea
S_{\beta}U_1(x,y)\Phi_{\beta} = (2\pi)^{\frac{1}{2}}
\sum_{i,j=0}^{\infty}\lambda^{\frac{1}{2}}_i S_{\beta}
\overline{\Psi_{ji}(x,y)} S_{\beta}\Psi_{ji}. \nonumber
\ena
Since
$ S_{\beta}\left[U_1(x,y)\Phi_{\beta}\right] = U_1(x,y)^{*}
\Phi_{\beta}$, then $ S_{\beta}\overline{\Psi_{ji}(x,y)}
= \Psi_{ji}(x,y).$ 
\noi Thus, 
\bea
S_{\beta}U_1(x,y)\Phi_{\beta} = (2\pi)^{\frac{1}{2}}
\sum_{i,j=0}^{\infty}\lambda^{\frac{1}{2}}_i\Psi_{ji}S_{\beta}\Psi_{ji},\nonumber
\ena
which rewrites 
\bea\label{quik}
S_{\beta}U_1(x,y)\Phi_{\beta} = U_1(x,y)^{*}\Phi_{\beta}
= (2\pi)^{\frac{1}{2}}\sum_{i,j=0}^{\infty}\lambda^{\frac{1}{2}}_i
\Psi_{ji}(x,y)S_{\beta}\Psi_{ji}.
\ena
\noi From the relations  (\ref{muon}) and  (\ref{quik}) together, it follows that
\bea\label{quak}
\lambda^{\frac{1}{2}}_i S_{\beta}\Psi_{ji} = \lambda^{\frac{1}{2}}_j
\Psi_{ij} \quad  \mbox{{i.e.}} \quad
S_{\beta}\Psi_{ji} = \left[\frac{\lambda_j}{\lambda_i}
\right]^{\frac{1}{2}}\Psi_{ij},
\ena
\noi for all$\,\Psi_{ij} \in \tilde{\mathfrak H},
\; i,j=0,1,2,\cdots,\infty$.
$\hfill{\square}$

\noi {\bf Proof of (\ref{antunit01})}. 
\noi Consider the operator $J_{\beta}$ with  $J_{\beta}\Psi_{ji} = \Psi_{ij}$. We have 
\bea
J^{2}_{\beta}\Psi_{ji} = J_{\beta}\Psi_{ij} = \Psi_{ji} ,
\forall i,j  \quad \mbox{i.e.} \quad 
J^{2}_{\beta} = I_{\tilde{\mathfrak H}}.\nonumber
\ena 
\noi Besides, 
\bea
\Phi_{\beta} = \sum_{i=0}^{\infty}\lambda^{\frac{1}{2}}_i\Psi_{ii} \quad
\mbox{i.e.} \quad J_{\beta}\Phi_{\beta} = \sum_{i=0}^{\infty}\lambda^{\frac{1}{2}}_i
(J_{\beta}\Psi_{ii}) = \sum_{i=0}^{\infty}\lambda^{\frac{1}{2}}_i
\Psi_{ii} = \Phi_{\beta}. \nonumber
\ena 
Thus
$J_{\beta}\Phi_{\beta} = \Phi_{\beta}$.
Therefore,  $J_{\beta}$ is such that
\bea
J_{\beta}\Psi_{nl} = \Psi_{ln}, \quad J^{2}_{\beta} = I_{\tilde
{\mathfrak H}}, \quad J_{\beta}\Phi_{\beta} = \Phi_{\beta}.\nonumber
\ena
\noi From  (\ref{antunit00}) and 
\bea
\Delta_{\beta} = \sum_{n,l=0}^{\infty} \frac{\lambda_n}{\lambda_l}
|\Psi_{nl}\rangle \langle\Psi_{nl}| =  e^{-\beta H},\; H = H_1 
- H_2, \nonumber
\ena
we get
\bea
S_{\beta} = J_{\beta}\Delta^{\frac{1}{2}}_{\beta} \quad \mbox{and}
\quad  \Delta_{\beta} = \sum_{n,l=0}^{\infty}\frac{\lambda_n}
{\lambda_l}|\Psi_{nl}\rangle \langle\Psi_{nl}|. \nonumber
\ena
\noi Using this above relations yields
\begin{eqnarray*}
\forall i,j, \quad (J_{\beta}\Delta^{\frac{1}{2}}_{\beta})|\Psi_{ji}\rangle
&=& J_{\beta}\left[\Delta^{\frac{1}{2}}_{\beta}|\Psi_{ji}\rangle\right]
= J_{\beta}\left[\sum_{n,l=0}^{\infty}\left(\frac{\lambda_n}
{\lambda_l}\right)^{\frac{1}{2}}|\Psi_{nl}\rangle \langle\Psi_{nl}|\Psi_{ji}\rangle\right]\\
&=& J_{\beta}\left[\sum_{n,l=0}^{\infty}\left(\frac{\lambda_n}
{\lambda_l}\right)^{\frac{1}{2}}|\Psi_{nl}\rangle\delta_{nj}\delta_{il}
\right] = J_{\beta}\left[\left(\frac{\lambda_j}{\lambda_i}
\right)^{\frac{1}{2}}|\Psi_{ji}\rangle\right] \\
&=&\left(\frac{\lambda_j}{\lambda_i}\right)^{\frac{1}{2}}
\left[J_{\beta}|\Psi_{ji}\rangle\right].
\end{eqnarray*}
\noi From (\ref{quak}), 
\bea
S_{\beta}|\Psi_{ji}\rangle = \left(\frac{\lambda_j}{\lambda_i}\right)^
{1/2}|\Psi_{ij}\rangle = (J_{\beta}\Delta^
{\frac{1}{2}}_{\beta})|\Psi_{ji}\rangle, \,\forall i,j
\ena
i.e.
\bea
S_{\beta} = J_{\beta}\Delta^{\frac{1}{2}}_{\beta}.
\ena
$\hfill{\square}$

\beprop\label{sepvector} (\cite{ali-bagarello})
\noi If  $\displaystyle\{\lambda_n\}_{n=0}^{\infty}$ is a sequence of non-zero positive numbers such that $\displaystyle\sum_{n=0}^{\infty}\lambda_n = 1$, then the vector
\bea
\Phi = \sum_{n=0}^{\infty}\lambda^{\frac{1}{2}}_n\Psi_{nn}
\ena
\noi is  cyclic (that is the set 
of vectors  $\{A\Phi / A \in
\mathfrak{A}_1\}$ is dense in  $\tilde{\mathfrak H}$) and separating 
(i.e.  if  $A\Phi=0$, for all
 $A \in \mathfrak{A}_1$ then $A=0$) for $\mathfrak{A}_1.$
  \enprop
\noi {\bf Proof.} 
\noi Let $X \in \mathcal{B}_{2}(\tilde{\mathfrak{H}}) $ and consider the operator $\mathcal{W}[U(x,y) \vee I_{\tilde{\mathfrak H}}]\mathcal{W}^{-1} \in \mathfrak{A}_{1}$. Taking $U(x,y) \vee I_{\tilde{\mathfrak H}}, $ we have, since $\mathcal{W}$ is unitary,
\bea\label{complx00}
\langle X|\mathcal{W}[U(x,y) \vee I_{\tilde{\mathfrak H}}]\mathcal{W}^{-1}\Phi\rangle_{\tilde{\mathfrak H}}
 &=&\langle X|(U(x,y) \vee I_{\tilde{\mathfrak H}})\Phi\rangle_{\mathcal{B}_{2}(\tilde{\mathfrak{H}})} \cr
&=& \mbox{Tr}[X^{*}(U(x,y) \vee I_{\tilde{\mathfrak H}})\Phi] \cr
&=& \mbox{Tr}[(X^{*}U(x,y))\Phi] \cr
&=& (2\pi)^{\frac{1}{2}}\overline{(\mathcal{W}X\Phi)(x,y)} \quad [\mbox{by}\, (\ref{map1})]
\ena
\noi and the complex conjugate of (\ref{complx00}) given by
\bea
\overline{\langle X|(U(x,y) \vee I_{\tilde{\mathfrak H}})\Phi\rangle_{\mathcal{B}_{2}(\tilde{\mathfrak{H}})}} = \mbox{Tr}[(X^{*}U(x,y))\Phi]^{*} = (2\pi)^{\frac{1}{2}}(\mathcal{W}X\Phi)(x,y).
\ena
\noi Then,   integrating over $\CN$ the modulus squared
\bea
\langle X|(U(x,y) \vee I_{\tilde{\mathfrak H}})\Phi\rangle_{\mathcal{B}_{2}(\tilde{\mathfrak{H}})}
\overline{\langle X|(U(x,y) \vee I_{\tilde{\mathfrak H}})\Phi\rangle_{\mathcal{B}_{2}(\tilde{\mathfrak{H}})}} = |\langle X|(U(x,y) \vee I_{\tilde{\mathfrak H}})\Phi\rangle_{\mathcal{B}_{2}(\tilde{\mathfrak{H}})}|^{2}
\ena
with respect to $x,y, $
we get  
\bea
&&\int_{\CN}\overline{((\mathcal{W}X\Phi)(x,y))}((\mathcal{W}X\Phi)(x,y))dxdy \cr
&& =  \int_{\CN}\sum_{i,j=0}^{\infty}\sum_{l,k=0}^{\infty}\langle\Psi_{kk}|\lambda^{\frac{1}{2}}_i\lambda^{\frac{1}{2}}_k  
\overline{\mathcal{W}\phi_{jk}(x,y)}\mathcal{W}\phi_{li}(x,y)dxdy|\Psi_{ii}\rangle_{\tilde{\mathfrak{H}}} \cr
&& = \sum_{i,j=0}^{\infty}\sum_{l,k=0}^{\infty}\lambda^{\frac{1}{2}}_i\lambda^{\frac{1}{2}}_k
 \langle\Psi_{kk}|\left\{
\int_{\CN}\overline{\mathcal{W}\phi_{jk}(x,y)}\mathcal{W}\phi_{li}(x,y)dxdy\right\}|\Psi_{ii}\rangle_{\tilde{\mathfrak{H}}} \cr
&& = \sum_{i,j=0}^{\infty}\sum_{l,k=0}^{\infty}\lambda^{\frac{1}{2}}_i\lambda^{\frac{1}{2}}_k
\langle\Psi_{kk}| \left\{
\int_{\CN}\overline{\mathcal{W}|\phi_{j}\rangle \langle \phi_{k}|(x,y)}\mathcal{W}|\phi_{l}\rangle \langle \phi_{i}|(x,y)dxdy\right\}|\Psi_{ii}\rangle_{\tilde{\mathfrak{H}}} \cr
&& =  \sum_{i,j=0}^{\infty}\sum_{l,k=0}^{\infty}\lambda^{\frac{1}{2}}_i\lambda^{\frac{1}{2}}_k\langle\Psi_{kk}|  \left\{
\int_{\CN}\overline{\Psi_{jk}(x,y)}\Psi_{li}(x,y)dxdy\right\}|\Psi_{ii}\rangle_{\tilde{\mathfrak{H}}}  \cr
&& = \sum_{i,j=0}^{\infty}\sum_{l,k=0}^{\infty}\lambda^{\frac{1}{2}}_i\lambda^{\frac{1}{2}}_k\langle\Psi_{kk}|  \left\{
\int_{\CN}\Psi_{jk}(x,y)^{*}\Psi_{li}(x,y)dxdy\right\}|\Psi_{ii}\rangle_{\tilde{\mathfrak{H}}}  \cr
&& = \sum_{i,j=0}^{\infty}\sum_{l,k=0}^{\infty}\lambda^{\frac{1}{2}}_i\lambda^{\frac{1}{2}}_k\langle\Psi_{kk}|\Psi_{ii}\rangle_{\tilde{\mathfrak{H}}} \delta_{ij}\delta_{kl}\cr
&& =  \sum_{i,k=0}^{\infty}\lambda^{\frac{1}{2}}_i\lambda^{\frac{1}{2}}_k\delta_{ik}\cr
&& = \sum_{i=0}^{\infty}\lambda_i = 1.
\ena
\noi Since $\mathcal{W}$ is unitary, we may write
\bea
\int_{\CN}\overline{((\mathcal{W}X\Phi)(x,y))}((\mathcal{W}X\Phi)(x,y))dxdy = 0 &\Longrightarrow  &
|\langle X|\mathcal{W}[U(x,y) \vee I_{\tilde{\mathfrak H}}]\mathcal{W}^{-1}\Phi\rangle_{\tilde{\mathfrak{H}}}|^{2}  =0 \cr
&\Longrightarrow & |\langle X|(U(x,y) \vee I_{\tilde{\mathfrak H}})\Phi\rangle_{\mathcal{B}_{2}(\tilde{\mathfrak{H}})}|^{2} =0 \cr
& \Longrightarrow & X = 0.
\ena
\noi This implies  that the set $\left\{\mathcal{W}[U(x,y) \vee I_{\tilde{\mathfrak H}}]\mathcal{W}^{-1}\Phi, 
\mathcal{W}[U(x,y) \vee I_{\tilde{\mathfrak H}}]\mathcal{W}^{-1} \in \mathfrak{A}_1\right\}$ is dense in $\tilde{\mathfrak{H}}$, 
proving from the Definition \ref{vecyclicsepar} that $\Phi$ is cyclic for $\mathfrak{A}_1.$
$\hfill{\square}$

\noi The fact that $\Phi$ is separating for $\mathfrak{A}_1$ is obtained through the relation 
\bea
&& \mathcal{W}(U(x,y) \vee I_{\tilde{\mathfrak H}})\mathcal{W}^{-1}\Phi = 
\mathcal{W}(U'(x,y) \vee I_{\tilde{\mathfrak H}})\mathcal{W}^{-1}\Phi \cr
 \Longleftrightarrow  && \mathcal{W}(U(x,y)
\vee I_{\tilde{\mathfrak H}})\mathcal{W}^{-1} = \mathcal{W}(U'(x,y) \vee I_{\tilde{\mathfrak H}})\mathcal{W}^{-1}. 
\nonumber \\
\ena
\noi {\bf Proof.}
\noi Let $U(x,y), U'(x,y)$ such that $\mathcal{W}(U(x,y) \vee I_{\tilde{\mathfrak H}})\mathcal{W}^{-1}\Phi = \mathcal{W}(U'(x,y) \vee I_{\tilde{\mathfrak H}})\mathcal{W}^{-1}\Phi.$ Take   $X \neq 0,\, X \in \mathcal{B}_{2}(\tilde{\mathfrak{H}})$ and set $\Phi = \displaystyle\sum^{N}_{i=1} \lambda^{\frac{1}{2}}_i |\zeta_i\rangle \langle \zeta_i|$. We have
\bea
 &&\langle X|\mathcal{W}\left\{(U(x,y) \vee I_{\tilde{\mathfrak H}})-(U'(x,y) 
\vee I_{\tilde{\mathfrak H}})\right\}\mathcal{W}^{-1}\Phi \rangle_{\tilde{\mathfrak H}} \cr
&& = 
\langle X|\left\{(U(x,y) \vee I_{\tilde{\mathfrak H}})-(U'(x,y) \vee I_{\tilde{\mathfrak H}})\right\}\Phi \rangle_{\mathcal{B}_{2}(\tilde{\mathfrak{H}})} \cr  
&& =   \mbox{Tr}\left[X^{*}\left\{(U(x,y) \vee I_{\tilde{\mathfrak H}})-(U'(x,y) \vee I_{\tilde{\mathfrak H}})\right\}\Phi\right]\cr
\cr
&& =    \sum^{N}_{k=1} \langle \zeta_k|X^{*}\left\{U(x,y)-U'(x,y)\right\}
\left\{\sum^{N}_{i=1} \lambda^{\frac{1}{2}}_i |\zeta_i\rangle \langle \zeta_i|\right\}|\zeta_k\rangle 
\cr
&& =  \sum^{N}_{k=1}\sum^{N}_{i=1} \lambda^{\frac{1}{2}}_i \langle \zeta_k|X^{*}\left\{U(x,y)-U'(x,y)\right\}\zeta_i\rangle \langle \zeta_i|\zeta_k\rangle \cr
&& = \sum^{N}_{k=1}\sum^{N}_{i=1} \lambda^{\frac{1}{2}}_i \langle \zeta_k|X^{*}\left\{U(x,y)-U'(x,y)\right\}\zeta_i\rangle \delta_{ik}\cr
&& =  \sum_{i=1}^{N}\lambda^{\frac{1}{2}}_i\langle \zeta_i| X^{*}\left\{U(x,y)-U'(x,y)\right\} \zeta_i\rangle.
\ena
\noi Then, we have 
\bea
 &&\langle X|\mathcal{W}\left\{(U(x,y) \vee I_{\tilde{\mathfrak H}})-(U'(x,y) \vee I_{\tilde{\mathfrak H}})\right\}
\mathcal{W}^{-1}\Phi \rangle_{\tilde{\mathfrak{H}}} \cr
&& =   
\langle X|\left\{(U(x,y) \vee I_{\tilde{\mathfrak H}})-(U'(x,y) \vee I_{\tilde{\mathfrak H}})\right\}
\Phi \rangle_{\mathcal{B}_{2}(\tilde{\mathfrak{H}})}  = 0 \cr
   \Longleftrightarrow  && \sum_{i=1}^{N}\lambda^{\frac{1}{2}}_i\langle \zeta_i| X^{*}\left\{U(x,y)-U'(x,y)\right\} \zeta_i\rangle = 0, \,\,
\lambda_i > 0, \, \forall i \cr
  \Longleftrightarrow &&   U(x,y) \vee I_{\tilde{\mathfrak H}} = U'(x,y) \vee I_{\tilde{\mathfrak H}}
\ena
\noi which completes the proof.
$\hfill{\square}$

\subsection{Modular theory-Thermal state}
Here, we give two examples  of  thermal states as known from  the literature. 
For more details, see  \cite{tomita, takesaki, connes1, bratelli, summers, bertozzini, ali-bagarello, ali-bagarello-honnouvo}.
\bee
\item 
Let $\alpha_i, i= 1,2,\dots, N$ be a sequence of non-zero, positive numbers, satisfying
$\displaystyle\sum_{i=1}^{N}\alpha_i = 1.$  Then, the thermal state is defined as:
\bea
\Phi := \sum_{i=1}^{N}\alpha^{\frac{1}{2}}_i \mathbb P_i = 
\sum_{i=1}^{N}\alpha^{\frac{1}{2}}_i 
X_{ii}  \in \mathcal B_2(\mathfrak H),
\ena
where $\mathbb P_i = X_{ii}= |\zeta_i\rangle \langle \zeta_i|$ is defined as in (\ref{vectphi00})-(\ref{vectphi01}), 
with $\{X_{ij} = |\zeta_i\rangle \langle \zeta_j|, i,j =1,2,\dots, N \}$ forming an orthonormal basis of $\mathcal{B}_2(\mathfrak{H})$. 
\item The thermal equilibrium state $\Phi$ at inverse temperature $\beta, $
corresponding to the harmonic oscillator  Hamiltonian 
\bea\label{oschamil}
H_{OSC} = \frac{1}{2}(P^2 + Q^2),\;\;    \mbox{with}\; \;  
H_{OSC} \phi_n = \omega(n + \frac{1}{2})\phi_n, n= 0, 1, 2, \dots, 
\ena
where the density matrix is
\bea
\rho_{\beta} = \frac{ e^{-\beta H_{\mbox{osc}}}}{Tr\left[ e^{-\beta
H_{\mbox{osc}}}\right]} = (1 -  e^{-\omega \beta})\sum_{n=0}^{\infty}
 e^{-n \omega \beta}|\phi_n\rangle \langle \phi_n|, \quad  \mbox{Tr}[e^{-\beta H_{OSC}}] = \frac{e^{-\frac{\beta \omega}{2}}}{1- e^{-\beta \omega}},
\ena
is 
\bea\label{vectcyclic}
\Phi = \left[1-e^{-\omega\beta}\right]^{\frac{1}{2}}\sum_{n=0}^{\infty}e^{-\frac{n}{2} \omega\beta}|\phi_n\rangle \langle \phi_n|.
\ena
\item 
Let the two von Neumann algebras be given by 
\bea
\mathfrak A_{l} = \left\{A_{l} = A \vee I| A \in \mathcal L(\mathfrak H)\right\}, \quad 
\mathfrak A_{r} = \left\{A_{r} = I \vee A| A \in \mathcal L(\mathfrak H)\right\}
\ena
where  $\mathfrak A_{l}$ corresponds to the operators given with $A$ in the left, and $\mathfrak A_{r}$ corresponds 
to the operators given with $A$ in the right of the identity operator $I_{\mathfrak{H}}$ on $\mathfrak H, $ respectively.
\noi   $\Phi$ defines a vector state $\varphi$,  called {\it KMS} state,  on the von Neumann algebra $\mathfrak A_l$. 
For any $A \vee I \in \mathfrak A_l, $ one has the state 
$\varphi$ on $\mathfrak A_l$  given by 
\bea\label{densmatrix}
\langle \varphi; A \vee I\rangle = \langle \Phi|(A \vee I)(\Phi)\rangle_2 =
 Tr[\Phi^{*}A\Phi] = Tr[\rho_\varphi A], \;\; \mbox{with} \;\;  \rho_\varphi = 
 \sum_{n=1}^{N}\alpha_n \PN_n 
\ena
with $\PN_n = |\phi_n\rangle \langle \phi_n|, $ where 
\bea
\rho_{\varphi}= \frac{e^{-\beta H_{OSC}}}{\mbox{Tr}[e^{-\beta H_{\varphi}}]} = (1- e^{-\omega \beta})\sum_{n=0}^{\infty}e^{-n \omega \beta}|\phi_n\rangle \langle \phi_n|
\ena
and 
\bea
H_\varphi = -\frac{1}{\beta}\sum_{n=0}^{\infty} (\ln \alpha_n)\PN_n, \quad 
\alpha_n = (1- e^{-\omega \beta})e^{-n \omega \beta}.
\ena 
\ene
\subsection{Coherent states built from the harmonic oscillator thermal state} 
 Before dealing with the CS construction, we  shall  first extract few facts and notations  about the modular structures emerging 
 for von Neumann algebras in the study of an electron in a magnetic field as needed for the developement of this paragraph. 
 For details see \cite{ali-bagarello} and references therein.
\subsubsection{Electron in a magnetic field} 
 Considering  the quantum  Hilbert space $\mathfrak H = L^2(\mathbb R)$ of the Hamiltonian $H_{\mbox{osc}}$ in (\ref{oschamil}), 
 take $\mathcal B_2(\mathfrak H)\simeq \mathfrak{H} \otimes \bar{\mathfrak{H}}$ the space of Hilbert-Schmidt operators on $\mathfrak{H}$ 
 with an orthonormal basis given by 
\bea\label{oschamilbasis}
\phi_{nl}:= |\phi_n\rangle \langle \phi_l|, \quad n,l = 0,1,2,\dots,\infty.
\ena 
Taking  the classical Hamiltonian describing an electron placed in the $xy$ plane and subjected to a constant magnetic field 
\cite{ali-bagarello},
\bea
H_{elec} = \frac{1}{2}(\overrightarrow{p}+ \overrightarrow{A})^2 = \frac{1}{2}\left(p_x + \frac{y}{2}\right)^2 + \frac{1}{2}\left(p_y -
\frac{x}{2}\right)^2
\ena
let the following quantum Hamiltonians 
\bea\label{quanthamil00}
H_1 = \frac{1}{2}(P^2_1 + Q^2_1), \quad [Q_1, P_1] = i I_{\tilde{\mathfrak{H}}}
\ena
with the magnetic field aligned along the negative $z$ axis, $\overrightarrow{A} = \frac{1}{2}(y, -x, 0), $ where 
the quantized observables given on $\tilde{\mathfrak{H}} = L^2(\mathbb R^2, dxdy)$
by
\bea\label{quanthamil01}
p_x + \frac{y}{2} \rightarrow Q_1 =-i \frac{\partial}{\partial x} + \frac{y}{2}; \quad p_y-\frac{x}{2} \rightarrow P_1 = -i 
\frac{\partial}{\partial y}- \frac{x}{2}
\ena
and 
\bea
H_2 = \frac{1}{2}(P^2_2 + Q^2_2), \quad [Q_2, P_2] = i I_{\tilde{\mathfrak{H}}}
\ena
with the magnetic field aligned along the positive $z$ axis, $\overrightarrow{A} = \frac{1}{2}(-y, x, 0), $ with 
the quantized observables given on $\tilde{\mathfrak{H}} = L^2(\mathbb R^2, dxdy)$
by
\bea
 p_y+\frac{x}{2}\rightarrow Q_2 =-i \frac{\partial}{\partial y} + \frac{x}{2}; 
 \quad p_x - \frac{y}{2} \rightarrow P_2 = -i \frac{\partial}{\partial x}- \frac{y}{2}. 
\ena
Since $[H_1, H_2] = 0$, the eigenvectors $\Psi_{nl}$ of $H_1$ can be so chosen that they are also the eigenvectors of $H_2$ as follows:
\bea\label{eigenvect}
H_1\Psi_{nl} = \omega \left(n+\frac{1}{2}\right)\Psi_{nl}, \qquad H_2\Psi_{nl} = \omega\left(l+\frac{
1}{2}\right)\Psi_{nl} 
\ena
so that $H_2$ lifts the degeneracy of $H_1$ and vice versa. Next, from the Definition \ref{wignermap}, 
it is established that \cite{ali-bagarello}
\bea\label{wigequal} 
\mathcal W\phi_{nl} = \mathcal W(|\phi_n\rangle \langle \phi_l|) = \Psi_{nl}
\ena
where the $\phi_{nl}$ are the basis vecrtors given in (\ref{oschamilbasis}) and the $\Psi_{nl}$ the normalized eigenvectors in 
(\ref{eigenvect}). Then, from (\ref{wigequal}),  the vectors $\Psi_{nl}, n, l=0, 1, 2, \dots, \infty$ form a basis of 
$\tilde{\mathfrak{H}} = L^2(\mathbb R^2, dxdy).$ 
 
\noi Note that, in the sequel, the CS will be constructed from the thermal state $\Phi_{\beta}, $ identified with
the vector $\Phi$ given in (\ref{vectcyclic}),  denoted as a ket state 
$|\Phi_{\beta}\rangle, $ the normalized eigenvectors (\ref{wigequal}) $\Psi_{nl}$  also denoted $|\Psi_{nl}\rangle$
as proceeded in \cite{ali-bagarello}. 
\subsubsection{Coherent states built from the thermal state}
Take the cyclic vector $\Phi$ of the von Neumann algebra $\mathfrak A_1$ 
generated by the unitary 
operator (\ref{unitop}) 
\bea\label{unitop00}
U_1(x,y) = \mathcal{W}\left[U(x,y) \vee I_{\mathfrak H}\right]
\mathcal{W}^{-1},
\ena
where   $\mathcal{W}$ and $U(x,y) =  e^{-i (xQ + y P)}$ are defined by  (\ref{map0}) and (\ref{map1}), and consider Proposition 
\ref{sepvector} with the thermal state $\Phi_{\beta}, $ instead of $\Phi, $ such that 
\bea\label{thermalst00}
\Phi_{\beta} = \left[1 -  e^{-\omega \beta}\right]^{\frac{1}{2}}
\sum_{n=0}^{\infty} e^{-n\frac{\omega \beta}{2}}\Psi_{nn}, \; \mbox{i.e.},\;
\lambda_n = (1- e^{-\omega \beta}) e^{-n \omega \beta}.
\ena
The CS, denoted  $|z,\bar{z},\beta\rangle^{\mbox{\tiny{KMS}}}, $ built from  the
thermal state in ket notation $|\Phi_{\beta}\rangle$ (see \cite{ali-bagarello}), are given by 
\bea\label{thermalcs00}
|z,\bar{z},\beta\rangle^{\mbox{\tiny{KMS}}} = U_1(z)|\Phi_{\beta}\rangle :=
 e^{zA^{\dag}_1-\bar{z}A_1}|\Phi_{\beta}\rangle
\ena
 \noi where  the annihilation and creation operators, 
$A_1$ and $A^{\dag}_1$ with \cite{ali-bagarello}
\bea
A_1 = \frac{1}{\sqrt{2}}(Q_1 + i P_1), \quad A^{\dag}_1 = \frac{1}{\sqrt{2}}(Q_1 - i P_1)
\ena
 act on $|\Psi_{n,l}\rangle$, eigenstates of the oscillator Hamiltonian in one 
 dimension  (\ref{quanthamil00}) with eigenvalues $E_n = \omega (n+\frac{1}{2})$, 
     as follows:
\bea
 A^{\dag}_1|\Psi_{n,l}\rangle =\sqrt{n+1} |\Psi_{n+1,l}\rangle, \qquad  A_1|\Psi_{n,l}\rangle =\sqrt{n} |\Psi_{n-1,l}\rangle
\ena
where  the states $|\Psi_{nl}\rangle$  are here  denoted $|\Psi_{n, l}\rangle$ by  commodity.
\beprop (\cite{ali-bagarello})
\noi The CS $|z,\bar{z},\beta\rangle^{\mbox{\tiny{KMS}}}$ satisfy the resolution  of the identity condition
\bea
\frac{1}{2\pi}\int_{\CN}|z,\bar{z},\beta\rangle^{\mbox{\tiny{KMS}}}
{^{\mbox{\tiny{KMS}}}}\langle z,\bar{z},\beta|dxdy = I_{\tilde{\mathfrak{H}}}, \;\;\; \tilde{\mathfrak{H}} = L^2(\RN^2, dxdy).
\ena 
\enprop
\noi {\bf Proof}.
Consider the unitary operator $U(x,y)=  e^{-i(xQ + yP)}, \forall(x,y)\in \RN^2 $.  Let
\bea
U(z)\phi:=\phi(x,y) =  e^{-i(xQ + yP)}\phi \;\;
\forall \phi \in \mathfrak H.
\ena
\noi Show that, for any normalized state $\phi \in \mathfrak{H}= L^2(\RN)$:
\bea 
\frac{1}{2\pi}\int_{\RN^2}|\phi(x,y)\rangle \langle\phi(x,y)|
dxdy = I_{\mathfrak H}.
\ena
\begin{eqnarray*}
\forall q \in \RN, \; \; \mbox{set}\;\; \phi(x,y)(q) = (\pi)^{-\frac{1}{4}}
e^{-i(\frac{x}{2}-q)y} e^{-\frac{(q-x)^2}{2}},
\end{eqnarray*}
such that $\forall \psi , \xi  \in \mathfrak H$, we have
\begin{eqnarray*}
\langle\xi|\phi(x,y)\rangle_{\mathfrak H} 
= (\pi)^{-\frac{1}{4}} e^{-i \frac{xy}{2}}\int_{\RN}
\overline{\xi(q)} e^{i yq} e^{-\frac{(q-x)^2}{2}}dq \\
\langle\phi(x,y)|\psi\rangle_{\mathfrak H} = (\pi)^{-\frac{1}{4}} e^{i\frac{xy}{2}}\int_{\RN} e^{-i yq'} e^{-\frac{(q'-x)^2}{2}}\psi(q')dq'.
\end{eqnarray*} 
\noi Then, 
\begin{eqnarray}
 &&\frac{1}{2\pi}\int_{\RN^2}\langle\xi|\phi(x,y)\rangle \langle\phi(x,y)|\psi\rangle  dxdy \cr
&=&\frac{1}{2\pi\sqrt{\pi}}\int_{\RN^2}dxdy\int_{\RN^2}dqdq'\overline{\xi(q)}
 e^{\imath y(q-q')} e^{\left[-\frac{(x-q)^2}{2}-\frac{(q'-x)^2}{2}\right]}
\psi(q')  \cr
&=& \frac{1}{\sqrt{\pi}}\int_{\RN}dx\int_{\RN^2}dqdq'\overline{\xi(q)}
\left[\frac{1}{2\pi}\int_{\RN} e^{i y(q-q')}dy\right]
 e^{\left[\frac{-(x-q)^2}{2} - \frac{(q'-x)^2}{2}\right]}\psi(q').
\end{eqnarray} 

\noi Since \quad
$ \frac{1}{2\pi}\int_{\RN} e^{i y(q-q')}dy = \delta(q-q') $,\quad we get 
\begin{eqnarray*}
&&\frac{1}{2\pi}\int_{\RN^2}\langle\xi|\phi(x,y)\rangle \langle\phi(x,y)|\psi\rangle dxdy \\
     &=&  \frac{1}{\sqrt{\pi}}\int_{\RN}dx\int_{\RN^2}dqdq'\overline{\xi(q)}\delta(q-q') e^{\left[-\frac{(x-q)^2}{2}-\frac{(q'-x)^2}{2}\right]}
\psi(q') \\
     &=& \frac{1}{\sqrt{\pi}}\int_{\RN}dx\int_{\RN}dq \, e^{-\frac{(x-q)^2}{2}}
     \overline{\xi(q)}\left[\int_{\RN}\psi(q')\delta(q-q') e^{-\frac{(q'-x)^2}{2}}dq'\right] \\
     &=& \frac{1}{\sqrt{\pi}}\int_{\RN^2} e^{-(x-q)^2}
     \overline{\xi(q)}\psi(q)dxdq =
     \int_{\RN}dq\left(\frac{1}{\sqrt{\pi}}\int_{\RN} e^{-u^2}du\right)
     \overline{\xi(q)}\psi(q) \\
     &=&    \langle\xi|\psi\rangle.
\end{eqnarray*}
\noi Thus
\begin{eqnarray}
\frac{1}{2\pi}\int_{\RN^2}|\phi(x,y)\rangle \langle\phi(x,y)|dxdy = I_{\mathfrak H} \qquad
\mbox{i.e.}\qquad \frac{1}{2\pi}\int_{\CN}|U(z)\phi\rangle \langle U(z)\phi|dxdy = I_{\mathfrak H}.
\end{eqnarray}
\noi Using the isometry property of $\mathcal{W}$, the states $|z,\bar{z},\beta\rangle^{\mbox{\tiny{KMS}}}$
satisfy the following resolution of the identity 
\begin{eqnarray}\label{hiho}
\frac{1}{2\pi}\int_{\CN}|z,\bar{z},\beta\rangle^{\mbox{\tiny{KMS}}}
{^{\mbox{\tiny{KMS}}}}\langle z,\bar{z},\beta|dxdy = I_{\tilde{\mathfrak{H}}}.
\end{eqnarray} 
\noi {\bf Proof}. 
By definition of the states $|z,\bar{z},\beta\rangle^{\mbox{\tiny{KMS}}}$ and using (\ref{phibeta00}),
 we have:
\begin{eqnarray}
|z,\bar{z},\beta\rangle^{\mbox{\tiny{KMS}}} &=& U_1(z)\Phi_{\beta}
=(2\pi)^{\frac{1}{2}}\sum_{i,j=0}^{\infty}\lambda_i^{\frac{1}{2}}\overline{\Psi_{ji}(x,y)}|\Psi_{ji}\rangle,\cr
^{\mbox{\tiny{KMS}}}\langle z,\bar{z},\beta| &=& U_1(z)^{*}\Phi_{\beta} = (2\pi)^{\frac{1}{2}}\sum_{l,k=0}^{\infty}\langle\Psi_{lk}|\lambda_k^{\frac{1}{2}}\Psi_{lk}(x,y). 
\end{eqnarray}
\noi Thereby
\bea\label{onyx}
|z,\bar{z},\beta\rangle^{\mbox{\tiny{KMS}}}{^{\mbox{\tiny{KMS}}}}\langle z,\bar{z},\beta|
= 2\pi\sum_{i,j=0}^{\infty}\sum_{l,k=0}^{\infty}\lambda^{\frac{1}{2}}_i
\lambda^{\frac{1}{2}}_k
\overline{\mathcal{W}\phi_{ji}(x,y)}\mathcal{W}\phi_{lk}(x,y)
|\Psi_{ji}\rangle\langle\Psi_{lk}|.
\ena
\noi Integrating  the two members of the Eq.(\ref{onyx}) over 
$\CN$, and using the Wigner map $\mathcal W, $ we get 
\bea
\frac{1}{2\pi}\int_{\CN}|z,\bar{z},\beta\rangle^{\mbox{\tiny{KMS}}}{^{\mbox{\tiny{KMS}}}}\langle z,\bar{z},\beta|dxdy 
&=& \int_{\CN}\sum_{i,j=0}^{\infty}\sum_{l,k=0}^{\infty}\lambda^{\frac{1}{2}}_i\lambda^{\frac{1}{2}}_k |\Psi_{ji}\rangle \langle\Psi_{lk}|
\overline{\mathcal{W}\phi_{ji}(x,y)}\mathcal{W}\phi_{lk}(x,y)dxdy \cr
&=& \sum_{i,j=0}^{\infty}\sum_{l,k=0}^{\infty}\lambda^{\frac{1}{2}}_i\lambda^{\frac{1}{2}}_k |\Psi_{ji}\rangle \langle\Psi_{lk}| 
\int_{\CN}\overline{\mathcal{W}\phi_{ji}(x,y)}\mathcal{W}\phi_{lk}(x,y)dxdy \cr
&=& \sum_{i,j=0}^{\infty}\sum_{l,k=0}^{\infty}\lambda^{\frac{1}{2}}_i\lambda^{\frac{1}{2}}_k|\Psi_{ji}\rangle \langle\Psi_{ji}| \delta_{lj}\delta_{ki}\cr
&=& \sum_{i,j=0}^{\infty}|\Psi_{ji}\rangle \langle\Psi_{ji}|\cr
&=& I_{\tilde{\mathfrak H}}.
\ena
$\hfill{\square}$

\noi From
$ (A^{\dag}_1)|\Psi_{0n}\rangle = \sqrt{n !}|\Psi_{nn}\rangle$
and $ U_1(z)|\Psi_{0n}\rangle
 =  e^{-\frac{|z|^2}{2}}\sum_{k=0}^{\infty} \frac{(zA^{\dag}_1)^k}{k !}
 |\Psi_{0n}\rangle$, it comes that:
\bea\label{operat000}
U_1(z)|\Psi_{nn}\rangle = \frac{1}{\sqrt{n !}}\left(A^{\dag}_1 - \bar{z}
I_{\tilde{\mathfrak H}}\right)^n U_1(z)|\Psi_{0n}\rangle = \frac{1}{\sqrt{n !}}
\left(-\frac{\partial}{\partial z} - \frac{\bar{z}}{2}
I_{\tilde{\mathfrak H}}\right)^nU_1(z)|\Psi_{0n}\rangle.\nonumber
\\
\ena
\noi {\bf Proof of the first equality of (\ref{operat000}).}
The operators $P$ and $Q$ verify the following relations:
\bea
\left[Q,P^n\right] = nP^{n-1}\left[Q,P\right]=\imath \hbar nP^{n-1}, \;\;
\left[Q^n,P\right] = nQ^{n-1}\left[Q,P\right]=\imath \hbar nQ^{n-1}.
\ena
\noi We establish that
\bea\label{operat001}
e^{-\imath Pu} Q  e^{\imath Pu} = Q - u, \qquad
 e^{-\imath Qu} P  e^{\imath Qu} = P + u, \, \forall u \in \RN.
 \ena
\noi Multiplying the first and second equalities of (\ref{operat001}), by
$\frac{1}{\sqrt{2}}$ and $\frac{-\imath}{\sqrt{2}}$, respectively, provides:
\bea\label{operat003}
 e^{-\imath Pu}\frac{Q}{\sqrt{2}} e^{\imath Pu} = \frac{Q}{\sqrt{2}} -
\frac{u}{\sqrt{2}}, \qquad
 e^{-\imath Qu}\left(\frac{-\imath P}{\sqrt{2}}\right) e^{\imath Qu} = \frac{-\imath P}{\sqrt{2}} - \frac{\imath u}{\sqrt{2}}.
\ena
\noi Setting $u = -x$ and $u=-y$, in the first and second relations of (\ref{operat003}), respectively, gives with replacing $P$ by $P_1$, and  $Q$ by  $Q_1$, respectively:
\bea\label{operat005}
 e^{\imath P_1 x} \frac{Q_1}{\sqrt{2}}  e^{-\imath P_1 x} = \frac{Q_1}{\sqrt{2}} + \frac{x}{\sqrt{2}}, \qquad
 e^{\imath Q_1 y} \left(-\frac{\imath P_1}{\sqrt{2}}\right)  e^{-\imath Q_1 y} = \frac{-\imath P_1}{\sqrt{2}} + \frac{\imath y}{\sqrt{2}}.
\ena
\noi From
$ A^{\dag}_1 = \frac{Q_1 - \imath P_1}{\sqrt{2}} $, set
$ z = \frac{-x + \imath y}{\sqrt{2}} $. Summing both equalities of (\ref{operat005}) gives:
\bea\label{operat007}
 e^{\imath P_1 x} \frac{Q_1}{\sqrt{2}}  e^{-\imath P_1 x} +  e^{\imath Q_1 y} \left(-\frac{\imath P_1}{\sqrt{2}}\right)  e^{-\imath Q_1 y}= A^{\dag}_1 - \bar{z}I_{\tilde{\mathfrak H}}.
\ena
\noi Since $U_1(z)=  e^{zA^{\dag}_1 - \bar{z}A_1}$, 
 with $ zA^{\dag}_1 - \bar{z}A_1 = \imath (P_1 x + Q_1 y)$, it follows:
$U_1(z) =  e^{\imath (P_1 x + Q_1 y)}$. This latter equality with
 (\ref{operat007}) together lead to:
\bea\label{operat009}
(A^{\dag}_1 - \bar{z}I_{\tilde{\mathfrak H}})U_1(z)
&=&  e^{\imath P_1 x}
\frac{Q_1}{\sqrt{2}}  e^{-\imath P_1 x}  e^{\imath(P_1 x + Q_1 y)} +
 e^{\imath Q_1 y}\left(-\imath\frac{P_1}{\sqrt{2}}\right)  e^{-\imath Q_1 y}  e^{\imath (P_1 x + Q_1 y)} \cr
						    &=&  e^{\imath(P_1 x + Q_1 y )} e^{-\imath \frac{xy}{2}} e^{\imath \frac{xy}{2}}\frac{Q_1}{\sqrt{2}} +  e^{\imath (Q_1 y + P_1 x)} e^{\imath \frac{xy}{2}} e^{-\imath \frac{xy}{2}}\left(-\imath \frac{P_1}{\sqrt{2}}\right) \cr
						    &=& e^{zA^{\dag}_1 - \bar{z}A_1}\left(\frac{Q_1}{\sqrt{2}} - \imath \frac{P_1}{\sqrt{2}}\right) = U_1(z)A^{\dag}_1.
\ena
From (\ref{operat009}), we have
\bea
U_1(z)|\Psi_{1n}\rangle = (A^{\dag}_1 - \bar{z}I_{\tilde{\mathfrak H}})
U_1(z)|\Psi_{0n}\rangle
\ena
such that, by recursion, we get 
\bea
U_1(z)|\Psi_{nn}\rangle = \frac{1}{\sqrt{n !}}\left(A^{\dag}_1 - \bar{z}
I_{\tilde{\mathfrak H}}\right)^n U_1(z)|\Psi_{0n}\rangle.
\ena
$\hfill{\square}$

\noi {\bf Proof of the second equality of (\ref{operat000}).}\\
Considering 
$ A^{\dag}_1 = \frac{1}{\sqrt{2}}(Q_1-\imath P_1)$, where $z=\frac{1}{\sqrt{2}}(y-\imath x)$, with
\bea\label{moka}
\frac{\partial}{\partial x} = \frac{-\imath}{\sqrt{2}}
\frac{\partial}{\partial z} + \frac{\imath}{\sqrt{2}}
\frac{\partial}{\partial \bar{z}}, \;\;
\frac{\partial}{\partial y} = \frac{1}{\sqrt{2}}
\frac{\partial}{\partial z} + \frac{1}{\sqrt{2}}
\frac{\partial}{\partial \bar{z}}
\ena
provides
\bea
A^{\dag}_1 = -\frac{\partial}{\partial z} + \frac{\bar{z}}{2}
I_{\tilde{\mathfrak H}}, \quad  \mbox{i.e.,} \quad 
 A^{\dag}_1 - \bar{z}I_{\tilde{\mathfrak H}}= -\frac{\partial}{\partial z} - \frac{\bar{z}}{2}I_{\tilde
{\mathfrak H}}
\ena
\noi which completes the proof.
$\hfill{\square}$

\noi  Since
\bea
|z,\bar{z},\beta\rangle^{\mbox{\tiny{KMS}}} = U_1(z)|\Phi_{\beta}\rangle = (1- e^{-\omega \beta})^{\frac{1}{2}}\sum_{n=0}^{\infty} e^{-n\frac
{\omega \beta}{2}}U_1(z)|\Psi_{nn}\rangle,
\ena
setting $|z;n\rangle = U_1(z)|\Psi_{0n}\rangle$ and from (\ref{operat000}),  it comes 
\bea\label{operat012}
|z,\bar{z},\beta\rangle^{\mbox{\tiny{KMS}}} = (1- e^{\omega \beta})^{\frac{1}{2}}\sum_{n=0}^{\infty}\frac{1}{\sqrt{n !}} e^{-n \frac{\omega \beta}{2}}\left(-\frac{\partial}{\partial z} - \frac{\bar{z}}{2}I_{\tilde{\mathfrak H}}\right)^n|z;n\rangle.
\ena
Set $|z;n\rangle = U_2(z)|\Psi_{n0}\rangle$ with
$(A^{\dag}_2)^n|\Psi_{n0}\rangle = \sqrt{n !}|\Psi_{nn}\rangle\,$
and
\bea
U_2(z)|\Psi_{n0}\rangle =  e^{-\frac{|z|^2}{2}}\sum_{k=0}^{\infty}
\frac{(zA^{\dag}_2)^k}{k !}|\Psi_{n0}\rangle.
\ena
Taking  $ |z;0\rangle = U_2(z)|\Psi_{00}\rangle $ leads to
\bea 
(A^{\dag}_2)^n|z;0\rangle = \sqrt{n !}|z;n\rangle \quad \mbox{with} \quad 
U_2(z)\left[(A^{\dag}_2)^n|\Psi_{00}\rangle\right] = \sqrt{n ! }U_2(z)|
\Psi_{n0}\rangle.
\ena 
 Then, $\;$
$|z;n\rangle = \frac{1}{\sqrt{n !}}(A^{\dag}_2)^n |z;0\rangle $.
In (\ref{operat012}), we get 
\bea\label{nino}
|z,\bar{z},\beta\rangle^{\mbox{\tiny{KMS}}} = (1- e^{-\omega \beta})^{\frac{1}{2}}\sum_{n=0}^{\infty}\frac{1}{n !} e^{-n \frac{\omega \beta}{2}}\left(-\frac{\partial}{\partial z} - \frac{\bar{z}}{2}\right)^n(A^{\dag}_2)^n|z;0\rangle.
\ena
Using the relation $ U_{1}(x,y)^{*} = U_{1}(-x,-y)$,
the CS 
$|-z,-\bar{z},\beta\rangle^{\mbox{\tiny{KMS}}}$ are obtained as follows:
\bea\label{xixi}
|-z,-\bar{z},\beta\rangle^{\mbox{\tiny{KMS}}} = S_{\beta}|z,\bar{z},\beta\rangle^{\mbox{\tiny{KMS}}}.
\ena
Indeed, by definition $\;$
$ |z,\bar{z},\beta\rangle^{\mbox{\tiny{KMS}}} = U_1(z)|\Phi_{\beta}\rangle := U_1(x,y)|\Phi_{\beta}\rangle$ such that 
\bea
S_{\beta}\left[U_1(x,y)|\Phi_{\beta}\rangle\right] = U_1(x,y)^{*}|\Phi_{\beta}\rangle = U_1(-x,-y)|\Phi_{\beta}\rangle,  \; \mbox{i.e.,}\;
U_1(-z)|\Phi_{\beta}\rangle =  e^{-zA^{\dag}_1 + \bar{z}A_1}|\Phi_{\beta}\rangle
\ena
where $ e^{-zA^{\dag}_1 + \bar{z}A_1}|\Phi_{\beta}\rangle := |-z,-\bar{z'},\beta\rangle^{\mbox{\tiny{KMS}}}$, leading to (\ref{xixi}).
\noi The CS  (\ref{xixi}) satisfy a resolution of the identity analogue to (\ref{hiho}), i.e., 
\bea\label{xaxa}
\frac {1}{2\pi}\int_{\CN}|-z,-\bar{z},\beta\rangle^{\mbox{\tiny{KMS}}}
{^{\mbox{\tiny{KMS}}}}\langle-z,-\bar{z},\beta| dxdy = I_{\tilde{\mathfrak H}}.
\ena
\noi {\bf Proof.} Similar to the proof of (\ref{hiho}).
$\hfill{\square}$

\section{Noncommutative quantum harmonic oscillator Hilbert space}
{ Without loss of generality, we restrict our developments to  the noncommutative quantum mechanics formalism
\cite{connes1, scholtz, ben-scholtz, a-hk} for the physical system of harmonic oscillator.} We focus on the application of Hilbert-Schmidt
operators, bounded operators on the noncommutative classical configuration space  denoted by
\bea
\mathcal H_c = \mbox{span}\left\{|n\rangle\  = \frac{1}{\sqrt{n !}}(a^{\dag})^{n}|0\rangle\right\}_{n=0}^{\infty}.
 \ena 
This space is isomorphic to the boson Fock space $\mathcal F = \displaystyle\{|n\rangle\}_{n=0}^{\infty}, $ 
where the annihilation and creation operators $a, a^{\dag}$ obey the Fock algebra $[a, a^{\dag}] = 1$.
\noi The physical states of the system  represented on $\mathcal H_{q}$ known  as the set of Hilbert-Schmidt operators,   
is equivalent to the Hilbert space of square integrable function, with 
\bea
\mathcal H_{q} = \left\{\psi(\hat x_{1}, \hat x_{2}): \psi(\hat x_{1}, \hat x_{2}) \in \mathcal B(\mathcal H_{c}),\,
tr_{c}(\psi(\hat x_{1}, \hat x_{2})^{\dag}, \psi(\hat x_{1}, \hat x_{2}))
 < \infty \right \}
\ena
where $\mathcal B(\mathcal H_{c})$ is the set of bounded operators on $\mathcal H_{c}$. 
$\mathcal H_{q}$ is defined  as the set of bounded operators, with the form $|\cdot\rangle \langle \cdot|$,  
acting on the classical configuration space $\mathcal H_c$, with  a general element of the quantum Hilbert space, in
''bra-ket'' notation given by
\bea
|\psi) = \sum_{n,m=0}^{\infty}c_{n,m}|n, m),  
\ena
with $ \left\{|n, m) := |n\rangle \langle m|\right\}_{n,m=0}^{\infty}$  a basis of $\mathcal H_{q}$  endowed with the 
inner product
\bea\label{quantinner}
(\tilde n, \tilde m|n, m)
= {\mbox tr}_{c}[(|\tilde n \rangle \langle \tilde m|)^{\ddag}| n \rangle \langle  m|]
= \delta_{\tilde n,  n}
\delta_{\tilde m,m}. 
\ena
\noi Considering the  
unitary {\it  Wigner map}
$\mathcal W:  \mathcal B_2(\mathfrak H) \rightarrow L^{2}(\RN^{2}, dxdy)$ 
 let us discuss a correspondence between $L^{2}(\RN^{2}, dxdy)$ 
and $\mathcal B_2(\mathfrak H)$.
\beprop
Given the Hilbert space  $\mathfrak H =  L^{2}(\RN)$, the inverse of the map $\mathcal W$ is defined on the dense set of vectors 
$f \in L^{2}(\RN^{2}, dxdy)$  as follows:
\begin{eqnarray}
&&\mathcal W^{-1}:L^{2}(\RN^{2}, dxdy) \rightarrow \mathfrak H \otimes \overline{\mathfrak H} \cr
&&\mathcal W^{-1} f =  \int_{\RN}\int_{\RN}U(x,y)\mathcal W(|\phi\rangle \langle \psi|)(x,y)dxdy, 
\end{eqnarray}
where the integral is 
defined weakly, 
 $|\phi\rangle \langle \psi|$ is an element of $\mathcal B_2(\mathfrak H) \simeq \mathfrak H \otimes \overline{\mathfrak H}$ and 
$f = \mathcal W(|\phi\rangle \langle \psi|)$.
\enprop
\noi {\bf Proof}. 
Let us derive the inverse of the map $\mathcal W$ on  $L^{2}(\RN^{2}, dxdy)$ where the group $G$ and the   Duflo-Moore  
operator $C$ with 
domain $\mathcal D(C^{-1})$ given in \cite{ali-antoine-gazeau} are identified here to  $\RN^{2}$, $I_{\mathfrak H}$, the identity
 operator  on $\mathfrak H = L^{2}(\RN)$, respectively, with $\mathcal D(C^{-1}) = \mathfrak H $ and 
$\mathcal D(C^{-1})^{\dag} = \overline{\mathfrak H}$.
\noi Consider an element in $\mathcal B_2(\mathfrak H) \simeq \mathfrak H \otimes \overline{\mathfrak H}$ 
of the type $|\phi\rangle \langle \psi|$, with $\phi, \psi\in \mathfrak H$ and let 
$f = \mathcal W(|\phi\rangle \langle \psi|)$.
\noi For $\phi', \psi' \in \mathfrak H$, we have from the definition of $\mathcal W$ in (\ref{map1})
\bea{\label{map03}}
\int_{\RN}\int_{\RN}\langle \phi'|U(x,y)\psi' \rangle \mathcal W(|\phi\rangle \langle \psi|)(x,y)dxdy
&=& \int_{\RN}\int_{\RN}\langle \phi'|U(x,y)\psi' \rangle Tr(U(x,y)^{*}|\phi\rangle \langle \psi|)dxdy \cr
&=& \int_{\RN}\int_{\RN}\overline{\langle \phi|U(x,y)\psi \rangle}\langle \phi'|U(x,y)\psi' \rangle dxdy.
\ena
\noi By the orthogonality relation,  we  get
\bea{\label{map3}}
\int_{\RN}\int_{\RN}\langle \phi'|U(x,y)\psi' \rangle \mathcal W(|\phi\rangle \langle \psi|)(x,y)dxdy  
= \langle \phi'|\phi \rangle \langle \psi|\psi' \rangle.
\ena
\noi The relation $|\langle \phi'|\phi \rangle \langle \psi|\psi' \rangle| \leq ||\phi'||||\psi'||||\phi||||\psi||$ implies 
\bea
\left|\int_{\RN}\int_{\RN}\langle \phi'|U(x,y)\psi' \rangle \mathcal W(|\phi\rangle
 \langle \psi|)(x,y)dxdy \right| \leq ||\phi'||||\psi'||||\phi||||\psi||.
\ena
\noi Then,   (\ref{map3}) holds for all $\phi',\psi' \in \mathfrak H$. Then,  we obtain
\bea
|\phi\rangle \langle \psi| = \int_{\RN}\int_{\RN}U(x,y)\mathcal W(|\phi\rangle \langle \psi|)(x,y)dxdy
\ena
\noi which completes the proof.
 $\hfill{\square}$

\section{Application}
 Consider the motion of an electron  in the $xy$-plane, subjected to a constant
magnetic field pointing along the positive $z-$direction, i.e., in the symmetric gauge 
${\bf A}^{\uparrow} = \left(-\frac{B}{2}y, \frac{B}{2}x \right),$ in the presence of a harmonic potential, described by the 
following Hamiltonian \cite{ahb}
\bea
H_\theta =   \frac{1}{2M}\left(p_x - \frac{eB}{2c}y\right)^2 + 
\frac{1}{2M}\left(p_y + \frac{eB}{2c}x\right)^2  + \frac{M\omega^2_0}{2}(x^2  + y^2)
\ena
where $\omega_c =\frac{eB}{2c}$ is the cyclotron frequency, $\Omega^2 =\omega^2_0 +\frac{\omega^2_c }{4}$, with the 
commutation relations:
\bea
[x^{i}, x^{j}]  =  i\theta\epsilon^{ij}, \quad [x^{i}, p^{j}]  =  i\hbar \delta^{ij}, \quad [p^{i}, p^{j}]  = 0, 
\quad i,j= 1,2, \quad \epsilon^{12} =  -  \epsilon^{21} = 1.
\ena
Next, introduce 
the dimensionless complex variables,  related to the chiral decomposition of the physical model,   given by 
\bea\label{var00}
z_+ =  \frac{1}{\sqrt{2}}(x^1_+ - i x^2_+), \qquad 
z_-=  \frac{1}{\sqrt{2}}(x^1_- + i x^2_-)
\ena
%
such that they satisfy, with 
\bea
\partial_{z_+} = \frac{1}{\sqrt{2}}[\partial_{x^1_+}+i\partial_{x^2_+}], \quad  
\partial_{z_-} = \frac{1}{\sqrt{2}}[\partial_{x^1_-}-i\partial_{x^2_-}],
\ena
the relations
\bea\label{opchi00}
[\partial_{z_\pm}, z_\pm] = 1 =  [\partial_{\bar z_\pm}, \bar z_\pm], 
\qquad [\partial_{z_\pm}, z_\mp] = 0 =  [\partial_{\bar z_\mp}, \bar z_\pm], 
\qquad [z_+, z_-] = 0 = [\partial_{z_+}, \partial_{z_-}].
\ena
\noi Set 
\bea\label{ladder00}
A_{+} &=& \zeta \frac{{\bar z}}{2} + \frac{\imath}{\zeta \hbar}p_{z}, \quad 
A^{\dag}_{+} = \zeta \frac{z}{2} - \frac{\imath}{\zeta \hbar}p_{\bar z} \crcr
A_{-} &=&  \zeta \frac{z}{2} +
\frac{\imath}{\zeta \hbar}p_{\bar z}  , \quad  A^{\dag}_{-} =  \zeta \frac{{\bar{z}}}{2} -
\frac{\imath}{\zeta \hbar}   p_{z}, \quad  \zeta  
= \sqrt[4]{\frac{(M \Omega/\hbar)^{2}}{1-\frac{M\omega_{c}}{2}\theta + \left(\frac{M \Omega}{4}\theta\right)^{2}}} 
\ena 
satisfying the commutation relations 
\bea\label{opchi04}
[A_{-}, A^{\dag}_{+}] = 0 = [A_{+}, A^{\dag}_{-}], \quad 
[A_{\pm}, A^{\dag}_{\pm}] = 1.
\ena
Then, taking $\tilde \Omega_\pm = \tilde \Omega \pm \frac{\tilde \omega_{c}}{2}, $ where \cite{a-hk} 
\bea
\tilde \Omega = \Omega \sqrt{1-\frac{M\omega_{c}}{2}\theta + \left(\frac{M \Omega}{4}\theta\right)^{2}} \quad \quad 
\tilde \omega_{c} = \omega_{c}\left(1 - \left(\frac{\omega_{c}}{4} + \frac{\omega^{2}_{0}}{ \omega_{c}}\right)M\theta\right), 
\ena
the Hamiltonian $H_\theta$ is obtained as follows:
\bea\label{hamiltheta}
H_\theta = \hbar \tilde \Omega_+\left(N_+ + \frac{1}{2}  \right) +  
\hbar \tilde \Omega_-\left(N_- + \frac{1}{2}  \right), \quad N_\pm  = A^{\dag}_{\pm}A_{\pm},  
\ena
$N_\pm$ being the number operators such that $H_\theta$ writes 
\bea\label{hamiltheta00}
H_\theta = H_+ \otimes \mathbb I_{\mathcal H_{q,-}} +   \mathbb I_{\mathcal H_{q,+}}   \otimes H_-, 
\quad H_{\pm} = \hbar \tilde \Omega_\pm \left(N_\pm + \frac{1}{2}  \right).
\ena
The Hilbert spaces $\mathcal H_{q,\pm}$ are given by  
 $\mathcal H_{q,\pm} = span\{|n_{\pm}\rangle \langle m_{\pm}|\}^{\infty}_{{n_\pm, m_\pm}=0}$, with 
 $\mathbb I_{\mathcal H_q,\pm}$ their corresponding identity operators.
The system $\{A_{\pm}, A^{\dag}_{\pm}\}$ forms an irreducible set of operators on the 
chiral boson Fock space $\mathcal F = \{|n_{\pm}\rangle \}^{\infty}_{{n_\pm}=0}$, and  has the following realization on the states 
\bea
|n_{+},  n_{-};m_{+},  m_{-}):= 
|n_{+}\rangle \langle m_{+}|\otimes |n_{-}\rangle \langle m_{-}|,  \;  n_{\pm},  m_{\pm} = 0, 1, 2, \dots,  
\ena
of  the Hilbert space 
$\mathcal H_{q,+} \otimes \mathcal H_{q,-}$:
\bea\label{opchi02}
A_{+}| n_{+},   n_{-};m_{+},  m_{-})  &:=& \sqrt{n_{+}}| n_{+}-1,  n_{-};m_{+},  m_{-}) \cr
  A^{\dag}_{+}|n_{+},  n_{-};m_{+},  m_{-}) &:=& \sqrt{ n_{+} + 1 }| n_{+}+1,  n_{-};m_{+},  m_{-}),
\ena
\bea\label{opchi03}
A_{-}| n_{+},  n_{-};m_{+},  m_{-}) &:=& 
\sqrt{ n{-}}|n_{+}, n_{-}-1;m_{+},  m_{-}) \cr
\quad \quad  A^{\dag}_{-}|n_{+}, n_{-};m_{+},  m_{-}) &:=&  \sqrt{n_{-} + 1} | n_{+},  n_{-}+1;m_{+},  m_{-}).
\ena
The operators  $A_{\pm}$ act on the right by $A^{\dag}_{\pm}$   by conjugation of 
(\ref{opchi02}) and (\ref{opchi03}). 
We have
\bea\label{eig00}
|n_{+},  n_{-};0,  0)
= \frac{1}{\sqrt{n_{+} ! n_{-} !}}\left(A^{\dag}_{+}\right)^{n_{+}}
\left(A^{\dag}_{-}\right)^{n_{-}}|0,0\rangle \langle 0,0|
\ena
with  $|0,0\rangle \langle 0,0|$  
standing for the vacuum state of $\mathcal H_{q,+} \otimes \mathcal H_{q,-}.$ 

\noi Then, the eigenvalues of the Hamiltonian $H_\theta$ are derived 
from the relation 
\bea
H_\theta \left(|n_{+}\rangle \langle m_{+}|\otimes |n_{-}\rangle \langle m_{-}| \right)= 
E_{n_+, n_-}\left(|n_{+}\rangle \langle m_{+}|\otimes |n_{-}\rangle \langle m_{-}|\right)
\ena
 as follows:
\bea\label{eigenval001}
E_{n_+, n_-} =  \hbar \tilde \Omega_+\left(n_+ + \frac{1}{2}  \right) +  
\hbar \tilde \Omega_-\left(n_- + \frac{1}{2}  \right).
\ena
\noi Given a   state $|m\rangle \langle n|$ on $\mathcal H_q \simeq  \mathfrak H \otimes \overline{\mathfrak H},$ 
 the  left $a_{L}$ and right $b_{R}$ annihilation operators act as follows:
\bea{\label{vec}}
a_{L}|m\rangle \langle n|=  (a \otimes 
I_{\bar{\mathfrak H}})|m\rangle \langle n|  &=& a|m\rangle  \langle n|I_{\bar{\mathfrak H}} \cr
&=& \sqrt{m}|m-1 \rangle \langle  n|\cr
 b_{R}|m\rangle \langle n| = (I_{\mathfrak H} \otimes b)|m\rangle \langle n|  &=& I_{\mathfrak H} |m\rangle  \langle n| b \cr
&=& \sqrt{n+1}|m \rangle \langle  n+1|,
\ena
where 
\bea
 a_{L}:=a \otimes 
I_{\bar{\mathfrak H}}=  \frac{1}{\sqrt{2}}\left[(Q +  iP) \otimes  I_{\bar{\mathfrak H}}\right] , \, 
a^{\dag}_{L}:=a^{\dag} \otimes 
I_{\bar{\mathfrak H}} = \frac{1}{\sqrt{2}}\left[(Q -  iP) \otimes  I_{\bar{\mathfrak H}})\right],
\ena
\bea
b_{R}:= I_{{\mathfrak H}} \otimes b = \frac{1}{\sqrt{2}}\left[I_{{\mathfrak H}} \otimes (iQ - P)\right], \, 
b^{\dag}_{R}:= I_{{\mathfrak H}} \otimes  b^{\dag} =  \frac{1}{\sqrt{2}}\left[I_{{\mathfrak H}} \otimes (-iQ - P)\right]. 
\ena
 $Q, P$ are the usual position and momentum operators given on 
$\mathfrak H  =  L^2(\RN)$, with $[Q, P] = i \IN_{\mathfrak H}.$
\subsection{Coherent states construction}
\noi Using  the operators $\{A_{\pm}, A^{\dag}_{\pm}\}$, the eigenstates $|z_{\pm}\rangle $ satisfy
\bea
A_{\pm}|z_{\pm}\rangle  = z_{\pm}|z_{\pm}\rangle, \qquad \langle z_{\pm}|A^{\dag}_{\pm} = \langle z_{\pm}|\bar z_{\pm}
\ena
\noi with the complex eigenvalues $z_{\pm}$,    
\bea\label{cohst00}
|z_{\pm}\rangle  = e^{-\frac{|z_{\pm}|^2}{2}}e^{\{z_{\pm} A^{\dag}_{\pm}\}} |0\rangle, 
\ena
\noi given in terms of the chiral Fock basis. Provided the  Baker-Campbell-Hausdorff identity 
\bea 
e^{\{z_{\pm} A^{\dag}_{\pm} - \bar z_{\pm} A_{\pm}\}} = e^{-\frac{|z_{\pm}|^2}{2}}e^{\{z_{\pm} A^{\dag}_{\pm}\}}
e^{\{-\bar z_{\pm} A_{\pm}\}},
\ena
\noi the  CS of the  Hamiltonian  (\ref{hamiltheta00}) in  the noncommutative plane, 
 denoted by $|z_{+}, z_{-}), $ are defined by 
\bea\label{vect00}
|z_{+}, z_{-}) =    
e^{- (|z_{+}|^{2} + |z_{-}|^{2})}
\sum_{{n}_{+}, {m}_{+}  = 0}^{\infty}\sum_{{n}_{-}, {m}_{-} = 0}^{\infty}
\frac{z_{+}^{{n}_{+}}\bar z_{+}^{{m}_{+}}z_{-}^{{n}_{-}}\bar z_{-}^{{m}_{-}}}{\sqrt{{n}_{+} !{m}_{+} !{n}_{-} !{m}_{-} !}}
|n_{+}\rangle \langle m_{+}|\otimes |n_{-}\rangle \langle m_{-}|
\ena
 where we have used \cite{ben-scholtz}:
\bea
| z_{+} \rangle \langle  z_{+} |  = D_R D_L \left(|0\rangle \langle  0|\right) = 
e^{- |z_{+}|^2}e^{z_{+} A^{\dag}_{+}}|0\rangle \langle  0|e^{\bar z_{+} A_{+}}
\ena
with $D_R  =  e_R^{-z_{+} A^{\dag}_{+}+\bar z_{+} A_{+}}$ and $D_L  =  e_L^{-\bar z_{+} A_{+}+ z_{+} A^{\dag}_{+}}. $ The lower indices 
$R, L$ of the exponential operators refer to the right and left actions, respectively.

\noi These CS satisfy the resolution of the identity \cite{a-hk}
\bea{\label{resolv}}
\frac{1}{\pi^{2}}\int_{\mathbb C^{2}}|z_{+}, z_{-})
(z_{+}, z_{-}|d^{2}z_{+}d^{2}z_{-}  = \mathbb I_{\mathcal H_{q,+} \otimes \mathcal H_{q,-}}:= \mathbb I_{q}\otimes \mathbb I_{q},
\ena
\noi where $\mathbb I_{q}$   stands  for the identity on  $\mathcal H_{q}$ given by \cite{scholtz}:
 \bea{\label{resolv01}}
\mathbb I_{q} = \frac{1}{\pi}\int_{\mathbb C}dzd\bar{z}|z)e^{\overleftarrow{\partial_{\bar z}}\overrightarrow{\partial_{z}}} (z|.
\ena
and the identity operator on $ \mathcal H_{q,+} \otimes \mathcal H_{q,-}$ is: 
\bea
\mathbb I_{\mathcal H_{q,+} \otimes \mathcal H_{q,-}} &=& \sum_{{n}_{+}, {m}_{+}  = 0}^{\infty}\sum_{{n}_{-}, {m}_{-} = 0}^{\infty}
|n_{+},  n_{-};m_{+},  m_{-})(n_{+},  n_{-};m_{+},  m_{-}|.
\ena
{\bf Proof.}

\noi In order to provide an equivalence between (\ref{resolv}) and (\ref{resolv01}), let us  consider the following relations
\bea{\label{resolv02}}
\mathbb I_{q}|\psi) &=& \frac{1}{\pi^{2}}\int_{\mathbb C^{2}}dzd\bar{z}dwd\bar{w}
|z\rangle \langle w| \langle z|\psi|w \rangle \cr
&=& \frac{1}{\pi^{2}}\int_{\mathbb C^{2}}dzd\bar{z}dud\bar{u}
|z\rangle \langle z+u| \langle z|\psi|z+u \rangle \cr
&=& \frac{1}{\pi}\int_{\mathbb C}dzd\bar{z}\frac{1}{\pi}\int_{\mathbb C}d^{2}ue^{-|u|^{2}}|z\rangle\langle z|
e^{\bar u\overleftarrow{\partial_{\bar z}} + u\overrightarrow{\partial_{z}}}\langle z|\psi|z\rangle 
\ena
where   $w = z + u$ with $d^{2}w = d^{2}u $,  and   $e^{u \partial_{z}} f(z) = f(z+u)$.  Then,  set
\bea
\frac{1}{\pi}\int_{\mathbb C}d^{2}ue^{-|u|^{2}}|z\rangle\langle z|
e^{\bar u\overleftarrow{\partial_{\bar z}} + u\overrightarrow{\partial_{z}}}\langle z|\psi|z\rangle =
\frac{1}{\pi}\int_{\mathbb C}d^{2}ue^{-|u|^{2}}|z\rangle\langle z|e^{\bar u\overleftarrow{\partial_{\bar z}}}e^{u \overrightarrow{\partial_{z}}}
\langle z|\psi|z\rangle
\ena
and
\bea
I = |z\rangle\langle z|e^{\bar u\overleftarrow{\partial_{\bar z}}}e^{u \overrightarrow{\partial_{z}}}
\langle z|\psi|z\rangle.
\ena
We have
\bea
I &=& \left[\sum_{n', m' = 0}^{\infty}|n'\rangle\langle m'|e^{-{\bar z}z}\frac{{\bar z}^{m'}}{\sqrt{m' !}}\frac{{ z}^{n'}}{\sqrt{n' !}}\right]
 e^{\bar u\overleftarrow{\partial_{\bar z}}}e^{u \overrightarrow{\partial_{z}}}
\left[\sum_{n, m = 0}^{\infty} \langle m|\psi|n\rangle e^{-{\bar z}z}\frac{{\bar z}^{n}}{\sqrt{n !}}\frac{{ z}^{m}}{\sqrt{m !}}\right] \crcr
 &=& \left[ \sum_{n, m = 0}^{\infty} \sum_{n', m' = 0}^{\infty}\frac{z^{n'}}{\sqrt{n' !}}\frac{{\bar z}^{n}}{\sqrt{n !}}|n'\rangle
\langle m'|\langle m|\psi|n\rangle \right]
\left(e^{-{\bar z}z} \frac{{\bar z}^{m'}}{\sqrt{m' !}}\right)
e^{\bar u\overleftarrow{\partial_{\bar z}}}e^{u \overrightarrow{\partial_{z}}}\left(e^{-{\bar z}z} \frac{{z}^{m}}{\sqrt{m !}}\right).\nonumber
\\
\ena
Let 
\bea
K(z) = \left(e^{-{\bar z}z} \frac{{\bar z}^{m'}}{\sqrt{m' !}}\right)
e^{\bar u\overleftarrow{\partial_{\bar z}}}e^{u \overrightarrow{\partial_{z}}}\left(e^{-{\bar z}z} \frac{{z}^{m}}{\sqrt{m !}}\right).
\ena
We obtain
\bea
K(z) &=& \frac{1}{\sqrt{m' !}} \frac{1}{\sqrt{m !}}\sum_{k=0}^{\infty}\sum_{l=0}^{\infty}\frac{1}{k !}
\left({\bar u}^{k}\partial^{k}_{{\bar z}}[{\bar z}^{m'}
e^{-{\bar z}z}]\right)\frac{1}{l !}\left({u}^{l}\partial^{l}_{{z}}\left[{ z}^{m}
e^{-{\bar z}z}\right]\right)
\ena
which supplies,   by performing a radial parametrization,   
\bea
&&\frac{1}{\pi}\int_{\mathbb C}d^{2}ue^{-|u|^{2}}K(z) \cr
&=& \frac{1}{\sqrt{m' !}} \frac{1}{\sqrt{m !}}\sum_{k=0}^{\infty}\sum_{l=0}^{\infty}
\frac{1}{\pi}\int_{\mathbb C}d^{2}ue^{-|u|^{2}}\frac{{\bar u}^{k}}{k !}\frac{u^{l}}{l !}\partial^{k}_{{\bar z}}[{\bar z}^{m'}
e^{-{\bar z}z}]\partial^{l}_{{z}}\left[{ z}^{m}
e^{-{\bar z}z}\right] \cr
&=&  \frac{1}{\sqrt{m' !}} \frac{1}{\sqrt{m !}}\sum_{k=0}^{\infty}\sum_{l=0}^{\infty}
\frac{1}{\pi}\int_{0}^{\infty}rdre^{-r^{2}}\frac{r^{k+l}}{k ! l !}\int_{0}^{2\pi}e^{-i (l-k)\phi}d\phi \cr
&& \times \partial^{k}_{{\bar z}}[{\bar z}^{m'}
e^{-{\bar z}z}]\partial^{l}_{{z}}\left[{ z}^{m}
e^{-{\bar z}z}\right]\cr
&=&   \frac{1}{\sqrt{m' !}} \frac{1}{\sqrt{m !}}\sum_{k=0}^{\infty}\left[\frac{1}{k !}\int_{0}^{\infty}2r^{2k+1}e^{-r^{2}}dr\right]
\left[\frac{1}{k !}\partial^{k}_{{\bar z}}[{\bar z}^{m'}
e^{-{\bar z}z}]\partial^{k}_{{z}}\left[{ z}^{m}
e^{-{\bar z}z}\right]\right] \cr
&=& \frac{1}{\sqrt{m' !}} \frac{1}{\sqrt{m !}}\sum_{k=0}^{\infty}\left[\frac{1}{k !}\partial^{k}_{{\bar z}}[{\bar z}^{m'}
e^{-{\bar z}z}]\partial^{k}_{{z}}\left[{ z}^{m}
e^{-{\bar z}z}\right]\right]. 
\ena
Besides, 
\bea
\left(e^{-{\bar z}z} \frac{{\bar z}^{m'}}{\sqrt{m' !}}\right)
e^{\overleftarrow{\partial_{\bar z}}\overrightarrow{\partial_{z}}}\left(e^{-{\bar z}z} \frac{{z}^{m}}{\sqrt{m !}}\right) = 
\frac{1}{\sqrt{m' !}} \frac{1}{\sqrt{m !}}\sum_{k=0}^{\infty}\left[\frac{1}{k !}\partial^{k}_{{\bar z}}[{\bar z}^{m'}
e^{-{\bar z}z}]\partial^{k}_{{z}}\left[{ z}^{m}
e^{-{\bar z}z}\right]\right]
\ena
 implying
\bea
\frac{1}{\pi}\int_{\mathbb C}d^{2}ue^{-|u|^{2}}K(z) =  \left(e^{-{\bar z}z} \frac{{\bar z}^{m'}}{\sqrt{m' !}}\right)
e^{\overleftarrow{\partial_{\bar z}}\overrightarrow{\partial_{z}}}\left(e^{-{\bar z}z} \frac{{z}^{m}}{\sqrt{m !}}\right).
\ena
Then, 
\bea
&&\frac{1}{\pi}\int_{\mathbb C}d^{2}ue^{-|u|^{2}}|z\rangle\langle z|
e^{\bar u\overleftarrow{\partial_{\bar z}} + u\overrightarrow{\partial_{z}}}\langle z|\psi|z\rangle \cr
&=& 
\left[ \sum_{n, m = 0}^{\infty} \sum_{n', m' = 0}^{\infty}\frac{z^{n'}}{\sqrt{n' !}}\frac{{\bar z}^{n}}{\sqrt{n !}}|n'\rangle
\langle m'|\langle m|\psi|n\rangle \right]
\left(e^{-{\bar z}z} \frac{{\bar z}^{m'}}{\sqrt{m' !}}\right)
e^{\overleftarrow{\partial_{\bar z}}\overrightarrow{\partial_{z}}}\left(e^{-{\bar z}z} \frac{{z}^{m}}{\sqrt{m !}}\right)\cr
&=& \left[\sum_{n', m' = 0}^{\infty}|n'\rangle\langle m'|e^{-{\bar z}z}\frac{{\bar z}^{m'}}{\sqrt{m' !}}\frac{{ z}^{n'}}{\sqrt{n' !}}\right]
e^{\overleftarrow{\partial_{\bar z}}\overrightarrow{\partial_{z}}}
\left[\sum_{n, m = 0}^{\infty}\langle m|\psi|n\rangle e^{-{\bar z}z}\frac{{\bar z}^{n}}{\sqrt{n !}}\frac{{ z}^{m}}{\sqrt{m !}}\right] \cr
&=& |z)e^{\overleftarrow{\partial_{\bar z}}\overrightarrow{\partial_{z}}} (z|\psi)
\ena
allowing to obtain (\ref{resolv02}) under the form:
\bea
\mathbb I_{q}|\psi) &=& \frac{1}{\pi}\int_{\mathbb C}dzd\bar{z}\frac{1}{\pi}\int_{\mathbb C}d^{2}ue^{-|u|^{2}}|z\rangle\langle z|
e^{\bar u\overleftarrow{\partial_{\bar z}} + u\overrightarrow{\partial_{z}}}\langle z|\psi|z\rangle \cr
&=& \frac{1}{\pi}\int_{\mathbb C}dzd\bar{z}|z)e^{\overleftarrow{\partial_{\bar z}}\overrightarrow{\partial_{z}}} (z|\psi)
\ena
which completes the proof.
$\hfill{\square}$

\subsection{Density matrix and diagonal elements}
Considering that the quantum system obeys  the  canonical distribution \cite{cahill-glauber, gazbook09, ahb}, 
let us take the partition function $\mathcal Z$ as that of 
a composite system made of two independent systems such that it is the product of the partition functions  of the components, i.e.    
$\mathcal Z = \mathcal Z_+ \mathcal Z_-$. The  diagonal elements   of the density operator 
$\hat \rho = \frac{1}{\mathcal Z} e^{-\beta H_\theta}$, where  $H_\theta$ is given in (\ref{hamiltheta}) with eigenvalues
$E_{{n}_{+}, {n}_{-}}$  in (\ref{eigenval001}), 
  are then  derived, in the CS 
$|z_{+}, z_{-})$  (\ref{vect00}) representation, as
\label{density00}
\bea
(z_{+}, z_{-}|\hat \rho|z_{+}, z_{-})
 &=& (z_{+}, z_{-}|\left\{\sum_{{n}_{+}, {m}_{+}  = 0}^{\infty}\sum_{{n}_{-}, {m}_{-} = 0}^{\infty}
\frac{e^{-\beta H_\theta}}{\mathcal Z}|n_{+},  n_{-};m_{+},  m_{-}) (n_{+},  n_{-};m_{+},  m_{-}|\right\}\cr
&&|z_{+}, z_{-})  \cr
&=& \left\{\frac{1}{\mathcal Z_+}e^{-\frac{\beta\hbar\tilde \Omega_+}{2}}e^{- |z_+|^{2}}
 \left[e^{e^{- \beta \hbar\tilde \Omega_+}|z_+|^{2}}\right]\right\}\left\{\frac{1}{\mathcal Z_-}
 e^{-\frac{\beta\hbar\tilde\Omega_-}{2}}e^{- |z_-|^{2}}
 \left[e^{e^{-\beta\hbar\tilde \Omega_-}|z_-|^{2}}\right]\right\}\nonumber \\
\ena
where $\mathcal Z = \mathcal Z_+ \mathcal Z_-$.

\noi Since
\bea
\frac{1}{\mathcal Z_+} = \left[\frac{e^{-\frac{\beta\hbar\tilde \Omega_+}{2}}}{1-e^{-\beta\hbar\tilde \Omega_+}}\right]^{-1}, 
\qquad \mbox{and} \qquad \frac{1}{\mathcal Z_-} = \left[\frac{e^{-\frac{\beta\hbar\tilde \Omega_-}{2}}}
{1-e^{-\beta\hbar\tilde \Omega_-}}\right]^{-1}
\ena
then, 
\bea
(z_{+}, z_{-}|\hat \rho|z_{+}, z_{-}) &=&  \left[\frac{e^{-\frac{\beta\hbar\tilde \Omega_+}{2}}}{1-e^{-\beta\hbar\tilde \Omega_+}}\right]^{-1} 
e^{-\frac{\beta\hbar\tilde \Omega_+}{2}}e^{- |z_+|^{2}}\left[e^{e^{-\beta\hbar\tilde \Omega_+}|z_+|^{2}}\right]\cr
&& \times \left[\frac{e^{-\frac{\beta\hbar\tilde \Omega_-}{2}}}{1-e^{-\beta\hbar\tilde \Omega_-}}\right]^{-1}
e^{-\frac{\beta\hbar\tilde \Omega_-}{2}}e^{- |z_-|^{2}}
 \left[e^{e^{-\beta\hbar\tilde \Omega_-}|z_-|^{2}}\right].
\ena
Thereby, 
\bea\label{husimidistr00}
(z_{+}, z_{-}|\hat \rho|z_{+}, z_{-})
&=&\left[1-e^{- \beta\hbar\tilde \Omega_+}\right]e^{- (1-e^{- \beta\hbar\tilde \Omega_+})|z_+|^2} \times 
\left[1-e^{- \beta\hbar\tilde \Omega_-}\right]e^{- (1-e^{- \beta\hbar\tilde \Omega_-})|z_-|^2}\cr
&=& \frac{1}{\bar{n}+1} e^{-\frac{1}{\bar{n}+1}|z_+|^2} \times \frac{1}{{\bar{n}}^{*}+1} e^{-\frac{1}{{\bar{n}}^{*}+1}|z_-|^2}\cr
&=& Q(|z_+|^2)Q(|z_-|^2). 
\ena 
\noi $\bar{n} = \left[e^{\beta\hbar\tilde \Omega_+}-1\right]^{-1}$ and ${\bar{n}}^{*} =\left[e^{\beta\hbar\tilde \Omega_-}-1\right]^{-1}$ are 
the 
corresponding thermal expectation values of the number operator (i. e. the Bose-Einstein distribution functions for oscillators with 
angular frequencies  $\tilde \Omega_+$ and $\tilde \Omega_-, $ respectively), also called  the thermal mean occupancy for harmonic oscillators with the 
angular frequencies  $\tilde \Omega_+$ and $\tilde \Omega_-, $ respectively.

\noi Performing the variable changes $r_+ = \left[1-e^{- \beta\hbar\tilde \Omega_+}\right]^{1/2}|z_+|$ and $r_- =
 \left[1-e^{- \beta\hbar\tilde \Omega_-}\right]^{1/2}|z_-|$ 
 with $\frac{d^2 z}{\pi}  =  rdr \frac{d\varphi}{\pi}, \, r \in [0, \infty), \varphi 
\in (0, 2\pi]
$, we obtain
\bea
 Tr \hat \rho   &=& \frac{1}{\pi^2}\int_{\mathbb C^{2}}   
d^{2}z_{+}d^{2}z_{-}(z_{+}, z_{-}|\hat \rho|z_{+}, z_{-}) \cr 
&=& \frac{1}{\pi^2}\int_{\mathbb C^{2}}   
d^{2}z_{+}d^{2}z_{-}
\left[1-e^{- \beta\hbar\tilde \Omega_+}\right]e^{- (1-e^{- \beta\hbar\tilde \Omega_+})|z_+|^2}\left[1-e^{- \beta\hbar\tilde \Omega_-}\right]
e^{- (1-e^{- \beta\hbar\tilde \Omega_-})|z_-|^2} \cr
&=&  \frac{1}{\pi^2}\int_{0}^{\infty}    
r_{+}dr_{+}\left[e^{- r^2_+ } \right]\int_{0}^{2\pi}d\theta_{+} \times 
  \int_{0}^{\infty} r_{-}dr_{-} \left[e^{- r^2_- } \right]  \int_{0}^{2\pi}d\theta_{-}\cr  
&=&  1,  
\ena
with here $n_{\pm}=0$, where we have used the following integral 
\bea
\int_{0}^{\infty} \frac{1}{n_{\pm} !}  2r^{2n_{\pm} +1}_{\pm}dr_{\pm} e^{-r^2_{\pm}} = 1, 
\ena
ensuring that the normalization condition of the density matrix is accomplished.  The 
 right-hand side of (\ref{husimidistr00}) corresponds to the product of two harmonic oscillators
 Husimi distributions \cite{husimi}.
\subsection{Lowest Landau levels and reproducing kernel} 
Let us make a relationship between the quantum numbers  $n_{\pm}  \in \NN$ 
which  label the energy levels per sector and the quantum numbers 
 $n, m$ where $n$ labels the levels and $m$ describes the degeneracy \cite{a-hk-landauexotic}. Fixing  $n_- = 0$ (resp.  $n=0$), one obtains a 
state corresponding to the quantum number $m$ in the lowest Landau level (LLL) given by 
\bea{\label{compl10}}
\phi_{n=0,m}(z_+,\bar{z}_+) = \frac{1}{\sqrt{2\pi l^{2}_{0}m !}}\left(\frac{z_+}{\sqrt{2}l_{0}}\right)^{m}e^{-|z_+|^{2}/4l^{2}_{0}}
\ena  
\noi where $l_0 = \sqrt{\frac{1}{eB}} \equiv 1$ (with $\hbar = 1, e = 1$) is the scale of lengths associated with the Landau problem.

\noi Equivalently,  fixing $n_+= 0$ (resp.  $m=0$), 
one gets a state centered at the origin ($m=0$) in the Landau level  $n$ given by  
\bea{\label{compl11}}
\phi_{n,m=0}(z_-,\bar{z}_-) = \frac{1}{\sqrt{2\pi l^{2}_{0}n !}}\left(\frac{\bar{z}_-}{\sqrt{2}l_{0}}\right)^{n}e^{-|z_-|^{2}/4l^{2}_{0}}.
\ena
\noi Consider the projector onto   the LLL  given by 
\bea
\mathbb P_0 = \sum_{m = 0}^{\infty}|0, m)(0, m|.
\ena
\noi In the LLL $|0,m)$, the state $|0, \bar{\tilde z}_+)$, where $z_+ = x^1_+ - i x^2_+$, 
  is
{ such that} 
\bea\label{statefac00}
(0, \bar{\tilde z}_+|0,m) =  e^{-\frac{|\tilde z_{+}|^2}{2}}\frac{\tilde z^{m}_+}{\sqrt{m !}}, \qquad 
\tilde z_+ =  \frac{z_+}{l_0\sqrt{2}}.
\ena
and also
\bea
\overline{(0, \bar{\tilde z}_+|0,m)} = (0,m |0, \bar{\tilde z}_+) =  e^{-\frac{|\tilde z_{+}|^2}{2}}\frac{\bar{\tilde z}^{m}_+}{\sqrt{m !}}.
\ena
\noi The matrix elements of the  projector $\mathbb P_0$ are obtained as
\bea
(0, \bar{\tilde z}_+|\mathbb P_0|0, \bar{\tilde z}'_+) = 
e^{-\frac{1}{2}[|\tilde z'_{+}|^2 + |\tilde z_{+}|^2 - 2\tilde z_+  \bar{\tilde z}'_+]}.
\ena
Let  $|\psi) \in \mathcal H_q$,  a  state given on the LLL by 
\bea
|\psi) = \sum_{m = 0}^{\infty}a_{m}|0, m),  \, a_{m} \in \CN.
\ena
\noi We obtain that $|\psi)$
is analytic up to the Landau gaussian factor $e^{-\frac{|\tilde z_{+}|^2}{2}}$ as follows:
\bea\label{func00}
(0, \bar{\tilde z}_+|\psi)  = e^{-\frac{|\tilde z_{+}|^2}{2}} f(\tilde z_+), \qquad 
f(\tilde z_+) = \sum_{m = 0}^{\infty}a_{m}\frac{\tilde z^{m}_+}{\sqrt{m !}}  \in L^{2}_{hol}(\CN, d\nu(z,\bar  z))
\ena
with $d\nu(z,\bar  z) = \frac{e^{-|z|^{2}}}{2\pi}\frac{d {z} \wedge d\bar z}{i}$.
\noi Next, let us define the projection operator 
\bea{\label{proj01}}
 \mathbb P_{hol}: L^{2}(\CN, d\nu(z,\bar  z)) \longrightarrow L^{2}_{hol}(\CN, d\nu(z,\bar  z))
\ena
\noi which is an integral operator with  the reproducing kernel
\bea\label{func01}
K(\tilde z_+, \bar{\tilde z}'_+)  = e^{\frac{1}{2}[|\tilde z'_{+}|^2 + |\tilde z_{+}|^2]}(0, \bar{\tilde z}_+|\mathbb P_0|0, \bar{\tilde z}'_+) 
= e^{\tilde  z_+ \bar{\tilde z}'_+}
\ena 
\noi for $L^{2}_{hol}(\CN, d\nu(z,\bar  z))$\cite{ali-antoine-gazeau}. $L^{2}_{hol}(\CN, d\nu(z,\bar  z))$ is  the  subspace of the Hilbert 
space  $L^{2}(\CN, d\nu(z,\bar  z))$  of $d\nu$-square integrable holomorphic functions on $\CN$ in the variable $z$. Then,  given an operator 
$\mathcal O$ on $L^{2}(\CN, d\nu(z,\bar  z))$ and $f \in L^{2}_{hol}(\CN, d\nu(z,\bar  z)), $ we have
\bea
\frac{1}{\pi}\int_{\CN} e^{\tilde  z_+ \bar{\tilde z}'_+}
 (\mathcal O f)(\tilde z'_+)e^{-|\tilde z'_{+}|^2} d^{2}\tilde z'_{+} &=& \int_{\CN}e^{\tilde  z_+ \bar{\tilde z}'_+}(\mathcal O f)(\tilde z'_+)  
e^{-|\tilde z'_{+}|^2}
\frac{d{\tilde z'_+} \wedge d\bar{\tilde z}'_+}{2i \pi }
\cr
& =: & (\mathbb P_{hol} \mathcal O   f)(\tilde z_+).
\ena
\subsection{Statistical properties}
 Let us consider the  operators given  on $\mathcal  H_{q} \otimes \mathcal  H_{q}$ by
\bea
\hat P_{X} = \frac{-i \hbar }{\sqrt{2 \theta}}[a_{R} - a^{\dag}_{R}, \ .]
\qquad
\hat P_{Y} =  \frac{-\hbar }{\sqrt{2 \theta}}[a_{R} + a^{\dag}_{R}, \ .]
\ena
\bea
\hat X  =  \sqrt{\frac{\theta}{2}}[a_{R}+a^{\dag}_{R}] \qquad
\hat Y = i \sqrt{\frac{\theta}{2}}[a^{\dag}_{R} - a_{R}].
\ena
 From (\ref{vec}), we obtain in a state 
$|\tilde n \rangle  \langle \tilde m| \otimes | m \rangle\langle n| \in \mathcal  H_{q} \otimes \mathcal  H_{q}$ 
\bea
[a_{R} - a^{\dag}_{R}, \ |\tilde n \rangle  \langle \tilde m| \otimes | m \rangle\langle n|] =
\sqrt{n+1}|\tilde n \rangle  \langle \tilde m| \otimes |m \rangle \langle n+1| -
\sqrt{n}|\tilde n \rangle  \langle \tilde m| \otimes  |m \rangle \langle n|.
\ena
 We get
the following expressions:
\bea
(\Delta \hat X)^{2}
 = \frac{\theta}{2}, \qquad 
(\Delta \hat Y)^{2}
= \frac{\theta}{2}
\ena
\bea{\label{quad003}}
(\Delta \hat P_{X} )^{2}
= \frac{\hbar^{2} }{\theta}, \qquad 
(\Delta \hat P_{Y} )^{2} = \frac{\hbar^{2} }{\theta}
\ena
leading to the following uncertainties:
\bea
[\Delta \hat X \Delta \hat Y]^{2} &=&  \frac{\theta^{2}}{4} =  
\frac{1}{4}|\langle [\hat X, \hat Y] \rangle|^{2},  \cr
[\Delta \hat X  \Delta \hat P_{X}]^{2}  &=&    \frac{\hbar^{2} }{2} \geq  
 \frac{1}{4}|\langle [\hat X, \hat P_{X}]  \rangle|^{2},  \cr
[\Delta \hat Y \Delta \hat P_{Y}]^{2}  & = &   \frac{\hbar^{2} }{2}  
 \geq  \frac{1}{4}|\langle [\hat Y, \hat P_{Y}]  \rangle|^{2}, \cr
[\Delta \hat P_{X} \Delta \hat P_{Y}]^{2} &= & \frac{\hbar^{4} }{4\theta^{2}} \geq 
\frac{1}{4}|\langle [\hat P_{X}, \hat P_{Y}]  \rangle|^{2} = 0.
\ena
\section*{Concluding remarks}
\noi We have first dealt with some preliminaries about definitions, and remarkable properties on Hilbert-Schmidt operators and the 
Tomita-Takesaki modular theory. Then, the construction of  CS built from the thermal state has been achieved and discussed, with the resolution 
of
the identity.  Besides, some detailed proofs have been provided in the study of the modular theory and Hilbert-Schmidt operators. 
The  relation  between the
noncommutative quantum mechanics formalism  and the modular theory, both using Hilbert-Schmidt operators, has been evidenced by the use  of the
Wigner map as an interplay between them.  The formalism has been illustrated with the  physical model of a charged particle
on the flat plane xy in the presence of a constant magnetic field along the
z-axis with a harmonic potential. CS have been constructed. 
Then, the density matrix, 
the projection onto the lowest Landau level (LLL) and main statistical properties  have been discussed on  the CS basis.  

 \section*{Acknowledgments} This work is supported by TWAS Research Grant RGA No. 17-542 RG/MATHS/AF/AC\_G -FR3240300147. The ICMPA-UNESCO Chair is in partnership with Daniel Iagolnitzer Foundation (DIF), France, supporting the development of mathematical physics in Africa.

\end{document}